\documentclass[twocolumn,usenatbib]{aastex63}

\usepackage{latexsym}
\usepackage{longtable}
\usepackage{epsf}

\usepackage{graphicx}	
\usepackage[caption=false]{subfig}
\usepackage{amsmath}	
\usepackage{amssymb}	
\usepackage[shortlabels]{enumitem}
\usepackage{xcolor}
\usepackage{url}    
\usepackage{hyperref}
\usepackage{tensor}

\definecolor{darkgreen}{RGB}{0,128,0}

\newcommand{\norm}[1]{\lVert #1 \rVert}
\DeclareMathOperator*{\argmin}{argmin} 

\shorttitle{EM signatures from a BH-NS tidal tail}
\shortauthors{Darbha et al.}

\graphicspath{{./}{figures/}}

\begin{document}

\title{Electromagnetic Signatures from the Tidal Tail of a Black Hole -- Neutron Star Merger}

\correspondingauthor{Siva Darbha}
\email{siva.darbha@berkeley.edu}

\author{Siva Darbha}
\affiliation{Department of Physics, University of California, Berkeley, Berkeley, CA 94720, USA}

\author{Daniel Kasen}
\affiliation{Department of Physics, University of California, Berkeley, Berkeley, CA 94720, USA}
\affiliation{Department of Astronomy and Theoretical Astrophysics Center, University of California, Berkeley, Berkeley, CA 94720, USA}
\affiliation{Nuclear Science Division, Lawrence Berkeley National Laboratory, Berkeley, CA 94720, USA}

\author{Francois Foucart}
\affiliation{Department of Physics and Astronomy, University of New Hampshire, Durham, NH 03824, USA}

\author{Daniel J. Price}
\affiliation{School of Physics and Astronomy, Monash University, Vic 3800, Australia}



\begin{abstract}
Black hole - neutron star (BH-NS) mergers are a major target for ground-based gravitational wave (GW) observatories. A merger can also produce an electromagnetic counterpart (a kilonova) if it ejects neutron-rich matter that assembles into heavy elements through r-process nucleosynthesis. We study the kilonova signatures of the unbound dynamical ejecta of a BH-NS merger. We take as our initial state the results from a numerical relativity simulation, and then use a general relativistic hydrodynamics code to study the evolution of the ejecta with parameterized r-process heating models. The unbound dynamical ejecta is initially a flattened, directed tidal tail largely confined to a plane. Heating from the r-process inflates the ejecta into a more spherical shape and smooths its small-scale structure, though the ejecta retains its bulk directed motion. We  calculate the electromagnetic signatures using a 3D radiative transfer code and a parameterized opacity model for lanthanide-rich matter. The light curve varies with viewing angle due to two effects: asphericity results in brighter emission for orientations with larger projected areas, while Doppler boosting results in brighter emission for viewing angles more aligned with the direction of bulk motion. For typical r-process heating rates, the peak bolometric luminosity varies by a factor of $\sim 3$ with orientation while the peak in the optical bands varies by $\sim 3$ magnitudes. The spectrum is blue-shifted at viewing angles along the bulk motion, which increases the $V$-band peak magnitude to $\sim -14$ despite the lanthanide-rich composition.
\end{abstract}



\section{Introduction}
\label{sec:intro}

Black hole (BH) - neutron star (NS) mergers are major targets for the growing network of ground-based gravitational wave (GW) interferometric detectors \citep{abbott20d}. 
Population synthesis models estimate the local BH-NS merger rate to be roughly $R \sim 1 - 100$ Gpc$^{-3}$ yr$^{-1}$ \citep{oshaughnessy08,abadie10,dominik15,mapelli18}. 
The first (O1) and second (O2) observing runs of aLIGO did not yield any detections \citep{abbott19,venumadhav19,venumadhav20}, leading to a merger rate upper bound of $\sim 610$ Gpc$^{-3}$ yr$^{-1}$ \citep{abbott19}. The third observing run (O3) with aLIGO and AdV has produced several candidate events \citep{abbott20a,abbott20b,abbott20c}, but the detections cannot be distinguished from BH-BH mergers due to missing information (e.g. tidal deformability). In the fourth observing run (O4), the HLVK network (aLIGO, AdV, and KAGRA) of 2nd generation detectors will operate at design sensitivity \citep{abbott20d}, 
and the GW detection rates are estimated to be $\dot{N}_{GW} \sim 1 - 100$ yr$^{-1}$ \citep{dominik15,abbott20d,zhu20b}, suggesting a forthcoming observation.

Like NS-NS mergers, BH-NS mergers can eject neutron-rich matter that assembles into heavy elements through rapid neutron capture (the r-process) \citep{lattimer74,symbalisty82,meyer89,eichler89,freiburghaus99}. The radioactive heating from the r-process elements powers an electromagnetic (EM) transient known as a kilonova \citep{li98,metzger10,metzger19}. Follow-up searches of BH-NS merger GW candidate events have not found any kilonova counterparts, imposing constraints on the kilonova luminosity function and ejecta mass \citep{kawaguchi20b,kasliwal20,anand20}. Current estimates place the follow-up detection rate for BH-NS dynamical kilonova at roughly $\dot{N}_{KN} \sim (0.1 - 0.5) \dot{N}_{GW}$ \citep{bhattacharya19,zhu20b}.

Numerical relativity (NR) simulations have studied the range of binary parameters over which BH-NS mergers will disperse matter and produce a kilonova counterpart \citep{etienne09,foucart12,foucart13,foucart17,foucart19,kyutoku13,kyutoku15,kawaguchi15}. In broad terms, if the BH mass is too large then the BH will absorb the NS. In more detail, we can derive simple qualitative relations by taking the BH spin aligned with the binary orbital angular momentum and treating the NS as a point mass \citep{foucart20}. The relevant parameters are then the dimensionless BH spin $\chi = a_{BH}/M_{BH}$, the BH-NS mass ratio $\mathcal{Q} = M_{BH}/M_{NS}$, and the NS compactness $\mathcal{C} = M_{NS}/R_{NS}$, where we use units $G = c = 1$. The BH will disrupt the NS at the tidal radius $r_t \sim M_{BH} \mathcal{Q}^{-2/3} \mathcal{C}^{-1}$. Matter on a quasi-circular orbit will plunge into the BH at the innermost stable circular orbit (ISCO); for the Kerr metric it can be written as $r_\mathrm{ISCO} \equiv M_{BH} f(\chi)$ where $f(\chi) \in [1,9]$ is a decreasing function of $\chi$ \citep{bardeen72}. The merger will disperse matter if $r_t / r_\mathrm{ISCO}  \sim f(\chi)^{-1} \mathcal{Q}^{-2/3} \mathcal{C}^{-1} \gtrsim 1$, i.e. if the NS is tidally disrupted before reaching the ISCO, which occurs for a combination of more prograde BH spin, smaller BH mass, and larger NS radius. 
The amount of dispersed mass can be estimated with more involved parameterizations \citep{kawaguchi16,foucart18,kruger20}. In the most common expected GW events, the NS plunges into the BH and no kilonova is produced \citep{zappa19,foucart20}.

Nevertheless, an accurate characterization of the kilonova counterpart can inform all types of candidate detections. In the absence of a kilonova, null EM observations can place constraints on the BH-NS binary properties (see above). 
If the mass ejection is limited, even a weak counterpart would help to distinguish between BH-BH and BH-NS systems in the ``mass gap'' \citep{abbott20c}. Due to the sizable GW detection rate, a fortuitous event may produce a robust mass outflow and kilonova signature. The wealth of information from an EM and/or GW detection makes BH-NS mergers, and the similar NS-NS mergers, powerful systems to study the rate of heavy element production \citep{goriely11,just15,kasen17}, constrain the NS equation of state and radius (for GW-only, see \citealt{thorne87,flanagan08,abbott18}; for combined GW-EM, see \citealt{bauswein13,bauswein17,shibata17,radice18,coughlin18}), and measure the Hubble parameter (for EM-only, see \citealt{kashyap19,coughlin20a,coughlin20b}; for GW-only, see \citealt{schutz86,delpozzo12,fishbach19}; for combined GW-EM, see \citealt{holz05,abbott17,doctor20}). BH-NS mergers may even be more profitable than NS-NS mergers as instruments to measure the Hubble parameter if the BH has spin and the merger rate is $\gtrsim 10$ Gpc$^{-3}$ yr$^{-1}$ \citep{vitale18}.

In BH-NS mergers, the ejected matter takes two general forms: (1) the dynamical outflow from the tidal disruption of the NS and (2) the wind from the post-merger accretion disk driven by viscous, magnetic, or neutrino processes. The dynamical ejecta has mass $M_d \sim (0.001 - 0.1) M_\odot$ and velocity $v_d \sim (0.1 - 0.4) c$, and the disk wind has mass $M_w \sim (0.001 - 0.1) M_\odot$ and velocity $v_w \sim (0.01 - 0.1) c$. The dynamical ejecta consists of an unbound component that is directed, asymmetric, neutron-rich ($Y_e \sim 0.1$), and highly concentrated in a plane. This ejecta differs from that of  NS-NS merger simulations, where only $M_d \sim (0.001 - 0.01) M_\odot$ is dynamically ejected and thus the disk wind is presumed to be the main contribution to the kilonova. The unbound dynamical component in NS-NS mergers is more bidirectional due to mutual tidal disruption, less neutron-rich due to neutrino irradiation from a proto-NS, and more spherical since the collision interface expels matter orthogonally.

NR simulations of BH-NS mergers are computationally intensive and typically conclude a few ms after the merger. To follow the post-merger dynamics, the unbound ejecta must be transferred to a hydrodynamics simulation that can accommodate the rapid expansion of the distance scale. Few studies have explored the long-term evolution of the unbound component while incorporating the effect of r-process radioactive heating, which can affect the evolution; as an estimate, the total radioactive energy deposited over the first $\sim 10^{-4}$ d is $E_\mathrm{rad} \sim 10^{49}$ erg \citep{metzger10}, nontrivial compared to the total kinetic energy $E_\mathrm{kin} \sim M_d v_d^2 / 2 \sim \mathrm{few} \times 10^{50}$ erg. 
\citet{fernandez15} studied the long-term behavior using Newtonian hydrodynamics with a prescription for r-process heating; they found that heating enlarges the ejecta and smooths the small-scale irregularities. 
This corroborated earlier work in the NS-NS merger context by \citet{rosswog14}, who used a Newtonian smoothed particle hydrodynamics (SPH) code with heating and a nuclear network, and additionally found that the ejecta reaches homology at the $\lesssim 1$ percent level by $\sim 10^2$ s and the abundance of nucleosynthesis products remains roughly unaffected by heating. 
These two studies used Newtonian merger simulations to initialize the unbound ejecta. 
\citet{roberts17} simulated the hydrodynamic evolution of the BH-NS merger unbound component to generate thermodynamic trajectories as inputs for a nuclear reaction network; since their main aim was to study r-process abundances, they did not study the back-reaction of the nuclear heating on the ejecta structure. 
\citet{kawaguchi20c} recently examined the NS-NS post-merger evolution using 2D axisymmetric hydrodynamics with radioactive heating. They found that heating only modestly affects the ejecta structure and hydrodynamics minimally impacts the nucleosynthesis, results dependent on the details of the NR handoff. The overall ejecta remains mildly prolate with a lanthanide-present torus and some matter falls back to the BH-disk system.

The signatures of the unbound dynamical ejecta in BH-NS mergers have been studied through numerical radiative transfer (RT) simulations \citep{roberts11,tanaka14,fernandez17,kawaguchi20a,darbha20}. End-to-end models that directly extract the output of merger simulations have examined the emission from the unbound component in isolation \citep{roberts11,tanaka14} and with the accretion disk and bound component included \citep{fernandez17}. The results show that the radiation in the UVOIR peaks at roughly $L \sim \mathrm{few} \times 10^{41} \mathrm{erg}/\mathrm{s}$ and is a factor of $\sim 2$ brighter from the pole than the direction of mass ejecta \citep{tanaka14}. The infrared light curves retain these properties when the disk and bound component are present, since the unbound component evolves largely independently, and the optical light curves are brighter from the equator than the pole due to Doppler shift effects \citep{fernandez17}.

Though the qualitative properties of previous kilonova models are generally robust, most end-to-end models have avoided using hydrodynamic simulations to evolve the ejecta, neglected the role of heating on the ejecta evolution, and made 2D smoothing approximations. Geometric models using numerical \citep{kawaguchi20a,darbha20} and (semi-)analytic \citep{kawaguchi16,barbieri19,barbieri20,zhu20a} methods have illuminated the global features of the emission; these are also based on the post-merger state, but have tunable geometric parameters. For instance, \citep{kawaguchi20a} found that global photon diffusion is subject to blocking, reprocessing, and funneling effects, which makes the dynamical ejecta brightest in the infrared bands and equatorial direction, and the post-merger wind brightest in the optical bands and polar direction. In the NS-NS merger case, \citet{grossman14} found that dynamical ejecta transients have peak bolometric luminosities of $\sim \mathrm{few} \times 10^{40}$ erg/s, are brighter for more massive tidal disruption, are a factor of $\sim 2$ brighter from the pole than the front, and for asymmetric mergers exhibit a factor of $\sim 2$ variation around the equator. Recently, \citet{kawaguchi20c} also found a factor of $\sim 2$ pole-to-equator variation in the bolometric luminosity, and that the optical emission is suppressed due to the prolate geometry, large opacity, and low heating rate. In an earlier study, \citet{roberts11} found that the R-band luminosities peak at $\sim \mathrm{few} \times 10^{41}$ erg/s and show a factor of $\sim 2$ variation with polar angle. 

In this paper, we calculate the EM emission from the unbound dynamical component of a BH-NS merger. We make several new contributions in our approach. In particular, we (1) examine a BH-NS binary with the initial BH spin misaligned with the initial binary orbital angular momentum; (2) evolve the hydrodynamics using a general relativistic (GR) SPH code, avoiding artifacts generated when converting the NR output to Newtonian SPH input; (3) include a prescription for r-process heating in the hydrodynamic stage and quantify its effect on the ejecta; and (4) perform a full 3D Monte Carlo radiative transfer calculation on the ejecta, avoiding 2D smoothing approximations, and obtain light curves and spectra over all viewing angles.

\section{Simulation Methods}
\label{sec:simulation_methods}

We run a sequence of simulations consisting of the following stages: \begin{enumerate}[(1), labelindent=0pt, leftmargin=*, itemsep=0pt]
    
    \item Numerical relativity snapshot ($t = t_\mathrm{nr,f}$): We interpolate the unbound post-merger mesh data from \textsc{spec} NR simulations into Lagrangian fluid parcels (``particles'').
    
    \item Hydrodynamics ($t_\mathrm{nr,f} < t \leq t_\mathrm{hd,f}$): We evolve the particles using the GRSPH code \textsc{phantom} with a prescription for r-process heating until the onset of homologous expansion.
    
    \item Radiative transfer ($t_\mathrm{hd,f} < t \leq t_\mathrm{rt,f}$): We interpolate the SPH particles onto a spatial grid and input it into the MCRT code \textsc{sedona} to calculate the EM emission assuming homologous expansion and r-process energy deposition.

\end{enumerate}
We find that homologous expansion $r = v t_\mathrm{homol}$ is sufficiently achieved for $t_\mathrm{homol} \simeq t_\mathrm{hd,f} = 10$ s (Section \ref{subsec:results_hydrodynamics}). We present some general features of our setup (Sections \ref{subsec:metric} - \ref{subsec:heating_rate}), then discuss each stage in detail (Section \ref{subsec:numerical_relativity} - \ref{subsec:methods_radiative_transfer}).

\subsection{Metric}
\label{subsec:metric}

Stages (1) and (2) require a choice of metric and coordinate system. We write the metric $g$ with signature $(-,+,+,+)$. We use the geometric units $G = c = 1$ unless otherwise noted. We label tensors in abstract index notation using early Latin indices $a,b,\hdots$, or by their symbols alone if the context is clear. We use Greek indices $\mu,\nu,\hdots$ to label components over the full spacetime $\mu=0,1,2,3$, and middle Latin indices $i,j,\hdots$ to refer to spatial components $i=1,2,3$.

The \textsc{spec} simulation dynamically evolves the metric during the merger. At several milliseconds after the merger, the remnant BH dictates the spacetime and BH spin effects are negligible (Section \ref{subsec:numerical_relativity}). We thus model the gravity of the post-merger BH using the Schwarzschild metric \citep{chandrasekhar83,wald84}. The associated line element is
\begin{equation}
    \begin{split}
        ds^2 = & -\left(1 - \frac{2M}{r}\right) dt^2 + \left(1 - \frac{2M}{r}\right)^{-1} dr^2 \\
        & + r^2 d\theta^2 + r^2 \sin^2\theta d\phi^2 ,
    \end{split}
\label{eq:schwarzschild_metric}
\end{equation}
where $M$ is the BH mass and $x^\mu = (t,r,\theta,\phi)$ are Schwarzschild coordinates. It is also convenient at times to use rectangular coordinates $x^\mu = (t,x,y,z)$ with spatial components
\begin{align}
x &= r \sin\theta \cos\phi , \\
y &= r \sin\theta \sin\phi , \\
z &= r \cos\theta .
\end{align}
In Appendix \ref{sec:3_plus_1}, we write the Schwarzschild metric as a 3+1 decomposition and express some relevant quantities in the framework, notably the 4-velocity $u^a$, the coordinate 3-velocity $v^i$, the Eulerian 3-velocity $\bar{v}^i$, and the Lorentz factor $\Gamma$.

In the asymptotic region $r \gg 2M$, the metric and 3-velocities can be written as
\begin{align}
    g_{\mu\nu} &= \eta_{\mu\nu} + O\left(\frac{1}{r}\right) , \\
    v^i &= \bar{v}^i + O\left(\frac{1}{r}\right) ,
\end{align}
where $\eta_{\mu\nu} = \operatorname{diag} \left( -1, 1, r^2, r^2 \sin^2 \theta \right)$ is the Minkowski metric in spherical coordinates $x^\mu = (t,r,\theta,\phi)$ and $\frac{1}{r} \ll 1$. In stage (3), the particles move in this asymptotic region and the zeroth order terms will suffice.

\subsection{Heating rate}
\label{subsec:heating_rate}

Stages (2) and (3) require a prescription for r-process heating. In the r-process, neutron-rich nuclei form until neutrons are depleted at $\sim 1 - 10$ s after merger and newly-formed nuclei decay through various pathways \citep{metzger10}. Detailed nuclear reaction network calculations show that the heating rate $Q(t)$ has the same general structure for ejecta with different thermodynamic conditions and nuclear models \citep{metzger10,lippuner15}. The overall heating scale is largely determined by the initial electron fraction $Y_e = n_e / (n_e + n_n)$, where $n_e$ and $n_n$ are the number densities of electrons/protons and neutrons, respectively. Networks with lower $Y_e$ have larger integrated heating since they deplete their higher neutron densities over a longer time and form more neutron-rich nuclei. The shape of $Q(t)$ also depends on $Y_e$. At late times $t \gtrsim 10^{-4}$ d, neutron captures have ceased and the total radioactive power is dominated by the $\beta$-decay of an ensemble of heavy nuclei, and can be roughly approximated as $Q(t) \propto t^{-1.3}$ for networks with $Y_e \lesssim 0.3$ \citep{metzger10,wanajo14,hotokezaka17}. For M14M5S9I60, the ejecta has $Y_e \sim 0.01 - 0.1$ (Section \ref{subsec:numerical_relativity}); we consider network calculations in this range. 

We adopt a parameterized function $Q(t)$ for the specific heating rate, which approximates the more rigorous numerical heating rates from network calculations and provides a simple means to modify the heating properties. We examine several heating models in stage (2) but use a single heating model in stage (3); though this approach is physically inconsistent, it allows us to study how heating dynamically affects the ejecta geometry in the hydrodynamic stage, and yet fix the heating that powers the light curves in the radiative transfer stage. We write the specific energy deposition rate as $q(t) = f(t) p(t) Q(t)$, where $p$ is the fraction of decay products that thermalize and $f$ is the efficiency of the thermalizing products. 

\begin{figure}
\includegraphics[width=0.47\textwidth]{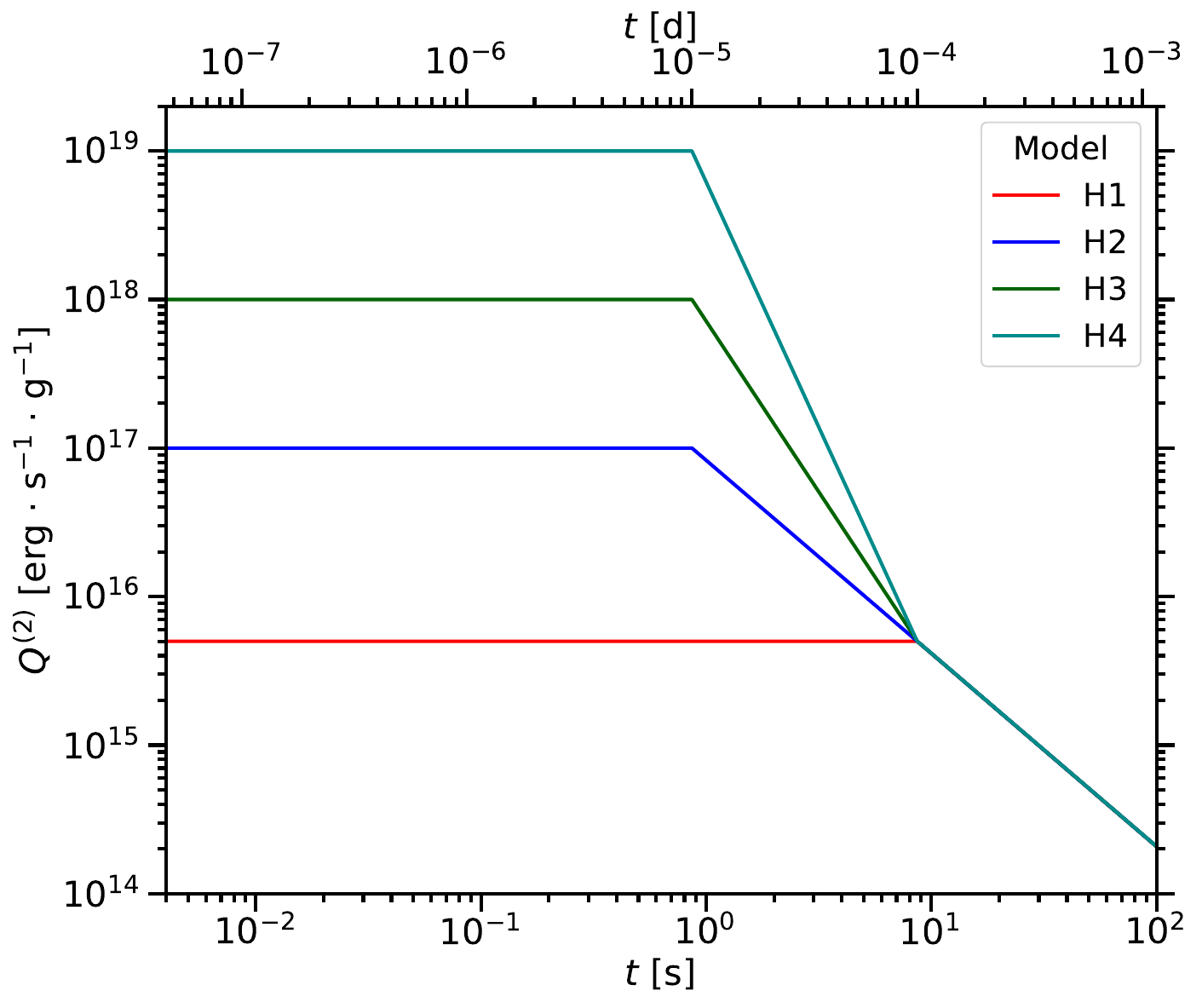}
\caption{The stage (2) analytic r-process heating rates $Q^{(2)}(t)$ for the models presented in Table \ref{tab:heating_models}. The heating rate is given in Equation \ref{eq:heating_rate_stage_2}. The model H0 (no heating) is not shown. The curves for H1 -- H4 all coincide for $t \geq t_{b2}$.}
\label{fig:heating_models}
\end{figure}

In stage (2), we use
\begin{equation}
    Q^{(2)}(t) = Q^{(2)}_0
    \begin{cases}
        1 & , \, t < t_{b1} ; \\
        \left(\frac{t}{t_{b1}}\right)^{\alpha_1} & , \, t_{b1} \leq t < t_{b2} ; \\
        \left(\frac{t_{b2}}{t_{b1}}\right)^{\alpha_1} \left(\frac{t}{t_{b2}}\right)^{\alpha_2} & , \, t_{b2} \leq t ;
    \end{cases}
    \label{eq:heating_rate_stage_2}
\end{equation}
where $Q^{(2)}_0$ is the specific heating scale, $t_{b1}$ and $t_{b2}$ are the times of the breaks, and $\alpha_1$ and $\alpha_2$ are the exponents. The break times fall in the range $t_{b1} \leq t_{b2} = 10^{-4} \,\mathrm{d} < t_\mathrm{homol} \simeq 10$ s. The exponents fall in the range $\alpha_1, \alpha_2 \leq 0$ with $\alpha_2 = -1.3$. Table \ref{tab:heating_models} summarizes several models and Figure \ref{fig:heating_models} presents a plot of the heating rates. We set $p^{(2)} = 0.5$, and $f^{(2)} \simeq 1$ since the ejecta is optically thick at these early times \citep{metzger10,barnes16,kasen19}, yielding $q^{(2)}(t) \simeq p^{(2)} Q^{(2)}(t)$. 

In stage (3), we use
\begin{equation}
    Q^{(3)}(t) = \hat{Q}^{(3)}_0 \left(\frac{t}{t_{b2}}\right)^{-\alpha_2} ,
\end{equation}
where $\hat{Q}^{(3)}_0 = 5 \times 10^{15}$ ergs s$^{-1}$ g$^{-1}$, which equals $Q^{(2)}(t_{b2})$ for H1 -- H4 (Table \ref{tab:heating_models}). We use $p^{(3)} = 0.5$ \citep{metzger10,barnes16,kasen19} and incorporate the thermalization efficiency with the ad hoc function \citep{kasen19}
\begin{equation}
    f^{(3)}(t) = \left( 1 + \frac{t}{t_e} \right)^{\alpha_3} ,
    \label{eq:thermalization_efficiency}
\end{equation}
which describes the late-time thermalization behavior of $\beta$-decay electrons. Here, $\alpha_3 \simeq -1.2$ and $t_e$ is the time at which electron thermalization becomes inefficient, which we approximate with the expression
\begin{equation}
    t_e \simeq 12.9 \left(\frac{M_\mathrm{ej}}{10^{-2} M_\odot}\right)^{2/3} \left(\frac{v_\mathrm{char}}{0.2 c}\right)^{-2} \zeta^{2/3} ,
\end{equation}
where $M_\mathrm{ej}$ is the ejecta mass, $v_\mathrm{char}$ is the ejecta characteristic velocity at $t = t_\mathrm{homol}$, and $\zeta \simeq 1$ is a constant that depends on nuclear physics. The value of $v_\mathrm{char}$ is obtained from the total relativistic kinetic energy $K_\mathrm{ej}(t_\mathrm{homol})$ by $v_\mathrm{char} \equiv (2 K_\mathrm{ej} / M_\mathrm{ej} )^{1/2}$, and depends on the NR model and $Q^{(2)}(t)$. In \citet{kasen19}, $t_e$ is defined with the maximum velocity $v_\mathrm{max}$ instead of $v_\mathrm{char}$. The deposition rate is then $q^{(3)}(t) = f^{(3)}(t) p^{(3)} Q^{(3)}(t)$.

\begin{table}
\centering
\begin{tabular}{|c|c|c|c|c|c|}
\hline
Model & $Q^{(2)}_0$ [ergs/s/g] & $t_{b1}$ [d] & $\alpha_1$ & $t_{b2}$ [d] & $\alpha_2$ \\
\hline
H0 & $0$ & $-$ & $-$ & $-$ & $-$ \\
H1 & $5 \times 10^{15}$ & $10^{-4}$ & $-$ & $10^{-4}$ & $-1.3$ \\
H2 & $1 \times 10^{17}$ & $10^{-5}$ & $-1.3$ & $10^{-4}$ & $-1.3$ \\
H3 & $1 \times 10^{18}$ & $10^{-5}$ & $-2.3$ & $10^{-4}$ & $-1.3$ \\
H4 & $1 \times 10^{19}$ & $10^{-5}$ & $-3.3$ & $10^{-4}$ & $-1.3$ \\
\hline
\end{tabular}
\caption{Models for the stage (2) analytic r-process heating rate $Q^{(2)}(t)$ given in Equation \ref{eq:heating_rate_stage_2}. Figure \ref{fig:heating_models} shows a plot of the heating rates. The model H0 (no heating) simply has $Q^{(2)}_0 = 0$. The model H4 corresponds to realistic heating for $Y_e \sim 0.1$ \citep{metzger10,lippuner15}. The models H1 -- H3 are intermediate between H0 and H4, and become more unrealistic for decreasing model number. In H1 -- H4, the late-time parameters $t_{b2} = 10^{-4}$ d and $\alpha_2 = -1.3$ are the same, and the heating scales all have $Q^{(2)}(t_{b2}) = 5 \times 10^{15}$ ergs s$^{-1}$ g$^{-1}$.}
\label{tab:heating_models}
\end{table}

\subsection{Numerical Relativity Snapshot (Stage 1)}
\label{subsec:numerical_relativity}

\citet{foucart17} (hereafter F17) performed NR simulations to study the dynamics of BH-NS mergers, obtaining the gravitational waveforms and the post-merger outflows. They systematically examined a range of merger parameters and used an equation of state (EOS) derived from the nuclear matter model DD2 \citep{typel10,hempel12}. We use the model M14M5S9I60 presented in that work, which initially has an NS mass $M_\mathrm{NS} = 1.4 M_\odot$, BH mass $M_\mathrm{BH} = 5 M_\odot$, dimensionless BH spin $\chi = 0.9$, and BH spin inclination $\iota = 60^\circ$ with respect to the orbital angular momentum of the binary. During inspiral, the BH spin and orbital angular momentum precess around the total angular momentum, and the angle between them remains roughly unchanged. At merger, the BH spin realigns due to accretion of the NS. After merger, the BH spin and the total angular momentum differ by $\lesssim 20^\circ$.

F17 simulated the mergers using the Spectral Einstein Code (\textsc{spec}; \citealt{spec}), whose functionality we briefly summarize. \textsc{spec} evolves the metric on a pseudospectral grid using the Generalized Harmonics formalism~\citep{Lindblom2006}, and the fluid equations on a finite volume grid using high-order shock capturing methods. The pseudospectral methods use adaptive mesh refinement, while the finite volume methods use nested grids focusing resolution close to the compact objects. A more detailed description of the methods used in \textsc{spec} to evolve BH-NS mergers can be found in earlier papers~\citep{Duez:2008rb,foucart13}. The simulations used here additionally include a treatment of neutrino transport in the leakage approximation \citep{deaton13}. They do not include magnetic fields. Neutrinos and magnetic fields are not expected to play a significant role in the production or properties of the dynamical ejecta.

The post-merger BH mass is roughly 
\begin{equation}
    M \simeq M_\mathrm{bin} - M_\mathrm{ej} - \Delta E ,
\end{equation}
where $M_\mathrm{bin} = M_\mathrm{BH} + M_\mathrm{NS}$ is the total mass of the binary, $M_\mathrm{ej}$ is the mass of the unbound (ejected) matter with $M_\mathrm{ej} \ll M_\mathrm{bin}, M$, and $\Delta E$ is the total energy emitted in gravitational radiation. In strict terms, the quantity $M$ that we label as the BH mass is more accurately the mass of the BH-disk system, which together determine the gravitational potential of the unbound ejecta. For M14M5S9I60, we find $M_\mathrm{ej} \simeq 0.014 M_\odot$, $\Delta E \simeq 0.27 M_\odot$, and $M \simeq 6.1 M_\odot$. In addition, the characteristic velocity of the unbound matter at $t = t_\mathrm{nr,f}$ is $v_\mathrm{c} \simeq 0.27 c$, obtained from the total relativistic kinetic energy $K_\mathrm{ej}(t_\mathrm{nr,f}) \simeq 5.6 \times 10^{-4} M_\odot c^2$ by $v_\mathrm{c} \equiv (2 K_\mathrm{ej} / M_\mathrm{ej})^{1/2}$.

\begin{figure*}
\gridline{
\fig{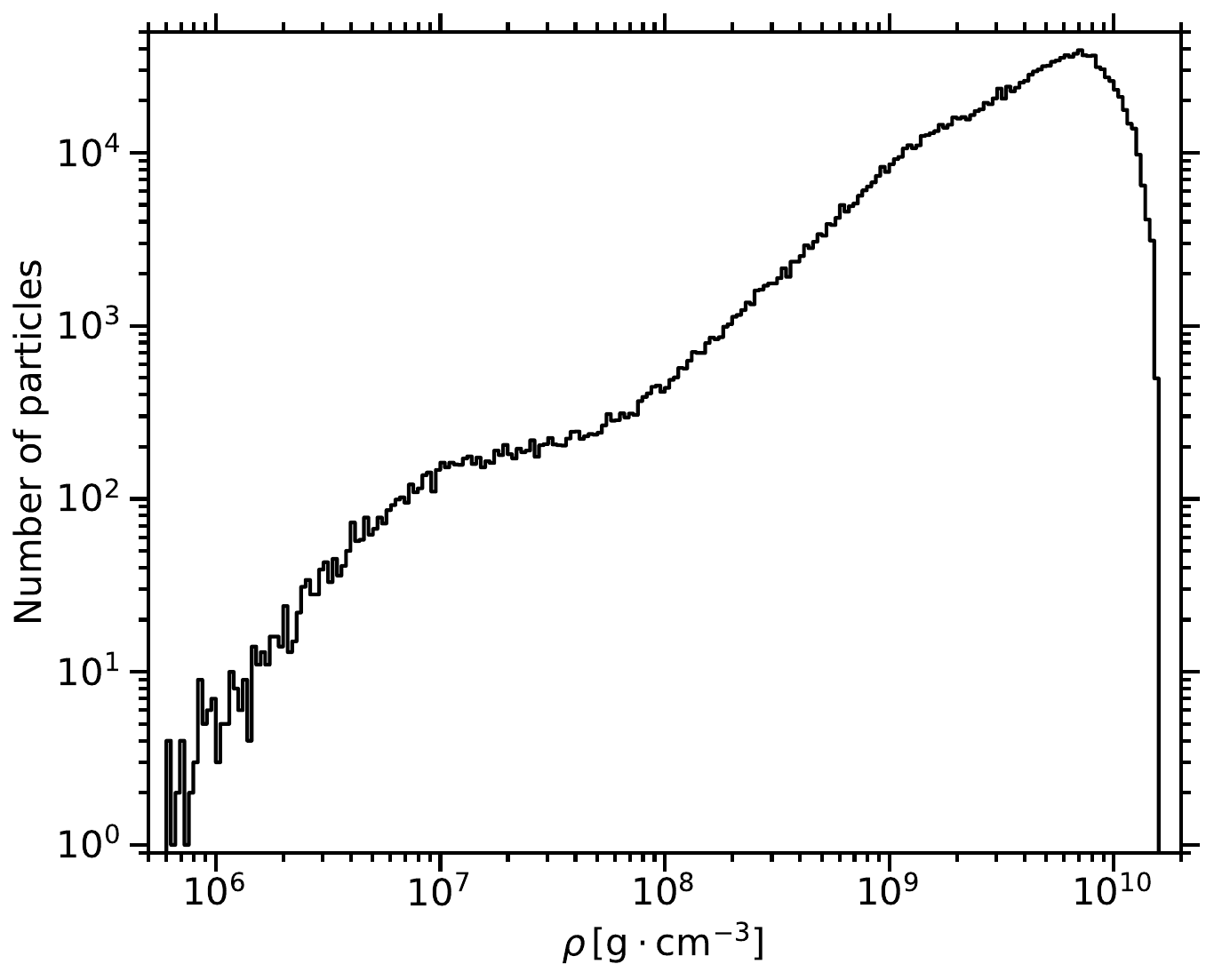}{0.32\textwidth}{(a) Rest-mass energy density $\rho$}
\fig{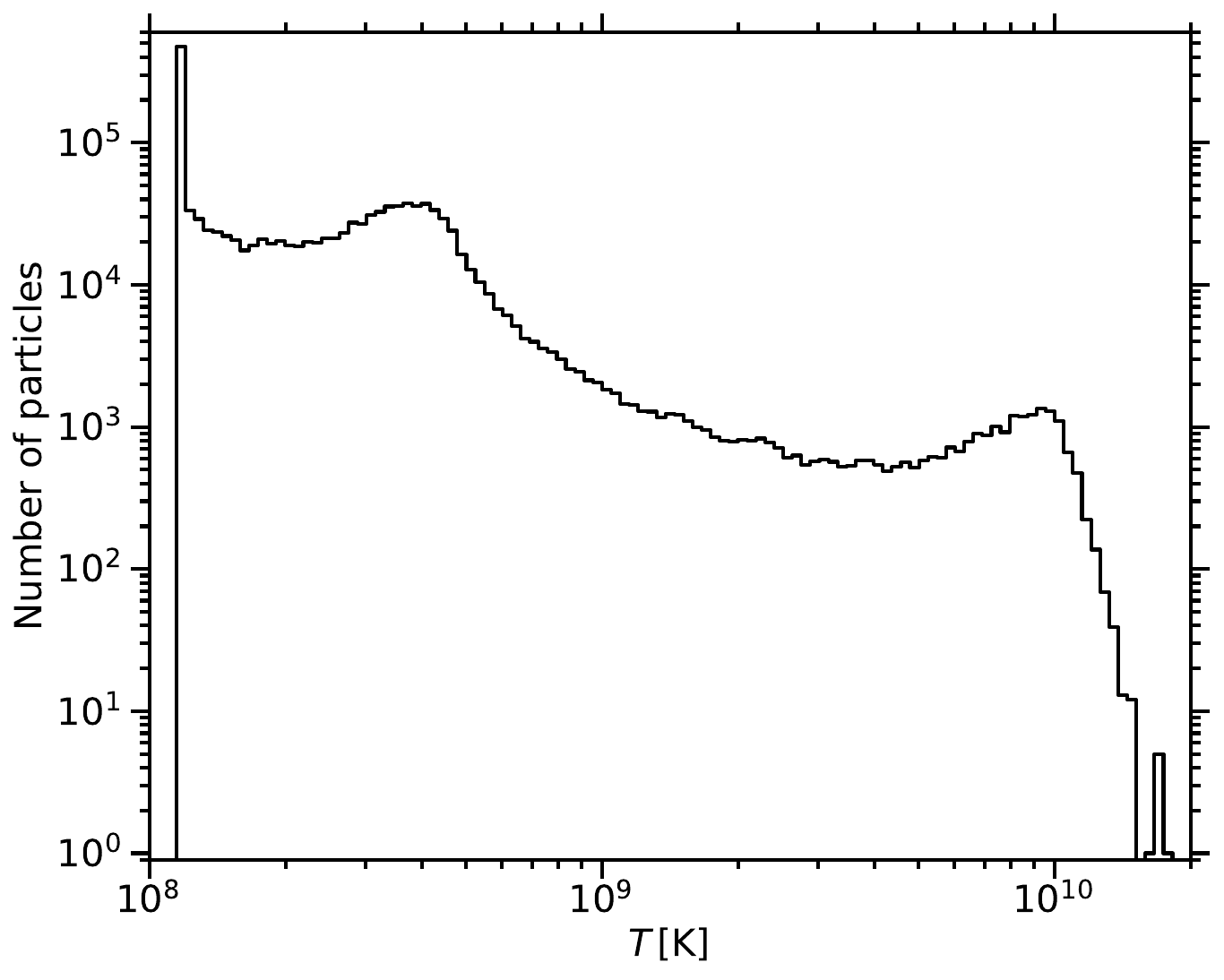}{0.32\textwidth}{(b) Rest frame temperature $T$}
\fig{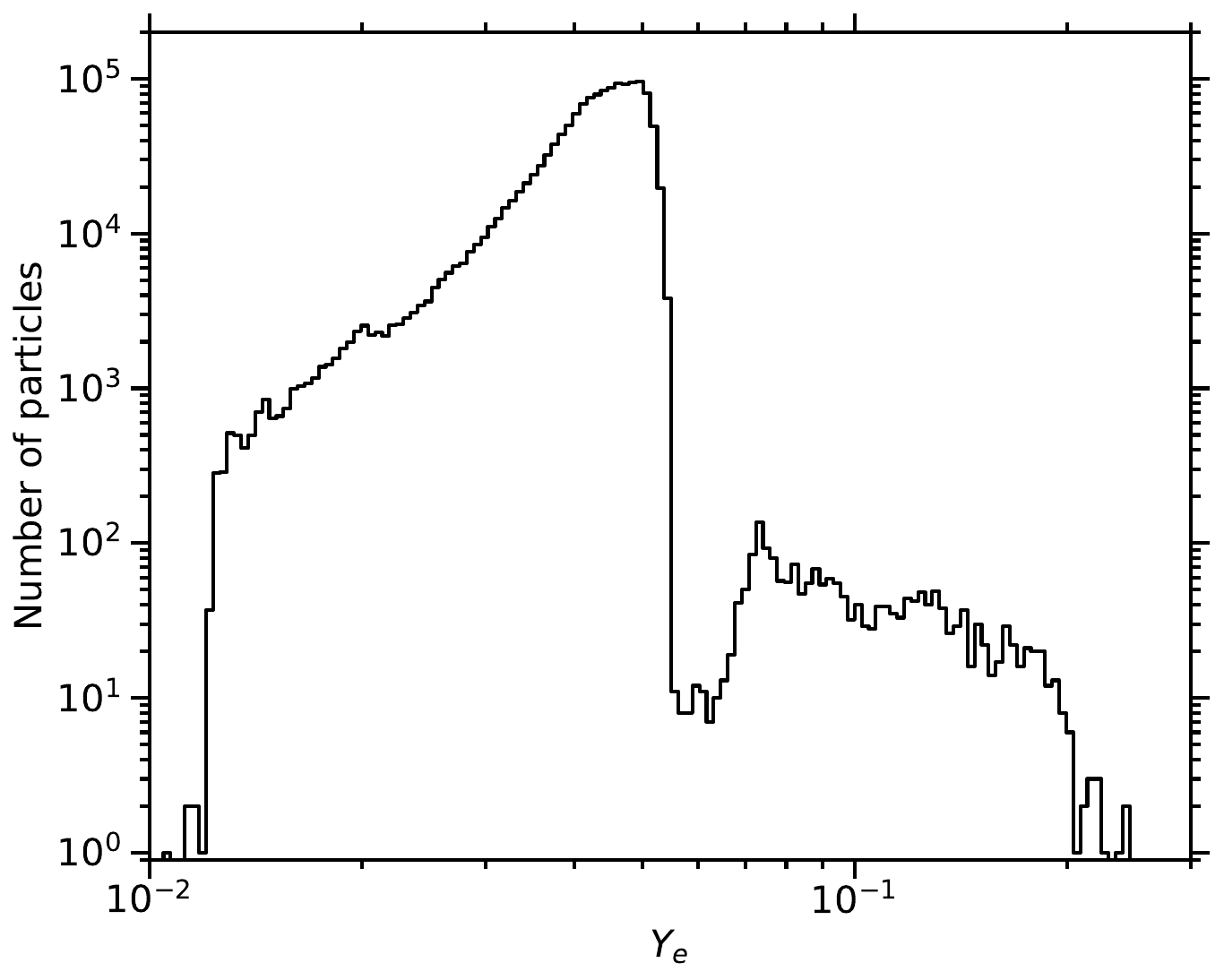}{0.32\textwidth}{(c) Electron fraction $Y_e$}
}
\caption{Distributions for the particles extracted from the \textsc{spec} simulation for merger model M14M5S9I60. The panels show (a) the rest-mass energy density $\rho$, (b) the rest frame temperature $T$, and (c) the electron fraction $Y_e$. The particles were extracted at $t_\mathrm{nr,f} = 4.5$ ms after merger. The temperature distribution has a lower bound of $T = 0.01 \mathrm{MeV} = 1.2 \times 10^8 \mathrm{K}$ which corresponds to the low-temperature cutoff in the EOS table used in the \textsc{spec} simulation. The electron fraction lies in the range $Y_e \lesssim 0.06$ for most particles and $Y_e \lesssim 0.2$ for all particles.}
\label{fig:particle_properties}
\end{figure*}

We extract the simulation data recorded at time $t_\mathrm{nr,f} = 4.5$ ms after the merger. The post-merger spacetime is still initially dynamical as it rings down before settling into a stable configuration. However, the metric is approximately spherically symmetric and static if we are far enough from the black hole and can be mapped onto the Schwarzschild metric (Section \ref{subsec:metric} and Equation \ref{eq:schwarzschild_metric}), as the BH mass dominates ($M \gg M_\mathrm{ej}$) and BH spin effects are small at the distances of the unbound debris ($2M/r < 0.1$). To any radius $r$ in the \textsc{spec} simulation, we can then associate a Schwarzschild radius $\tilde r$ by requiring that the area of the coordinate sphere of constant radius $r$ is $A=4\pi \tilde r^2$. We produce a set of Lagrangian particles $i = 1, \hdots, N$ with equal mass $10^{-8}M_\odot$ from the finite volume data by randomly drawing particles that each represent $10^{-8}M_\odot$ of matter. 
If a finite volume cell contains a mass $m_\mathrm{cell}$ of ejecta, then we produce $10^8 m_\mathrm{cell}/M_\odot$ Lagrangian particles randomly distributed in that cell. This approach easily handles fractional particle numbers; if a cell needs to create 2.3 particles, it has a $30\%$ chance of creating 3 particles and a $70\%$ chance of creating 2 particles. A particle is assigned a coordinate radius corresponding to its approximate Schwarzschild radius, and the same angular position as in the \textsc{spec} code. For each particle, we record the coordinates, four-velocity, density, temperature, entropy, and electron fraction. The speed of the ejecta is chosen so that the asymptotic kinetic energy of the particles is $-u_t$, with $u_t$ the time-component of the 4-velocity one-form in the \textsc{spec} simulation. We emphasize that the ejecta we extract consists of the unbound debris only, $E \equiv -u_t > 1$; we do not extract the accretion disk or bound debris. 
For M14M5S9I60, we obtain $N \simeq 1.4 \times 10^6$ particles. Figure \ref{fig:particle_properties} shows the density, temperature and compositional distribution of several particle quantities. Notably, the electron fraction lies in the range $Y_e \lesssim 0.06$ for most particles. The distribution has a negligible amount of mass in the range $0.06 \lesssim Y_e \lesssim 0.2$; this component is likely due to shocks generated at the disk-tail interface or noise from numerical viscosity at the edge of the tail.

The unbound post-merger \textsc{spec} particles are primarily concentrated in a plane. In general, this plane does not coincide with the plane of the post-merger BH spin or the total angular momentum, though it is within $\sim 20^\circ$. The orientation of the post-merger ejecta plane can be obtained from the \textsc{spec} simulation, but we instead calculate it from the extracted particles (Appendix \ref{sec:ejecta_plane}). For the model M14M5S9I60, we find that the optimal normal to the ejecta plane is in the direction $(\beta^\mathrm{opt}, \alpha^\mathrm{opt}) = (0.861, 1.86)$, where $(\beta^\mathrm{opt}, \alpha^\mathrm{opt})$ are the polar and azimuthal angles with respect to the orbital plane at the beginning of the \textsc{spec} simulation.

In \textsc{sedona}, we use a three-dimensional (3D) Cartesian grid, and the grid resolution can be optimized if the ejecta lies primarily in the $xy$-plane. We thus perform an active coordinate transformation to rotate the \textsc{spec} particles from their post-merger plane to the $xy$-plane, i.e. to align the direction $(\beta^\mathrm{opt}, \alpha^\mathrm{opt})$ to the $z$-axis (Appendix \ref{sec:active_rotation}).

\subsection{Hydrodynamics (Stage 2)}
\label{subsec:methods_hydrodynamics}

We evolve the post-merger ejecta using the smoothed particle hydrodynamics (SPH) code \textsc{phantom} \citep{price18}, which has been expanded to treat hydrodynamics in general relativity (GR) \citep{liptai19}. The GR hydrodynamic equations are written in a conservative Lagrangian form, which permits an SPH numerical approach with the same structure as the nonrelativistic case, including interpolation of the conserved variables using the flat space volume element \citep{siegler00,monaghan01}. The code can accommodate analytic metrics written in 3+1 form.

We describe the BH gravity using the Schwarzschild metric (Section \ref{subsec:metric} and Equation \ref{eq:schwarzschild_metric}). Since $M_\mathrm{ej} \ll M$, we ignore the ejecta self-gravity and the back-reaction on the BH. We disregard the effects of BH spin since the input particles are at sufficiently large distances ($2M/r_{(i)} < 0.1$) and are expanding outwards ($u_{(i)}^r > 0$), and we are not interested in fallback accretion at later times.

We load the equal-mass particles from the NR snapshot into the SPH simulation. For M14M5S9I60, we thus have $N \simeq 1.4 \times 10^6$ SPH particles. To initialize the SPH simulation, \textsc{phantom} only requires the particle coordinates, four-velocity, and internal energy density. The \textsc{spec} particles carry interpolated values for the thermodynamic variables, though \textsc{phantom} recomputes these from the SPH equations for self-consistency, particularly to accommodate the new and simpler EOS (Equation \ref{eq:eos_hydro}). The recomputed values are very close to the extracted values, as expected, with only minor differences. We use the $M_4$ cubic spline as our SPH kernel function \citep{price18}.

We model the ejecta as a perfect fluid with stress-energy tensor
\begin{equation}
    T_{ab} = (\rho + \epsilon) u_a u_b + p h_{ab} ,
    \label{eq:stress_energy_tensor_hydro}
\end{equation}
where $h_{ab} = g_{ab} + u_a u_b$ is the projection tensor Lorentz-orthogonal to the four-velocity $u^a$, and the quantities $\rho$, $\epsilon$, and $p$ are the rest-mass energy density, the internal energy density, and the isotropic pressure, all in the rest frame. The specific internal energy in the rest frame is then $\epsilon_s = \epsilon/\rho$. We can convert the primitive variables $(\rho, p, \epsilon_s, v^i)$ to the conservative variables $(D^*, S_i, \mathcal{E})$ given by \citep{siegler00,liptai19}
\begin{align}
    D^* &= \sqrt{-g} \frac{\Gamma}{\alpha} \rho , \\
    S_i &= w \Gamma \bar{v}_i , \\
    \mathcal{E} &= S_i v^i + \frac{\alpha (1 + \epsilon_s)}{\Gamma} ,
\end{align}
which are the relativistic conserved density, specific momentum, and specific energy, respectively. 
Here, $g$ is the metric determinant and $w = 1 + \epsilon_s + \frac{p}{\rho}$ is the specific enthalpy. 
In SPH, the variable $D^*$ (and thus $\rho$) is not set directly, but achieved by adjusting the particle placement. 

We adopt a $\gamma$-law EOS 
\begin{equation}
    p = (\gamma_\mathrm{eos} - 1) \epsilon ,
    \label{eq:eos_hydro}
\end{equation}
where $\gamma_\mathrm{eos}$ is the $\gamma$-law index. We treat the ejecta as radiation dominated and use $\gamma_\mathrm{eos} = 4/3$. 
We do not track or evolve the composition. 
We assume that the electron fraction $Y_e$ remains unchanged throughout the simulation and thus determines the nuclei formed by the end; \citet{roberts17} found that $Y_e$ is not significantly altered by weak interactions in this phase. 
This simplified EOS is sufficient to capture the coupling between the internal and kinetic degrees of freedom of the expanding ejecta. In contrast, F17 used a more detailed EOS derived from nuclear theory, which was needed to evolve the composition and to treat the large gradients in the merger simulation.

We modified \textsc{phantom} to include the stage (2) specific energy deposition rate $q^{(2)}$(t) (Section \ref{subsec:heating_rate}). 
In lieu of evolving 
$\mathcal{E}$, the code evolves an entropy-like variable $K$ defined by \citep{springel02,liptai19}
\begin{equation}
    K = \frac{p}{\rho^{\gamma_\mathrm{eos}}} ,
\end{equation}
to ensure that the internal energy remains positive. The evolution equation for $K$ is
\begin{equation}
    \frac{dK}{dt} = \frac{\gamma_\mathrm{eos} - 1}{\rho^{\gamma_\mathrm{eos} - 1}} \left( \frac{d\epsilon_s}{dt} - \frac{p}{\rho^2} \frac{d\rho}{dt} \right) ,
\end{equation}
where we set the first term to
\begin{equation}
    \frac{d\epsilon_s}{dt} = q^{(2)}(t) ,
\end{equation}
which incorporates the energy deposition. 
The function $q^{(2)}(t)$ describes the heating in the rest frame of each particle, though we parameterize it using the coordinate time $t$ instead of the proper time $\tau$, which introduces a small deviation of size $\frac{dt}{d\tau} = \Gamma \left( 1 - \frac{2M}{r} \right)^{-1/2}$. We set the cooling timestep $dt_\mathrm{cool} \rightarrow \infty$, allowing the other timesteps to dominate (force computation, Courant, etc.). The heating rate decreases rapidly, so this choice leads to slightly more energy deposited than the analytic expression, but the additional contribution is small.

We evolve the hydrodynamic simulation until the particles (1) are effectively in flat spacetime with $2M/r_{(i)} < 10^{-3}$ and (2) reach homologous expansion. 
We find that the particles satisfy these conditions at an end time of $t_\mathrm{hd,f} = 10$ s, and obtain the homologous expansion time $t_\mathrm{homol} \simeq t_\mathrm{hd,f}$ by fitting $r = v t_\mathrm{homol}$.

\subsection{Radiative Transfer (Stage 3)}
\label{subsec:methods_radiative_transfer}

We calculate the emission using the time-dependent MCRT code \textsc{sedona} \citep{kasen06}. We model photon-matter interactions assuming local thermodynamic equilibrium (LTE) and treat the ejecta as radiation dominated; these assumptions are valid at early times when the ejecta is optically thick, but break down at late times. We use the stage (3) specific energy deposition rate $q^{(3)}$(t) (Section \ref{subsec:heating_rate}). The ejecta is adiabatically expanded, with heating included, to the start time $t_\mathrm{rt,i} \geq 0.1$ days. This reduces the initial density, which allows the photon transport to occur over a viable computational time, and the temperature, so the light curve is then powered primarily by the stage (3) heating.

We interpolate the SPH data to a 3D Cartesian grid with dimensions $(n_x,n_y,n_z) = (80,80,80)$. A Cartesian grid conforms to the shape of the ejecta due to the active rotation on the particles (Section \ref{subsec:numerical_relativity}). We crop the grid such that it has tightly fitting limits, contains the bulk of the ejecta ($\gtrsim 99 \%$ of the mass), and has low-density outer regions. The grid cells have homologous velocities, $v = r/t_\mathrm{homol}$ where $t_\mathrm{homol} \simeq 10$ s. We interpolate using the S-normed SPH binning (SNSB) technique developed by \citet{rottgers18}, which conserves integrated quantities and can maintain high resolution. We compared this to the analytic technique developed by \citet{petkova18} as implemented in the \textsc{splash} visualization software \citep{price07}, which precisely interpolates SPH data to a general Voronoi grid, by interpolating the conserved density $D^*$ and computing the total mass, and found excellent agreement. In GRSPH, the interpolation integral is performed in a computational frame intended for the conserved variables. For simplicity, we interpolate the primitive variables $\rho$ and $\epsilon$ directly. This introduces a small deviation that is negligible in the asymptotic region; indeed, the total mass on the grid remains accurate.

By the end time $t_\mathrm{hd,f} = 10$ s of stage (2), the free nucleons will have formed robust quantities of r-process nuclei. The distribution of nuclei at the start of stage (3) is determined by the electron fraction $Y_e$ in the earlier stages. For M14M5S9I60, the post-merger electron fraction in stage (1) is in the range $Y_e \sim 0.01 - 0.2$ (Figure \ref{fig:particle_properties}), and we assumed that it remained unchanged in stage (2). In the homologous expansion phase, \citet{roberts17} found that the composition is not significantly altered by neutrino irradiation from the post-merger accretion disk and nucleosynthesis produces robust quantities of nuclei beyond the second r-process peak. 

The ejecta opacity is dominated by the atomic lines of these newly formed r-process nuclei. 
We use a parameterized analytic function to replicate this opacity. 
In a medium expanding rapidly and homologously, the opacity due to bound-bound line transitions can be conveniently expressed using the line expansion opacity formalism \citep{karp77,eastman93,kasen13}, in which the lines can be collected into a set of wavelength bins. The expansion opacity can be written as \citep{eastman93}
\begin{equation}
    \kappa_\mathrm{exp}(\lambda,t) = \frac{1}{ct\rho} \sum_i \frac{\lambda_i}{\Delta \lambda} \left[ 1 - e^{-\tau_{s,i}} \right] ,
\end{equation}
where the sum runs over all lines $i$ which have (rest frame) transition wavelengths $\lambda_i$ inside the bin with center $\lambda$ and width $\Delta \lambda$, and Sobolev optical depths $\tau_{s,i}$ given by \citep{sobolev60}
\begin{equation}
    \tau_{s,i} = \frac{\pi e^2}{m_e c} f_i n_1 t \lambda_i ,
\end{equation}
where $e$ is the electron charge, $m_e$ is the electron mass, $f_i$ is the oscillator strength of the transition, and $n_1$ is the number density of the lower level. Atomic structure calculations suggest that the expansion opacity has a common general shape for media with different $Y_e$ \citep{kasen13,tanaka20}, which at $t_1 = 1$ d we approximate with the piecewise function
\begin{equation}
    \kappa(\lambda,t_1) = \kappa_0(t_1) 
    \begin{cases}
        1 &, \, \lambda \leq \lambda_b ; \\
        \left(\frac{\lambda}{\lambda_b}\right)^{\sigma_1} &, \, \lambda > \lambda_b ;
    \end{cases}
    \label{eq:analytic_opacity}
\end{equation}
where $\lambda_b$ is the break wavelength and $\sigma_1$ is the exponent of the power law. For $Y_e \simeq 0.1$, we take $\kappa_0(t_1) = 10^2$ cm$^2$/g, $\lambda_b = 3 \times 10^3$ \AA, and $\sigma_1 = -2$.

\section{Results}
\label{sec:results}

\subsection{Hydrodynamics (Stage 2)}
\label{subsec:results_hydrodynamics}

\begin{figure}
\gridline{
\fig{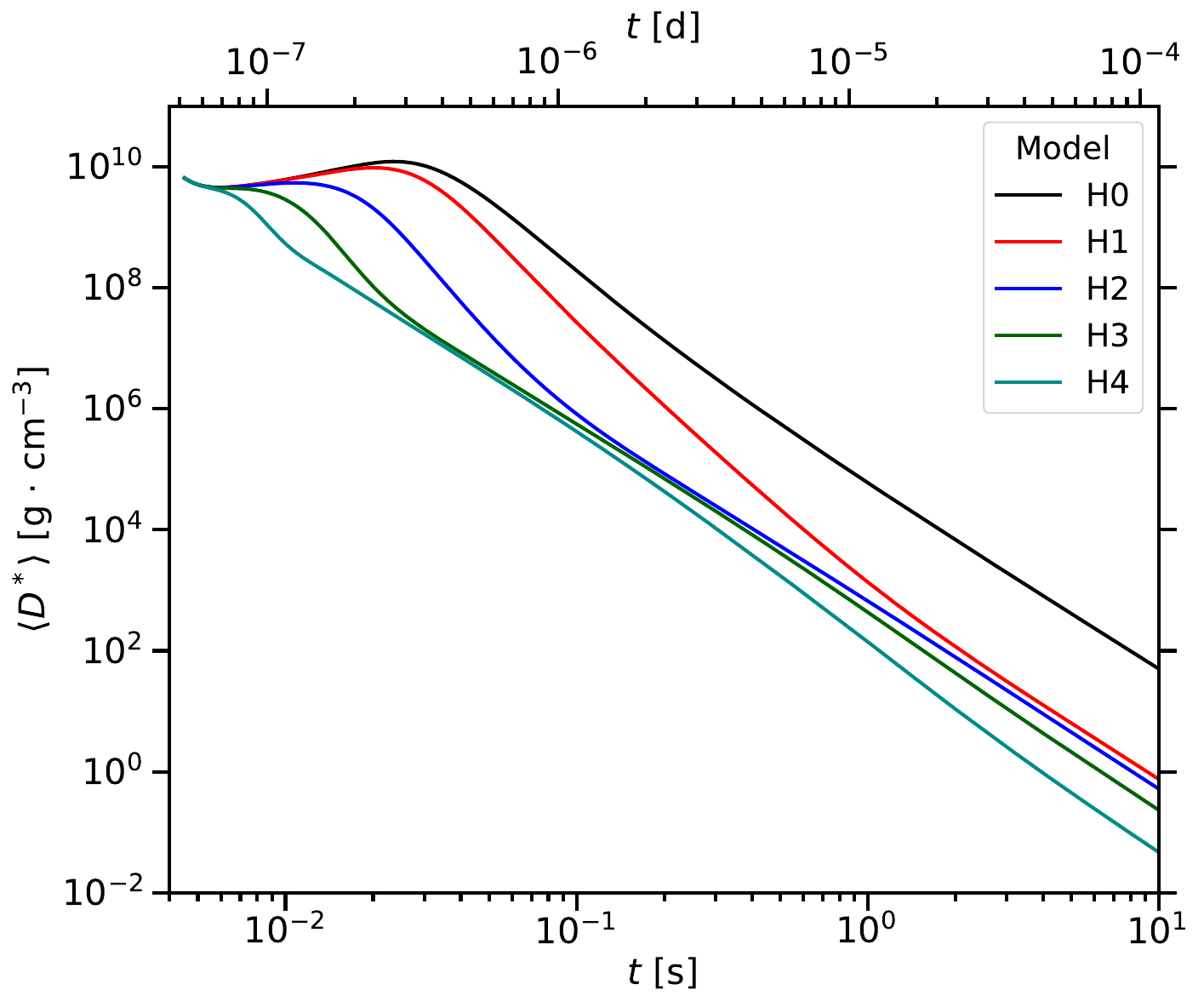}{0.46\textwidth}{(a) Mean relativistic conserved density}
}
\gridline{
\fig{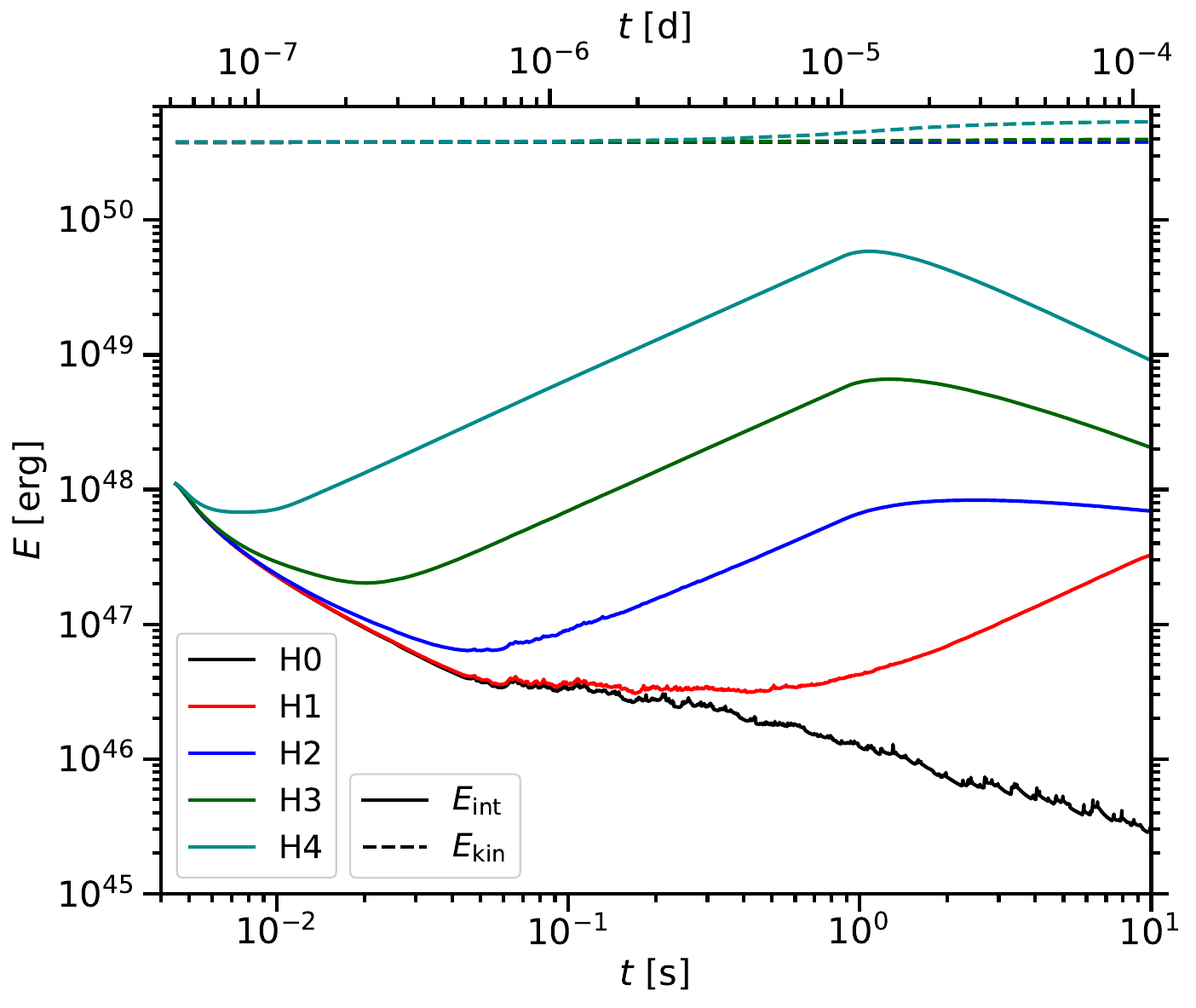}{0.46\textwidth}{(b) Total energies}
}
\caption{Hydrodynamic evolution of the ejecta parameters for merger model M14M5S9I60. The panels show (a) the mean relativistic conserved density $D^*$ and (b) the total internal ($E_\mathrm{int}$, solid) and kinetic ($E_\mathrm{kin}$, dashed) energies. The colors show the different heating models $Q^{(2)}(t)$ (Table \ref{tab:heating_models}), ranging from H0 (no heating) to H4 (the greatest heating, for $Y_e \sim 0.1$).}
\label{fig:hydrodynamic_evolution}
\end{figure}

Figure \ref{fig:hydrodynamic_evolution} shows the hydrodynamic evolution of several quantities for the various heating models. After $t \simeq 5$~s, all the ejecta have essentially reached the homologous expansion phase and scale as $\langle D^* \rangle \propto t^{-3}$; the ejecta with lower deposited heating reach this phase at earlier times. The mean density shows an early bounce as parts of the ejecta partially collide and compress; models with greater heating expand the ejecta more rapidly and exhibit a smaller bounce. The total internal energy decreases initially, then increases as heating dominates over adiabatic losses, then decreases at late times as adiabatic losses dominate. The total kinetic energy is roughly constant for H0 -- H2 since there is no or negligible heating, and it increases to a constant value for H3 and H4 as the increasing thermal pressure accelerates the ejecta until homologous expansion. We find characteristic velocities $v_\mathrm{char} = 0.17c$ for H0 and $v_\mathrm{char} = 0.20c$ for H4. 

\begin{figure*}
\gridline{
\fig{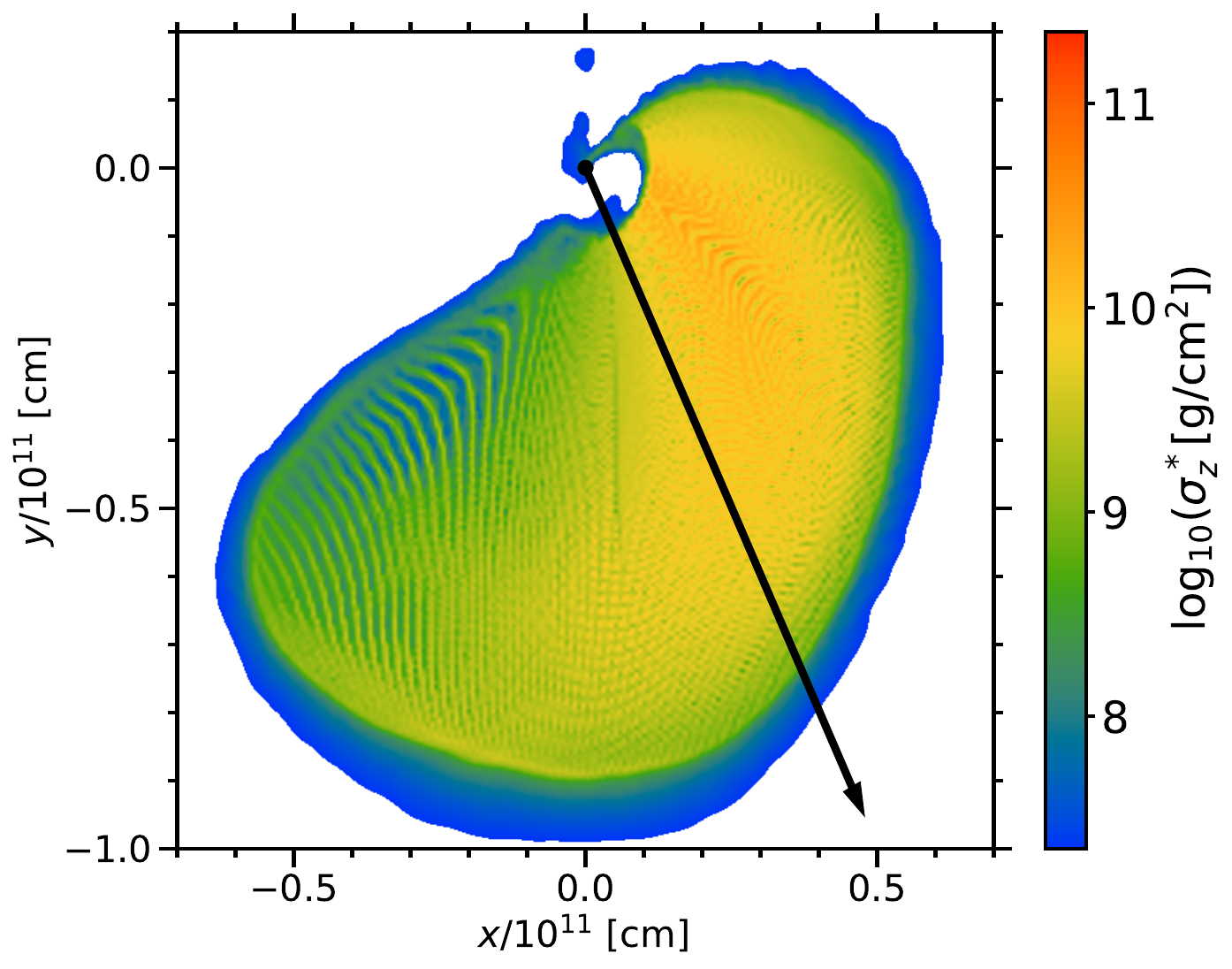}{0.32\textwidth}{(a) H0, $xy$}
\fig{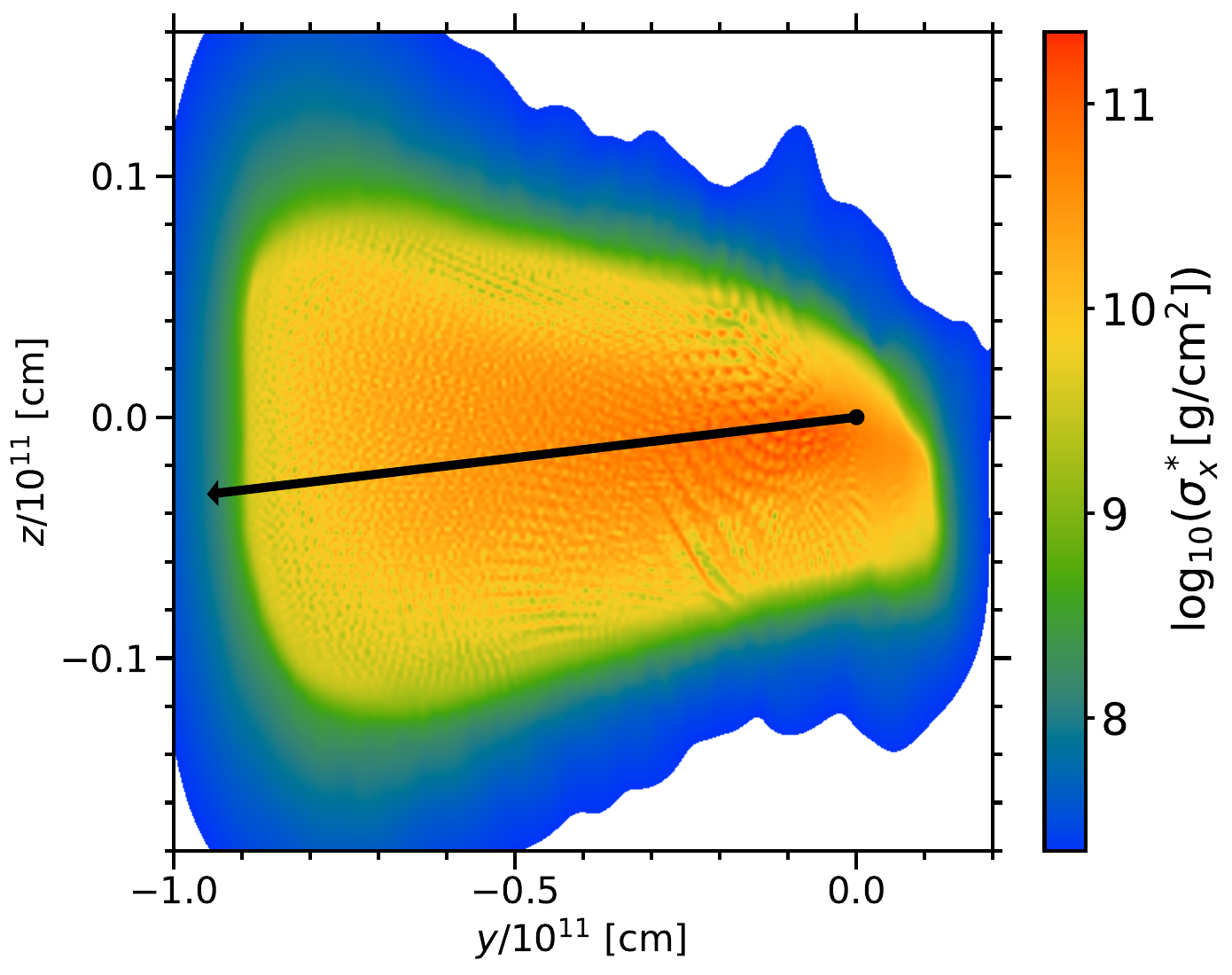}{0.32\textwidth}{(b) H0, $yz$}
\fig{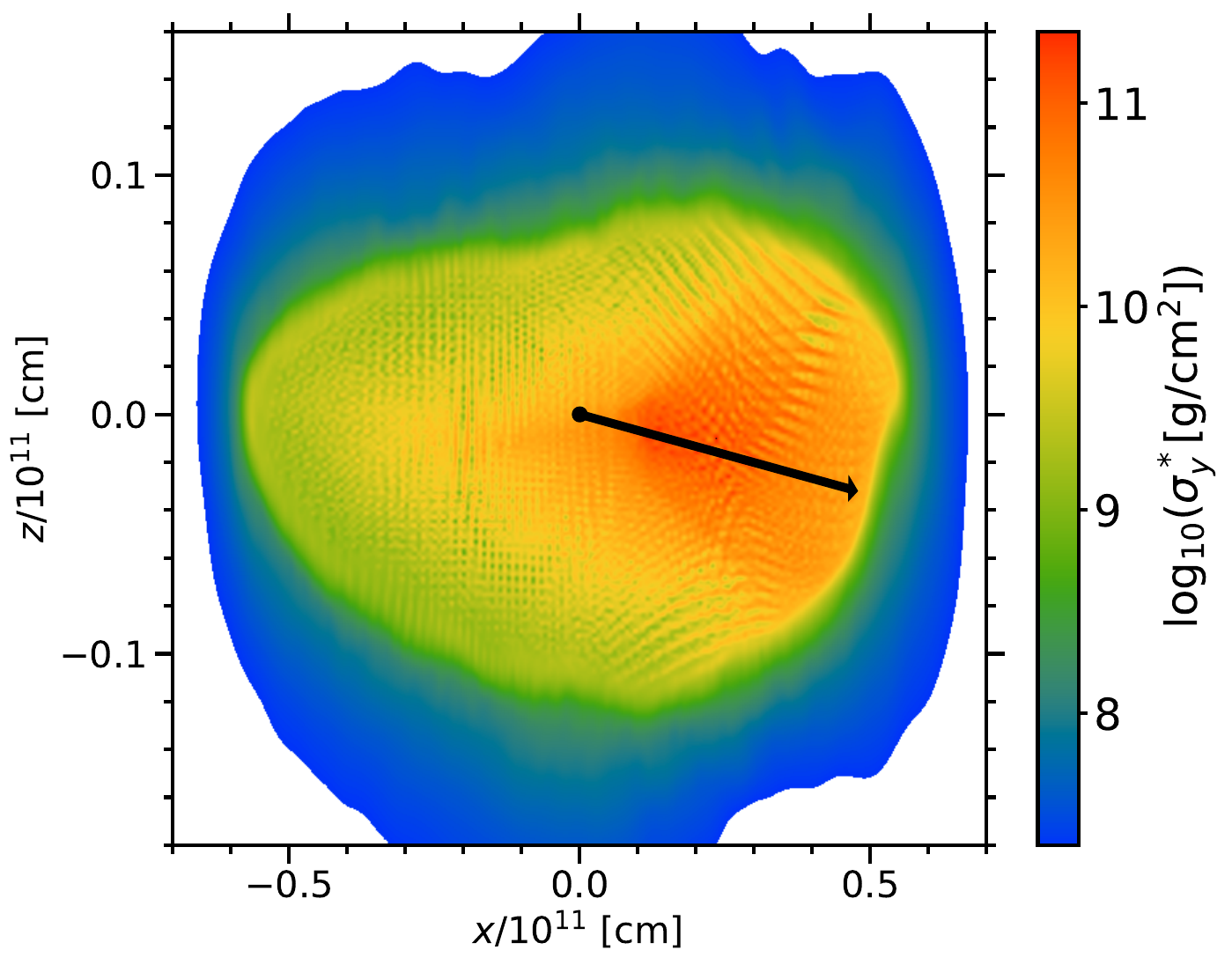}{0.32\textwidth}{(c) H0, $xz$}
}
\gridline{
\fig{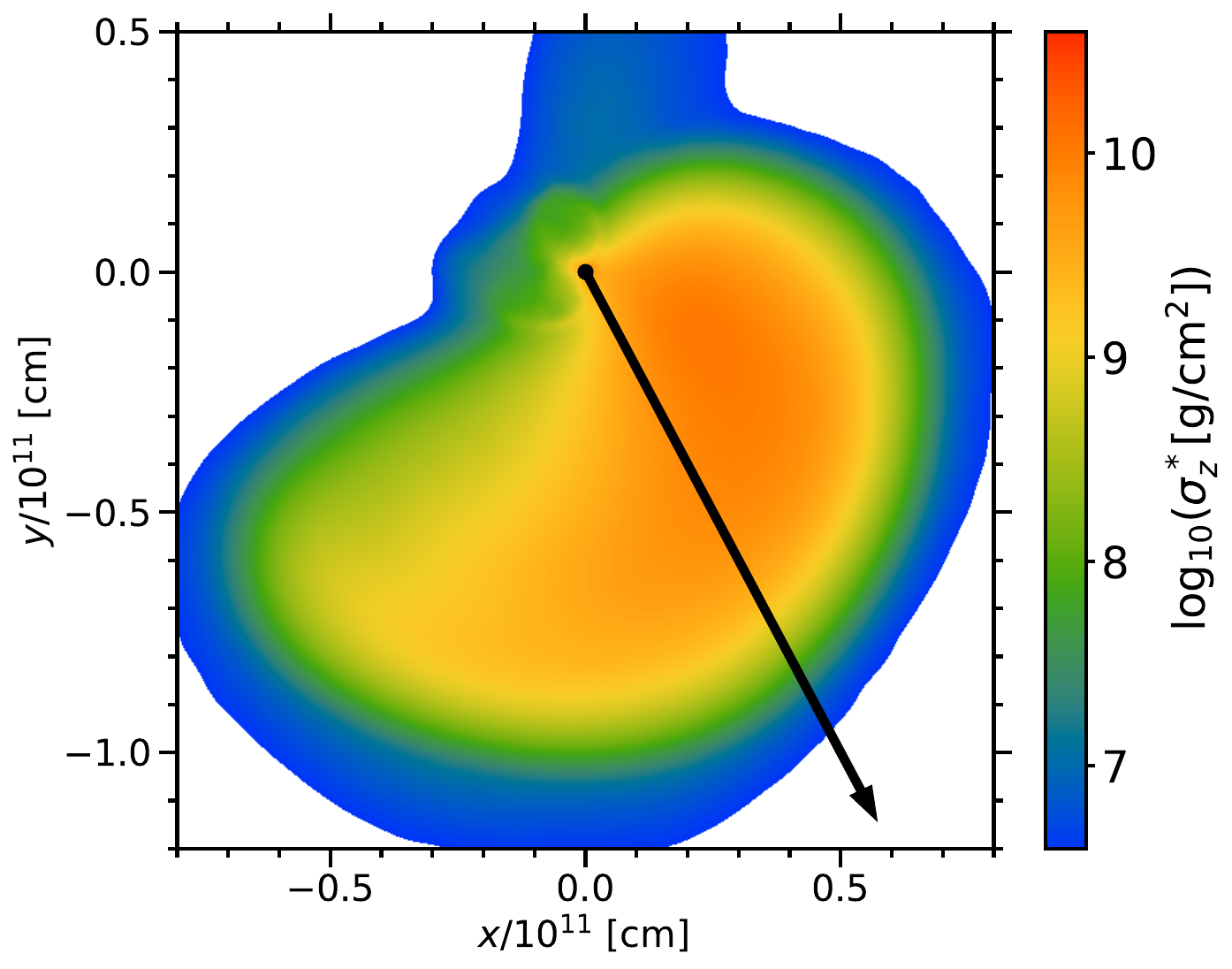}{0.32\textwidth}{(d) H3, $xy$}
\fig{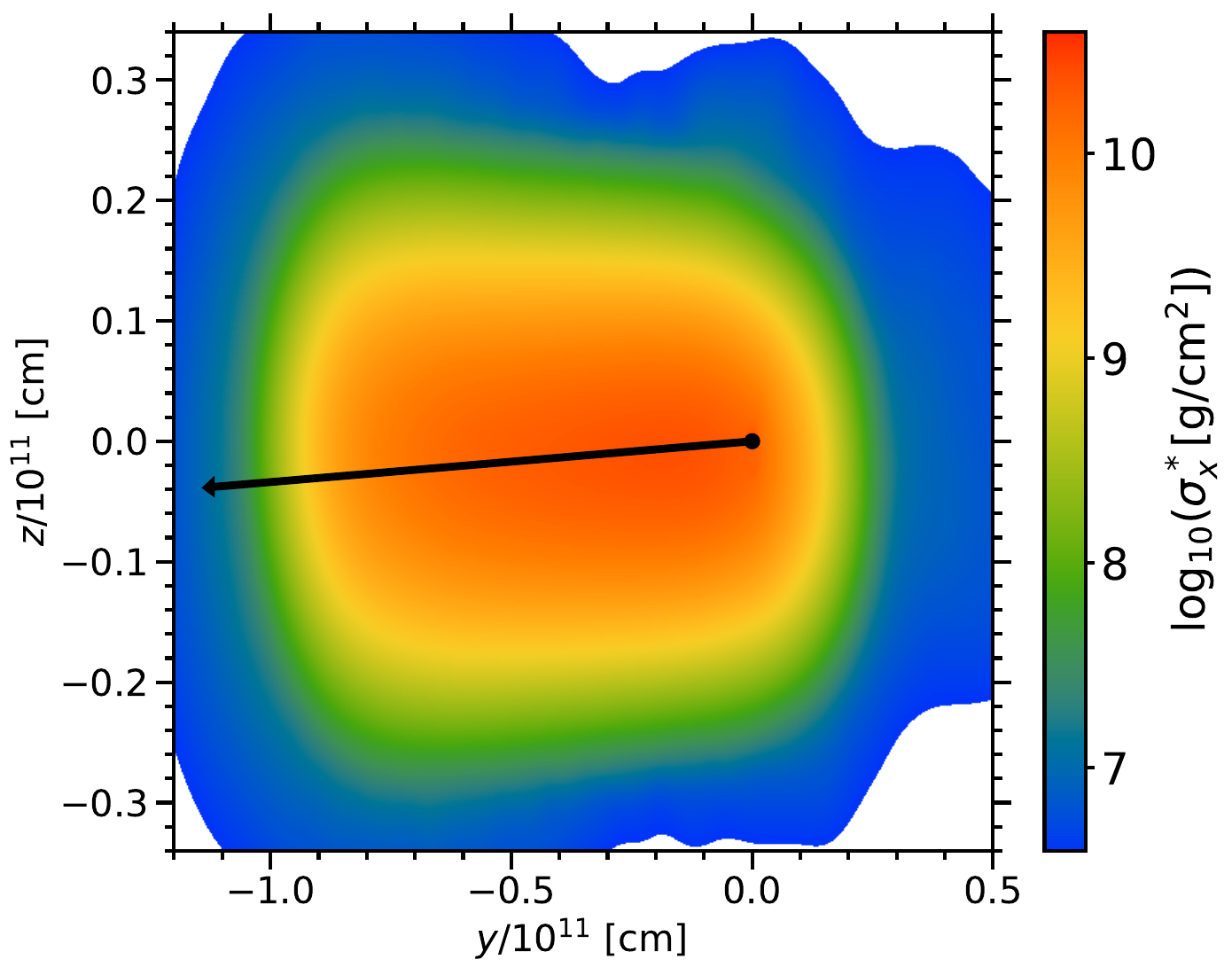}{0.32\textwidth}{(e) H3, $yz$}
\fig{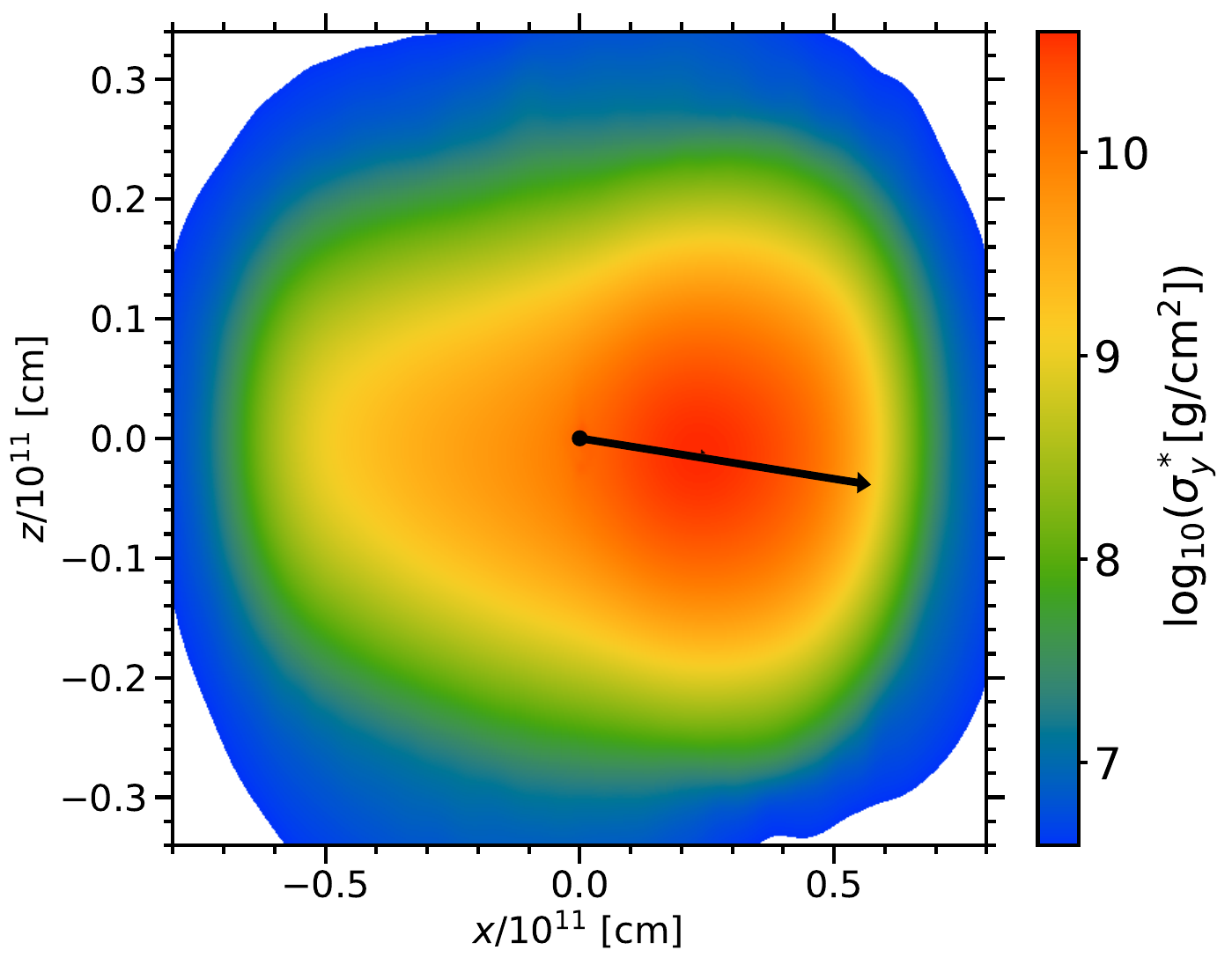}{0.32\textwidth}{(f) H3, $xz$}
}
\gridline{
\fig{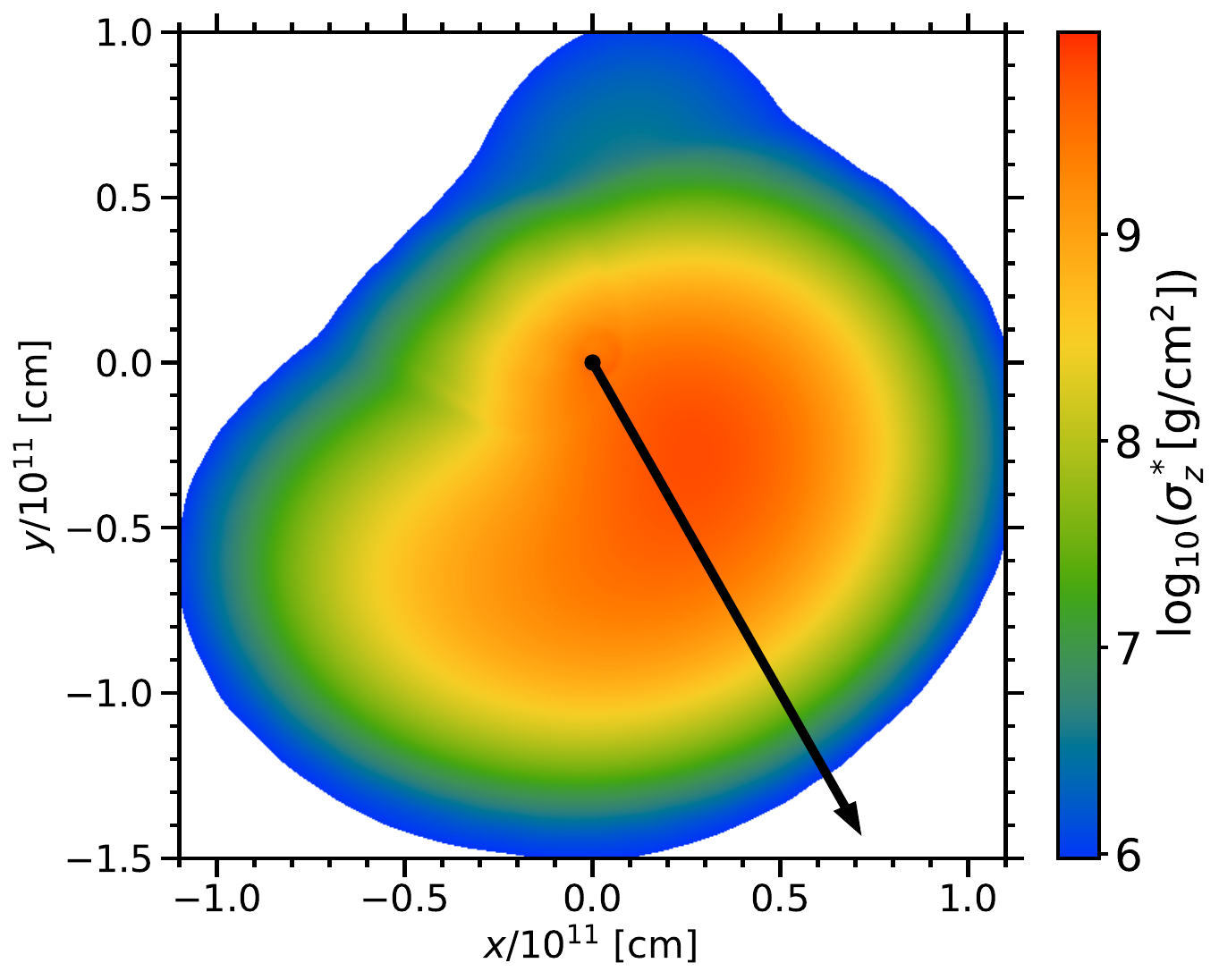}{0.32\textwidth}{(g) H4, $xy$}
\fig{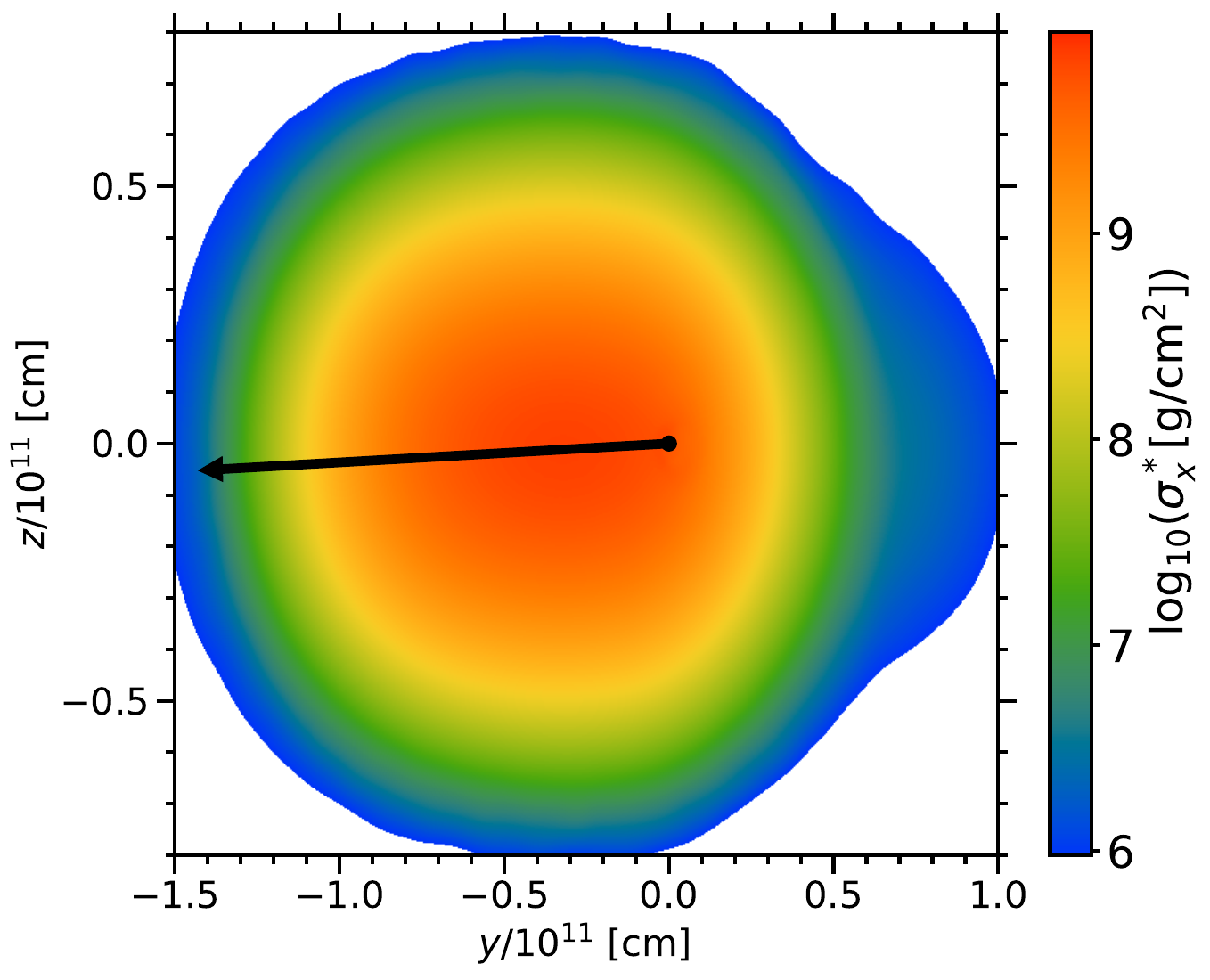}{0.32\textwidth}{(h) H4, $yz$}
\fig{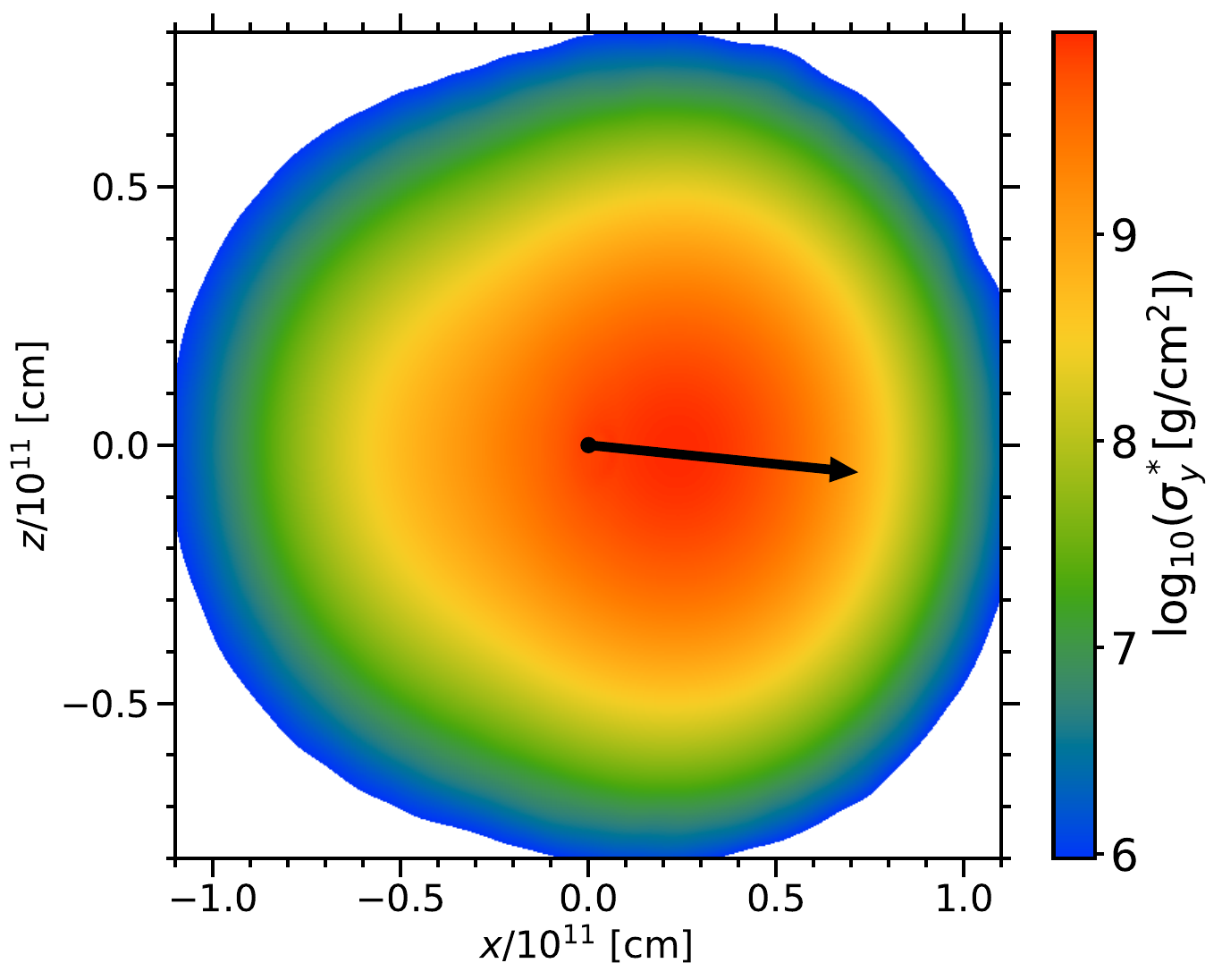}{0.32\textwidth}{(i) H4, $xz$}
}
\caption{The column density $\sigma^*_i = \int D^* dx_i$ at $t_\mathrm{hd,f} = 10$ s. The rows correspond to the stage (2) heating models H0 (top), H3 (middle), and H4 (bottom) (Table \ref{tab:heating_models}). The columns correspond to the projections $xy$ (left), $yz$ (middle), and $xz$ (right). 
For each model, the color bar has an upper limit at the highest column density and a lower limit at four orders of magnitude below. The white regions are those with column densities below the lower limit of the color bar. The black dots show the origin of the coordinate system. The black arrows show the projection of the total momentum $P_i = \sum_{j=1}^N m S_{(j)i}$. 
The results for H1 and H2 are similar to H0, simply with additional smoothing.}
\label{fig:column_density}
\end{figure*}

Figure \ref{fig:column_density} shows the column density at $t_\mathrm{hd,f} = 10$ s for the various heating models. In all cases, the ejecta is roughly a spiral arc in the $xy$-plane that subtends an angle $\Delta \phi \sim \pi$. This was the same general structure immediately after merger \citep{kyutoku15,foucart17}. The ejecta morphology for H0 is wedge-like with wedge half-opening angle $\tan \xi \sim 0.1$; for H4 it is more spherical. The density structure for H0 exhibits small-scale filaments, which are numerical artifacts from the distribution and interpolation of particles in SPH; these are smoothed in the heating models. 
The presence of heating thus inflates the matter in the direction perpendicular to the ejecta plane, smooths the small-scale inhomogeneities, and isotropizes the momentum in the rest frame. The ejecta with greater heating are more inflated and have lower densities. The direction $(\theta_P, \phi_P)$ of the total momentum is largely insensitive to the heating model, remaining at $(\theta_P, \phi_P) \simeq (1.6, 5.2)$. The column densities are roughly symmetric about $z = 0$; this is because we performed an active rotation on the particles in stage (1) to align the post-merger ejecta plane with the $xy$-plane (Section \ref{subsec:numerical_relativity}).

In \citet{darbha20}, we used 2D axisymmetric geometries to model the global deviations from spherical symmetry in the various kilonova ejecta components. Though BH-NS merger tidal tails are asymmetric, we can roughly map them to 2D geometries to capture this global asphericity. For its versatility, we map each outflow to a 2D oblate ellipsoid with axial ratio $R = a_x / a_z$, where $a_x$ and $a_z$ are the semi-major axes in the $x$- and $z$-directions. We find $R \simeq 5$ for H0, $R \simeq 3$ for H3, and $R \simeq 1.4$ for H4.

The simulation results complement and extend earlier work. 
\citet{fernandez15} performed a Newtonian hydrodynamic simulation with r-process heating, as did \citet{rosswog14} in the NS-NS merger context; both found results similar to ours for the influence of heating on the dynamical evolution of the unbound component. 
In lieu of hydrodynamic simulations, some previous end-to-end studies extrapolated the dynamical ejecta from the post-merger phase to later times assuming ballistic motion or homologous expansion \citep{roberts11,tanaka14,fernandez17}, potentially overestimating the degree of asymmetry. 
For comparison, we also considered the ballistic evolution of the ejecta along geodesics. We find that the ballistic approximation is comparable to model H0 and thus, when no heating is included, the ejecta shape and density structure are not appreciably modified by pressure forces alone, 
in agreement with earlier work \citep{fernandez15,roberts17}.

\subsection{Radiative Transfer (Stage 3)}
\label{subsec:results_radiative_transfer}

In this section, we shift notation and specify the polar angle with the direction cosine $\mu = \cos\theta$. The direction of the total momentum is then $(\mu_P, \phi_P) \simeq (-0.030, 5.2)$. This direction is an important reference for understanding the viewing angle dependence of the light curves.

\begin{figure*}
\gridline{
\fig{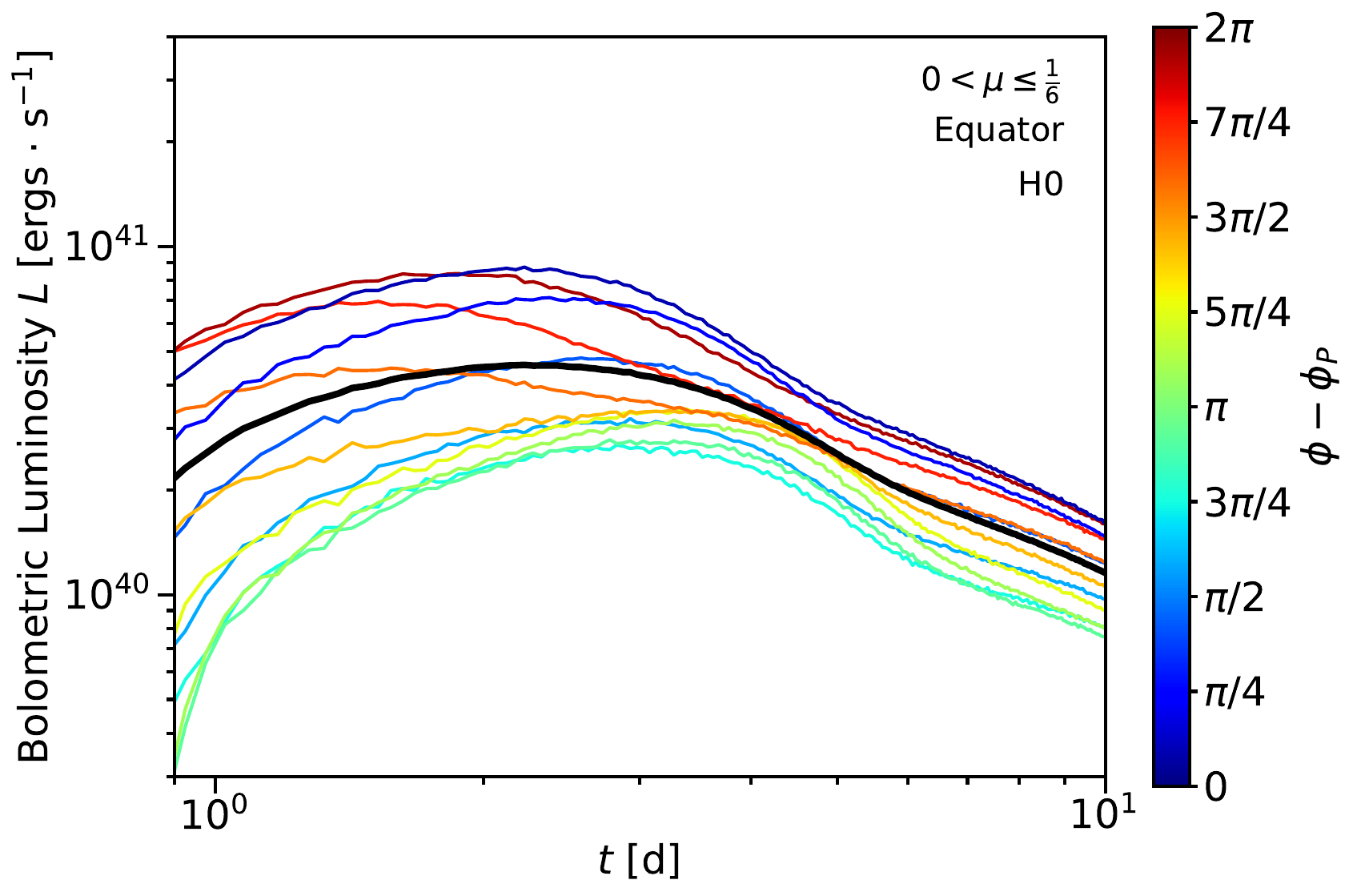}{0.49\textwidth}{(a) H0, $0 < \mu \leq \frac{1}{6}$ (Equator)}
\fig{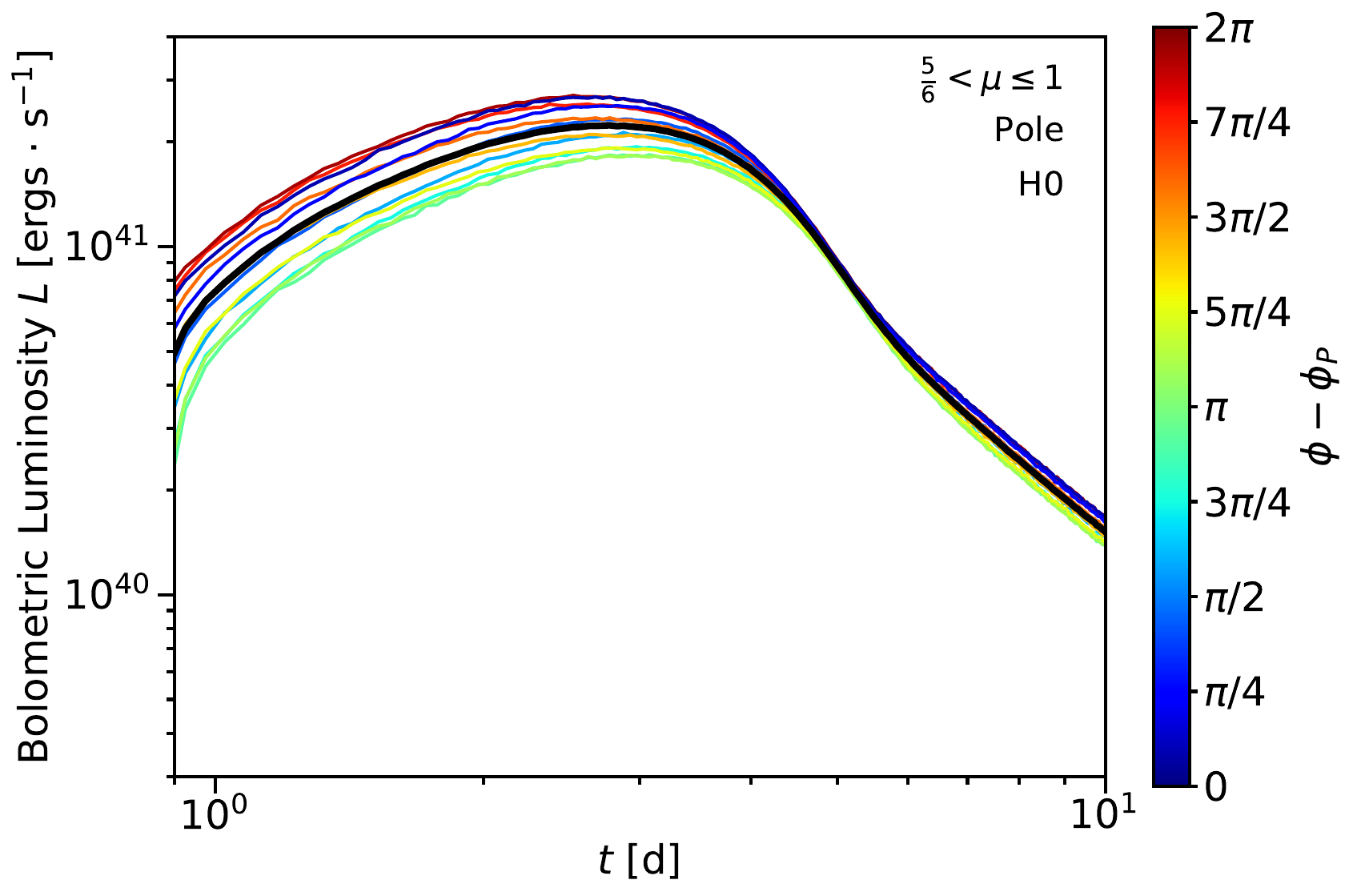}{0.49\textwidth}{(b) H0, $\frac{5}{6} < \mu \leq 1$ (Pole)}
}
\gridline{
\fig{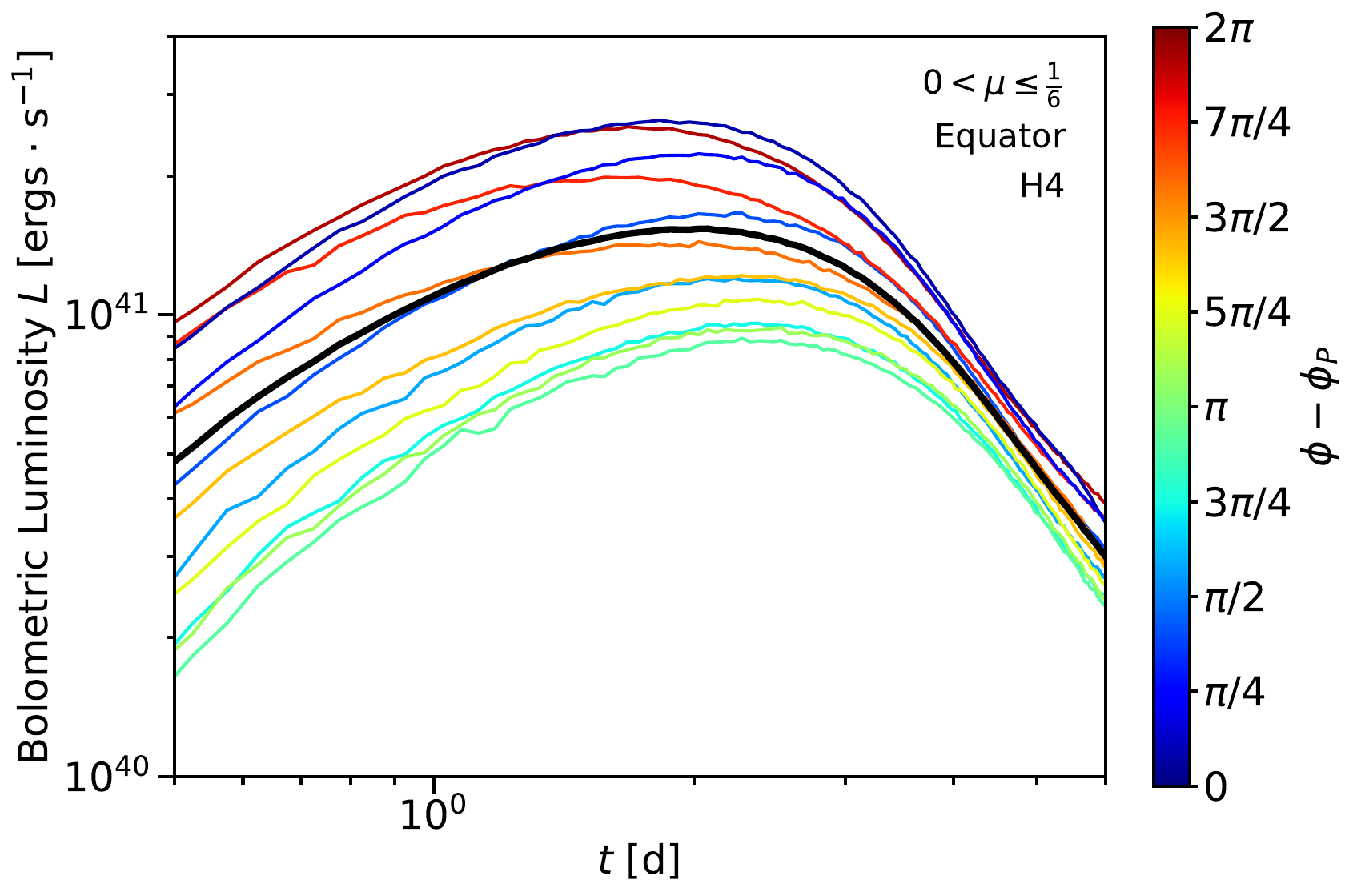}{0.49\textwidth}{(c) H4, $0 < \mu \leq \frac{1}{6}$ (Equator)}
\fig{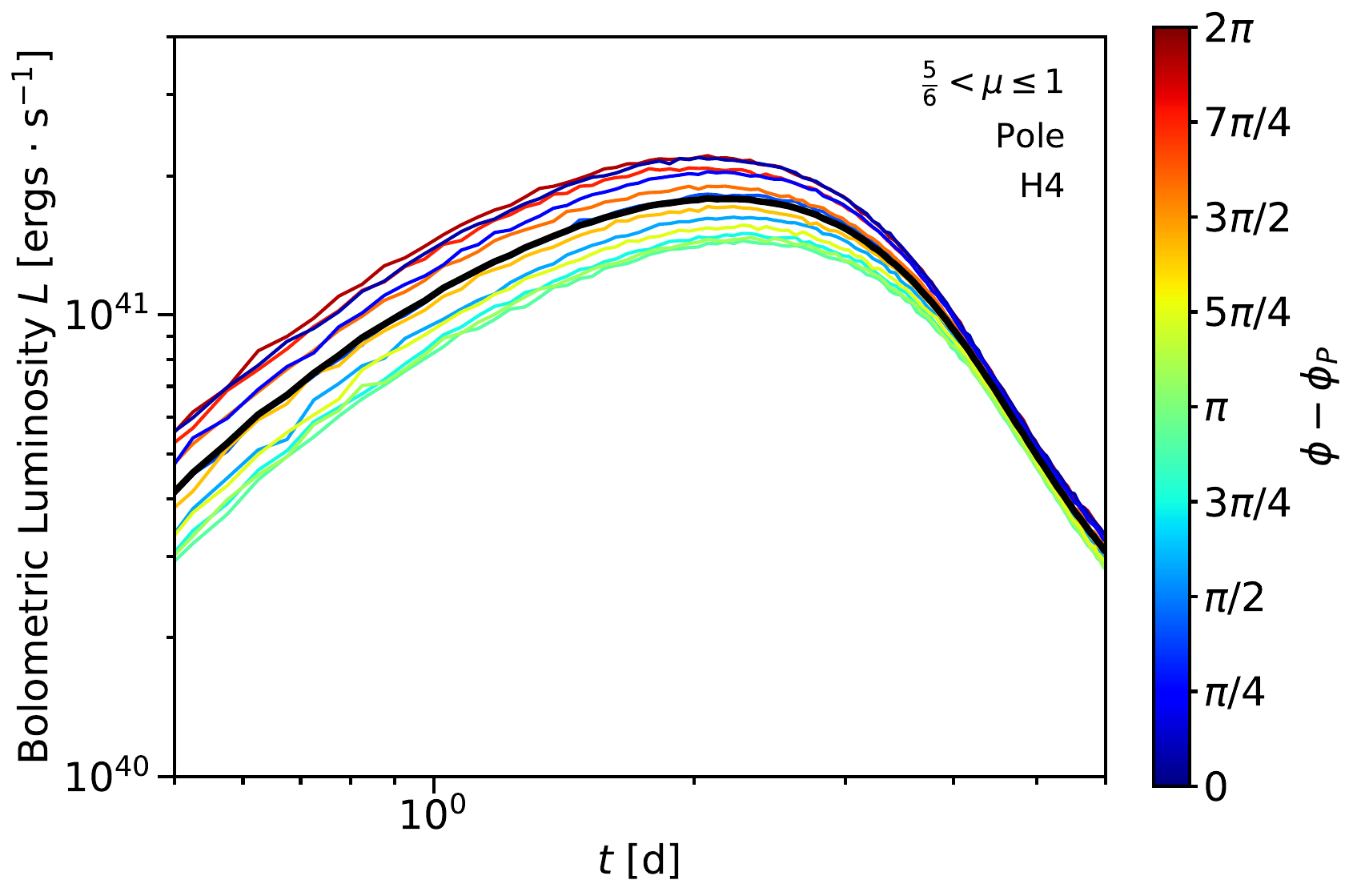}{0.49\textwidth}{(d) H4, $\frac{5}{6} < \mu \leq 1$ (Pole)}
}
\caption{
Isotropic-equivalent bolometric light curves. The rows correspond to the heating models H0 (top) and H4 (bottom). The columns correspond to different bins of $\mu = \cos\theta$, where $\theta$ is the polar angle. The color bar shows the centers of the 
$\phi - \phi_P$ bins. The thick black curve shows the $\phi$-averaged light curve. In each panel, the brightest curve corresponds to the azimuthal direction $\phi_P \simeq 5.2$ of the total momentum and the dimmest curve corresponds to the opposite direction $\phi_P - \pi$. The light curves are roughly the same for $\mu \rightarrow -\mu$ because the ejecta are roughly symmetric about $z = 0$ after the active rotation in stage (1) (Section \ref{subsec:numerical_relativity}).
}
\label{fig:bolometric_light_curves}
\end{figure*}

Figure \ref{fig:bolometric_light_curves}, Panels (a) and (b) show the bolometric light curves for the heating model H0. For any polar angle $\mu$, the light curves are brightest in the direction $\phi = \phi_P$ (i.e. when the velocity of the bulk ejecta points towards the observer) and dimmest in the opposite direction $\phi = \phi_P - \pi$. In each $\mu$ bin, the $\phi$-averaged light curve describes the offset of the set of $\phi$-dependent light curves. The offset is highest in the polar directions $\mu \sim \pm 1$ (Panel b) and lowest in the equatorial direction $\mu \sim 0$ (Panel a). The variation with $\phi$ is smallest in the polar direction $\mu \sim \pm 1$ and largest in the equatorial direction $\mu \sim 0$. The light curves exhibit a bend after the peak because the ejecta has a wedge-shaped morphology.

Panels (c) and (d) show the bolometric light curves for the heating model H4. In agreement with H0, the light curves are brightest in the direction $\phi = \phi_P$ and dimmest in the direction $\phi = \phi_P - \pi$ for any $\mu$, and the variation with $\phi$ is smallest in the polar directions and largest in the equatorial direction. In a departure from H0, the offset is roughly constant for each $\mu$ and the light curves decrease smoothly after the peak because the ejecta has a more spherical shape. The luminosity may either increase or decrease with $\mu$ at a constant $\phi$ due to the combination of a roughly constant offset and a smaller variation towards the poles.

\begin{figure*}
\gridline{
\fig{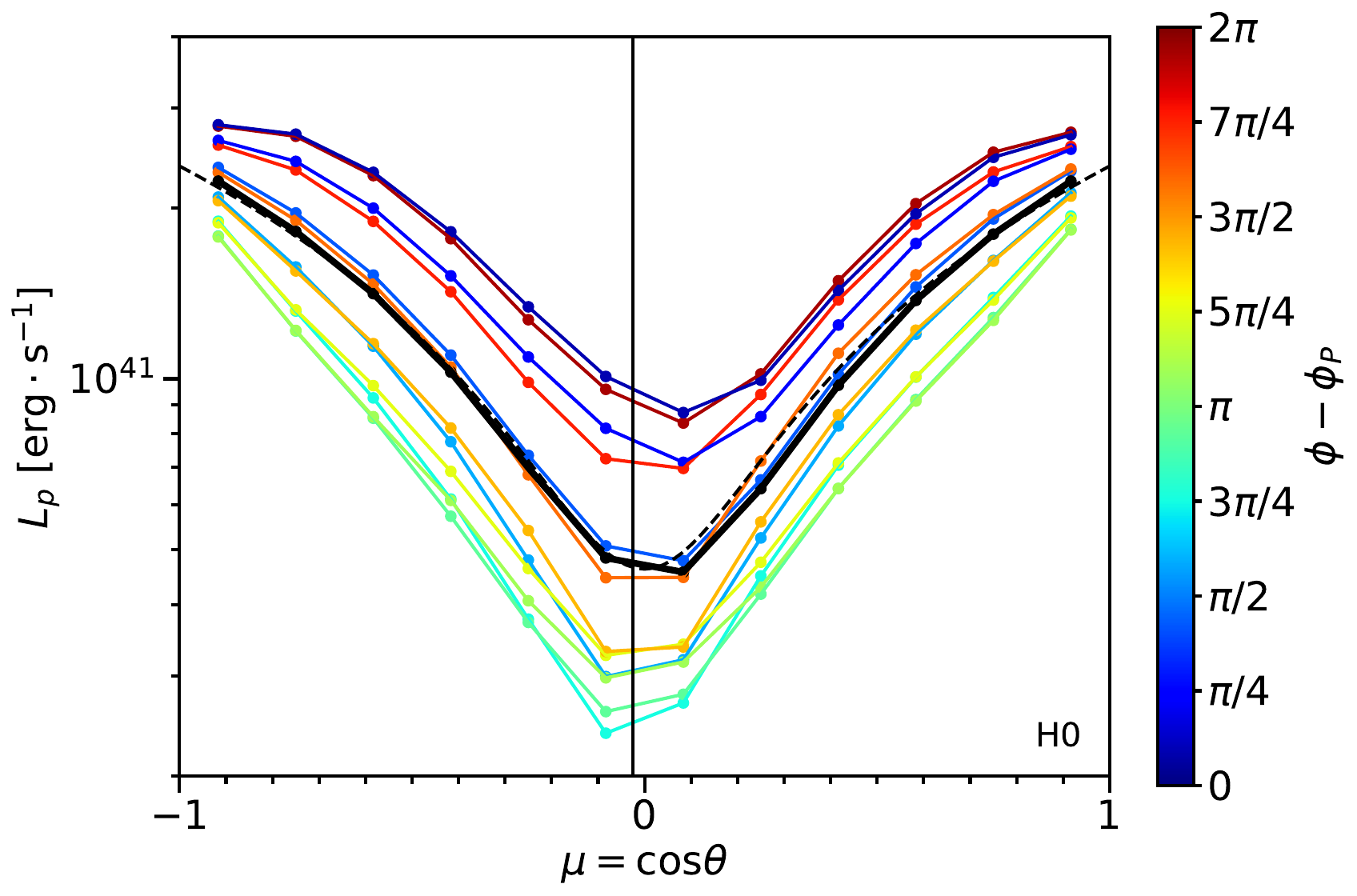}{0.49\textwidth}{(a) H0}
\fig{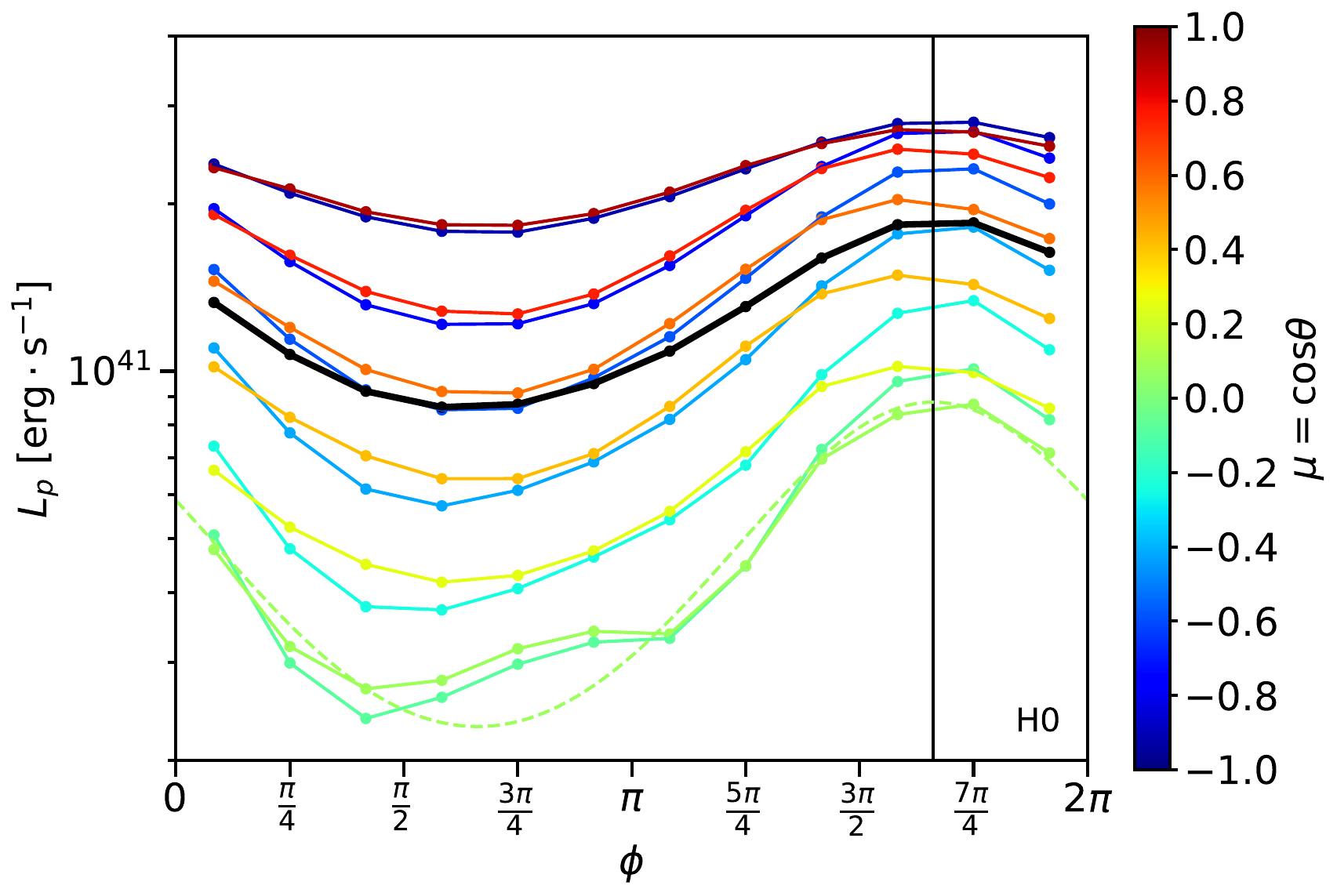}{0.49\textwidth}{(b) H0}
}
\gridline{
\fig{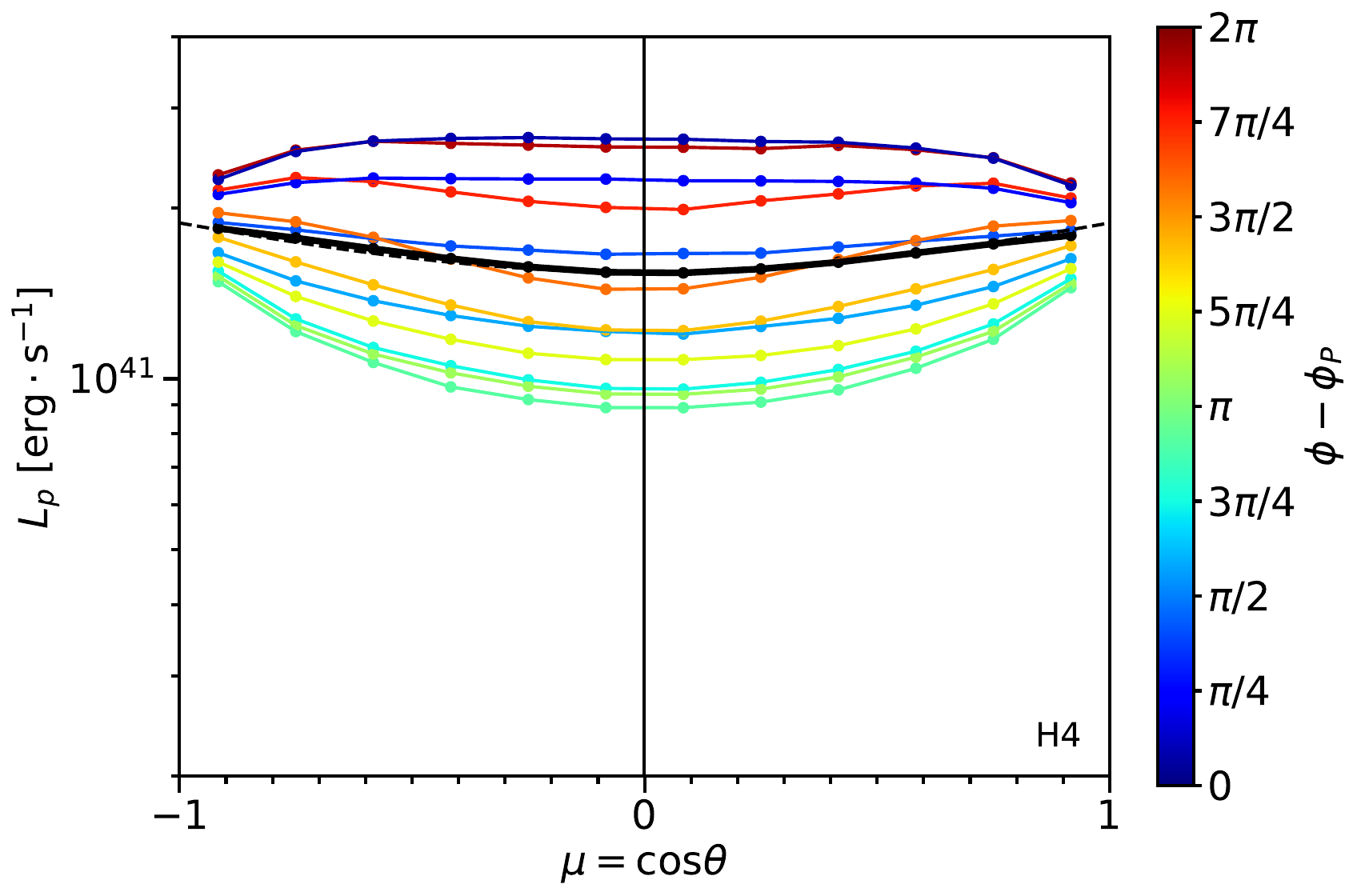}{0.49\textwidth}{(c) H4}
\fig{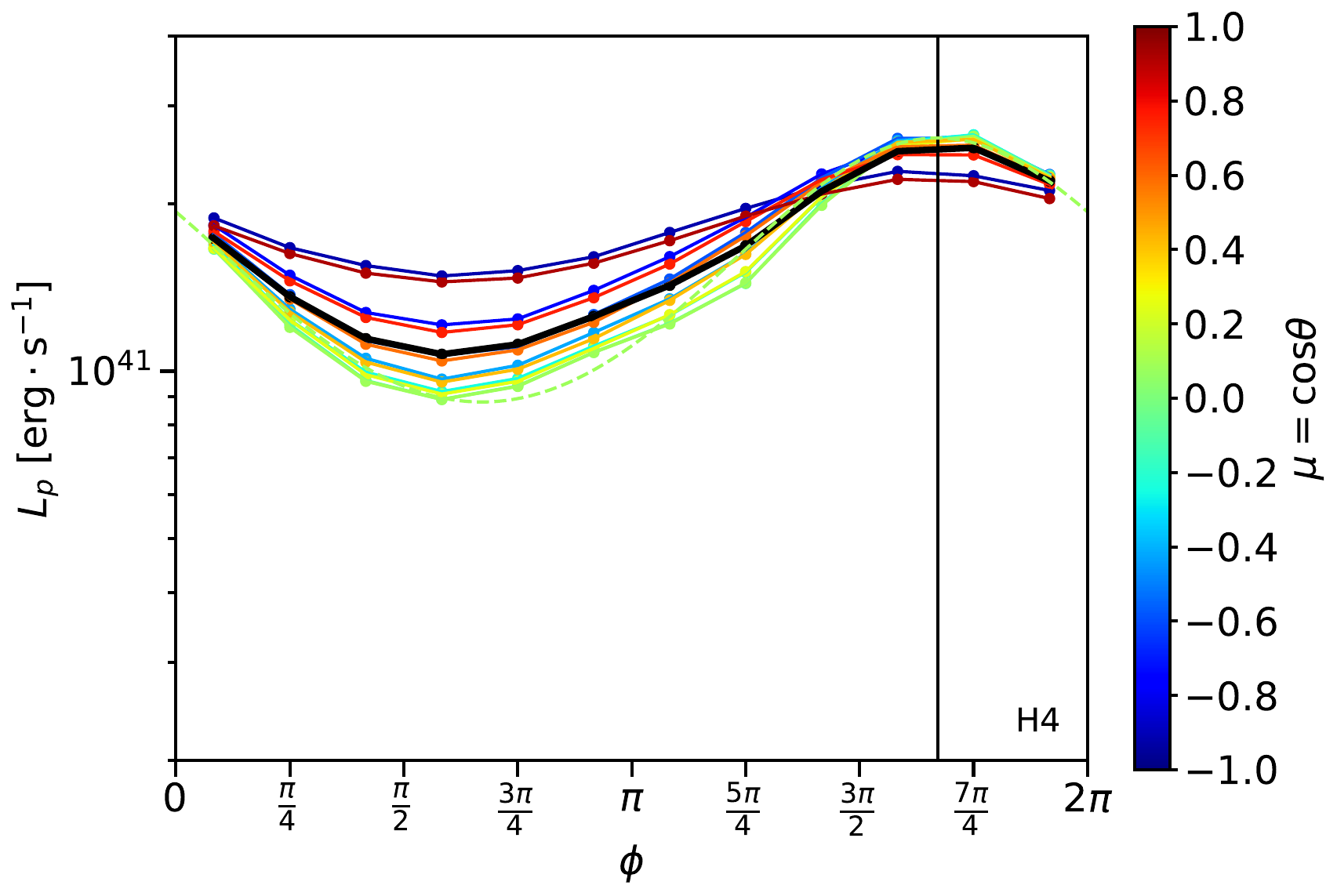}{0.49\textwidth}{(d) H4}
}
\caption{
The peak luminosities of the isotropic-equivalent bolometric light curves at each viewing angle. 
The rows correspond to the heating models H0 (top) and H4 (bottom). The columns correspond to $\mu$ on the $x$-axis and 
$\phi - \phi_P$ on the color bar (left) and $\phi$ on the $x$-axis and $\mu$ on the color bar (right). The color bar shows the centers of the angle bins. The black vertical line shows the direction $(\mu_P,\phi_P)$ of the total momentum. The thick black curve shows the peaks for the $\phi$-averaged light curves (left) and the $\mu$-averaged light curves (right). The dashed curve shows the fit of the projected area formula (Equation \ref{eq:Lpeak_of_mu_analytic}) to the peaks of the $\phi$-averaged light curves (left) and the Doppler enhancement formula (Equation \ref{eq:L_Doppler_analytic}) to the peaks of the equatorial light curves (right). The fit parameters have some degeneracy.
}
\label{fig:Lpeak_plots}
\end{figure*}

Figure \ref{fig:Lpeak_plots} shows the peak luminosities $L_p$ of the bolometric light curves over the full range of directions. The models with larger heating show smaller variation in the peak luminosity. For H0, the peaks lie in the range $\sim (0.3 - 3) \times 10^{41}$ erg$\cdot$s$^{-1}$ for an overall variation of $\sim 10$; for H4, they lie in the range $\sim (0.9 - 3) \times 10^{41}$ erg$\cdot$s$^{-1}$ for an overall variation of $\sim 3$. The variation at a fixed $\phi$ is also larger for H0 than H4. 
Panels (a) and (b) show the results for H0. The peak luminosities are roughly symmetric about $\mu = 0$ because the ejecta have rough reflection symmetry about $z = 0$ due to the active rotation performed in stage (1) (Section \ref{subsec:numerical_relativity}). The pole-to-equator variation at a fixed azimuthal angle $\phi$ is smallest for $\phi = \phi_P$ in the direction of the total momentum and is largest for $\phi = \phi_P - \pi$ in the opposite direction. The peak luminosity is largest in the polar directions $\mu \sim \pm 1$ and decreases towards the equatorial direction $\mu \sim 0$. Panels (c) and (d) show the results for H4. The peaks exhibit similar trends to H0, but with an important inversion around $\phi = \phi_P$; here, the peaks are higher in the equatorial direction than the polar directions. 

The viewing angle dependence of the light curves can generally be explained by two effects. The first effect is from the parallel projected area of the ejecta: the light curves are brighter at viewing angles that perceive larger projected areas. The second effect is from Doppler enhancement/reduction due to the bulk motion of the ejecta: the luminosity is larger in viewing directions more closely aligned to the bulk motion. We quantify each of these in turn.

In the first effect, the luminosity scales with the perceived projected area. In our models, the projected area is primarily a function of $\mu$ because the ejecta are roughly axisymmetric about some symmetry axis $(x_0,y_0)$; is roughly symmetric about $\mu = 0$ due to the active rotation performed in stage (1) (Section \ref{subsec:numerical_relativity}); and is a monotonically increasing function of $| \mu |$ because the ejecta are oblate. In Section \ref{subsec:results_hydrodynamics}, we roughly mapped the ejecta to 2D oblate ellipsoids. In \citet{darbha20}, we studied the light curves from 2D axisymmetric geometries and quantified the dependence on projected area. For a 2D oblate ellipsoid, the projected area $A_\mathrm{proj}(\mu)$ is
\begin{equation}
    A_\mathrm{proj}(\mu) = \pi R a_z^2 [ (R^2 - 1) \mu^2 + 1 ]^{1/2} ,
    \label{eq:projected_area-ellipsoid}
\end{equation}
where $R = a_x / a_z$ is the axial ratio and $a_x$ ($a_z$) is the semi-major axis in the $x$ ($z$) direction. The size of the pole-to-equator luminosity ratio around peak is a factor $\sim (1 - 2)$ of the pole-to-equator area ratio $A_\mathrm{pole} / A_\mathrm{eq} = R$. Using a more detailed parameterization, the peak luminosity $L_p(\mu)$ can roughly be written as
\begin{equation}
    L_p(\mu) \simeq L_0 \left[ 1 + k \left( \frac{A_\mathrm{proj}(\mu)}{A_\mathrm{proj}(\mu_\mathrm{ref})} - 1 \right) \right] ,
    \label{eq:Lpeak_of_mu_analytic}
\end{equation}
where $L_0$ is a reference luminosity, $\mu_\mathrm{ref} = 0.55$ is a geometry-dependent reference direction, 
and $k$ is an order unity fitting parameter.

In Figure \ref{fig:Lpeak_plots}, Panels (a) and (c), the thick black curve shows $L_p$ for the $\phi$-averaged light curves, which roughly isolates the projected area contribution. The curve is fit to Equation \ref{eq:Lpeak_of_mu_analytic} in each case. In Panel (a), which show the results for H0, this curve exhibits a factor of $\sim 5$ variation with $\mu$, in close agreement with the pole-to-equator projected area ratio $A_\mathrm{pole} / A_\mathrm{eq} = R \sim 5$.  In Panel (c), which shows the results for H4, the peaks of the $\phi$-averaged light curves exhibit a factor of $\sim 1.2$ variation with $\mu$, once again comparable to $A_\mathrm{pole} / A_\mathrm{eq} = R \sim 1.4$.

In the second effect, the luminosity is larger when the bulk ejecta moves in the direction of the observer. In particular, if $\vec{\beta} = \beta \hat{v}$ is the bulk velocity of the ejecta (over $c$) and $\hat{n}$ is the viewing direction, both measured in the laboratory frame, then the laboratory frame luminosity is roughly scaled by an increasing function of $\vec{\beta} \cdot \hat{n}$ compared to the bulk frame luminosity. In our models, $(\mu_P, \phi_P) \simeq (-0.030, 5.2)$, so the bulk velocity is largely confined to the $xy$ plane, $\hat{v} \simeq \cos(\phi_P) \hat{x} + \sin(\phi_P) \hat{y}$.

We can quantify this with a simple model. Let the ejecta be a point source moving at the bulk velocity and emitting isotropically in the bulk frame. Let unprimed quantities denote those in the laboratory frame seen by an observer in the direction $\hat{n}$ and primed quantities denote those in the bulk frame seen by an observer in the direction $\hat{n}'$. The kinematic Doppler factor in the lab frame is 
\begin{equation}
    \delta = [ \gamma ( 1 - \vec{\beta} \cdot \hat{n} ) ]^{-1} ,
\end{equation}
where $\gamma = (1 - \beta^2)^{-1/2}$ is the Lorentz factor. The frequencies are related by $\nu = \delta \nu'$. The emissivity in the bulk frame is $j'(\vec{x}',t';\hat{n}',\nu') = J(t';\nu') \delta(\vec{x}') / 4\pi$ and in the lab frame is $j(\vec{x},t;\hat{n},\nu) = \mathcal{J}(t;\hat{n},\nu) \delta(\vec{x} - \vec{v} t)$. The emissivities are related by 
$j = \delta^2 j'$ 
since the quantity 
$j / \nu^2$ 
is Lorentz invariant \citep{mihalas84}. The luminosity is 
$\mathcal{L}(t;\hat{n}) = \int j dV d\nu = \int \mathcal{J} d\nu$ 
and can be related to the bulk frame luminosity by 
$\mathcal{L} = \delta^4 \mathcal{L}'$, 
where we used $dV = \delta dV'$ for the volume elements. The emission is isotropic in the bulk frame, so $\mathcal{L}'(t';\hat{n}') = L'(t')/4\pi$. The isotropic equivalent luminosity is $L(t;\hat{n}) = 4\pi \mathcal{L}(t;\hat{n}) = \delta^4 L'(t')$. We can roughly adapt this to the peak luminosity to obtain
\begin{equation}
    L_p(t,\hat{n}) \simeq \delta^4 L'_p .
    \label{eq:L_Doppler_analytic}
\end{equation}
For $\beta = 0.15$, $L_p(t;\hat{n}) \sim 2.3 L'_p$ when $\hat{n}$ is aligned with $\vec{\beta}$, and $L_p(t;\hat{n}) \sim 0.95 L'_p$ when $\hat{n}$ is orthogonal to $\vec{\beta}$. Though this is a simple model, it captures the dominant influence of the bulk motion. Doppler modification is observed in other relativistic systems, most prominently radio jets \citep{lind85,urry95}.

In Figure \ref{fig:Lpeak_plots}, Panels (b) and (d), the bright green curve at the bottom roughly shows $L_p$ for the light curves in the equatorial plane $\mu \simeq 0$, which roughly isolates the Doppler contribution since $\vec{\beta} \cdot \hat{n} \simeq \beta \cos(\phi - \phi_P)$ for $\hat{n} = \cos\phi \hat{x} + \sin\phi \hat{y}$ in the equatorial plane. The curve is fit to Equation \ref{eq:L_Doppler_analytic} in each case. In Panel (b), which shows the results for H0, this curve exhibits a factor of $\sim 3$ variation from $\phi = \phi_P$ to $\phi = \phi_P - \pi/2$. In Panel (d), which shows the results for H4, the peaks of the equatorial light curves exhibit a factor of $\sim 2$ variation from $\phi = \phi_P$ to $\phi = \phi_P - \pi/2$. The model H4 has a slightly higher bulk velocity than H0, but exhibits a smaller equatorial variation with $\phi$. This is because the ejecta in H0 shows a larger deviation from axisymmetry, and thus has a small additional contribution from the projected area variation with $\phi$ in the equatorial plane.

In more detail, the ejecta is an extended source that has an outward expanding velocity gradient. We examined a toy model to determine if these features introduce additional corrections. The toy model was constructed by (1) generating an outflow with spherical symmetry, homologous expansion, and a broken power-law density profile, and (2) boosting the outflow in the $x$-direction with a center-of-mass velocity $\beta$ to obtain a directed ejecta. The spherical shape removes the projected area effect. We selected the parameters to match the mass and kinetic energy of the model H0. We found that the viewing angle variation due to Doppler modification is comparable in the toy model and the model H0, confirming that additional corrections are subdominant.

The viewing angle trends are roughly determined by the interplay between these two effects. To illustrate this, we consider the variation with $\mu$ in the directions $\phi = \phi_P - \pi$ and $\phi = \phi_P$ (Figure \ref{fig:Lpeak_plots}, Panels b and d). In the direction $\phi = \phi_P - \pi$, the projected area and the Doppler modification effects both enhance $L_p$ towards the poles, leading to a large variation with $\mu$. In the direction $\phi = \phi_P$, though, the projected area effect enhances $L_p$ towards the poles and the Doppler modification reduces it. In model H0, the ejecta is highly oblate, so the projected area effect dominates and the peaks increase towards the poles, but show a smaller variation with $\mu$ than the direction $\phi = \phi_P - \pi$. In model H4, the ejecta is nearly spherical, so the Doppler reduction dominates and the peaks decrease towards the poles, albeit by a small amount.

In Appendix \ref{sec:light_curve_peaks}, Figure \ref{fig:Lpeak_grids} reproduces the peak bolometric luminosities $L_p$ in a different format to highlight the overall trends, and adds the heating model H3 and the peak times. The peak times show a very rough symmetry about $\mu = 0$, similar to but less prevalent than the trend in the peak luminosities. The peak times are earlier for $5\pi/6 \lesssim \phi \lesssim 13\pi/6$; this is because the bulk of the ejecta is moving towards the observer in this range, since the total momentum is in the direction $\phi = \phi_P \simeq 5.2$ and the spiral arc subtends an angle $\Delta \phi \sim \pi$. The peak times are later for $\pi/6 \lesssim \phi \lesssim 5\pi/6$ because the bulk of the ejecta is moving away from the observer. There does not appear to be a clear trend for the variation with $\mu$ at a fixed $\phi$.

\begin{figure*}
\gridline{
\fig{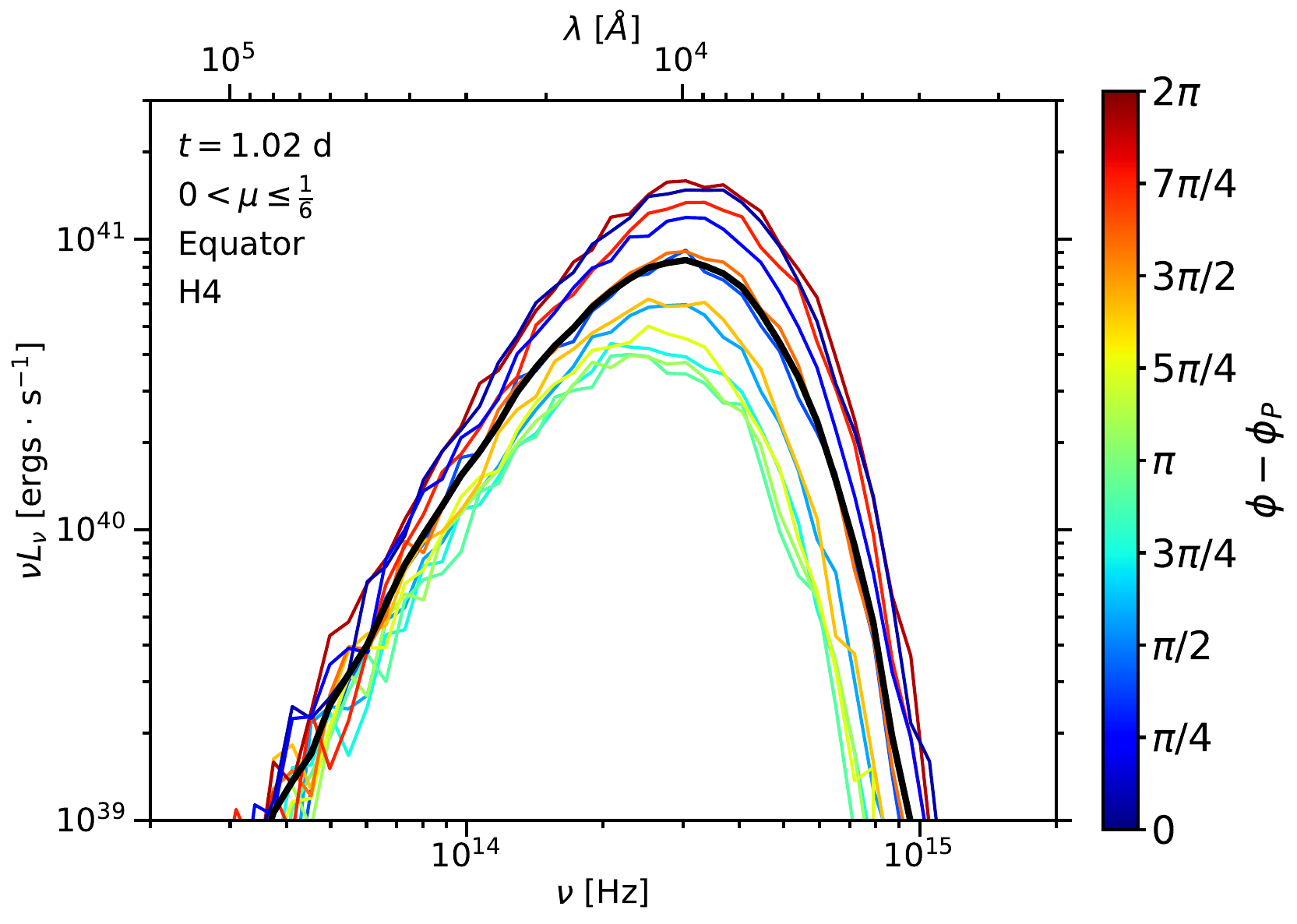}{0.49\textwidth}{(a) $t \sim 1$ d, $0 < \mu \leq \frac{1}{6}$ (Equator)}
\fig{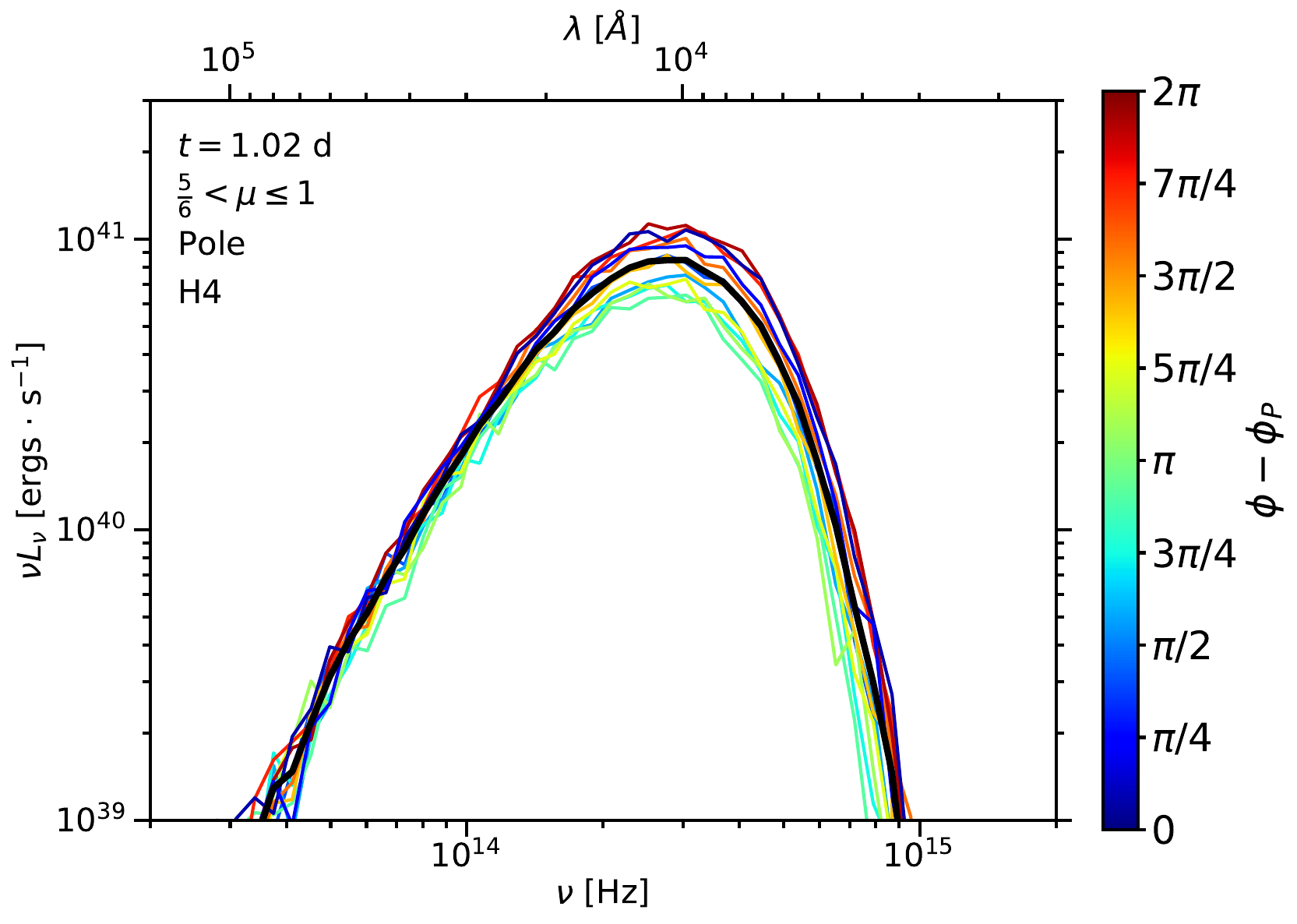}{0.49\textwidth}{(b) $t \sim 1$ d, $\frac{5}{6} < \mu \leq 1$ (Pole)}
}
\caption{
Spectra for the heating model H4 at $t \sim 1$ d, slightly before peak. The columns correspond to different bins of $\mu = \cos\theta$, where $\theta$ is the polar angle. The color bar shows the centers of the 
$\phi - \phi_P$ bins. The thick black curve shows the $\phi$-averaged spectrum. In each panel, the brightest curve corresponds to the azimuthal direction $\phi_P \simeq 5.2$ of the total momentum and the dimmest curve corresponds to the opposite direction $\phi_P - \pi$. The spectra are roughly the same for $\mu \rightarrow -\mu$ because the ejecta are roughly symmetric about $z = 0$ after the active rotation in stage (1) (Section \ref{subsec:numerical_relativity}).
}
\label{fig:spectra}
\end{figure*}

Figure \ref{fig:spectra} shows the spectra for heating model H4. At all times, the spectral intensity lies primarily in the range 
$10^{13} \, \mathrm{Hz} \lesssim \nu \lesssim 2 \times 10^{15} \, \mathrm{Hz}$, 
with most in the infrared (IR), $\nu \lesssim 4 \times 10^{14}$ Hz. The time evolution is straightforward. At early times before peak, the spectra have a blackbody shape because the ejecta is optically thick at all frequencies. At intermediate times near and after peak, the lower frequencies become optically thin more rapidly due to the form of the analytic opacity and are suppressed. At late times after peak, the ejecta becomes optically thin at all frequencies and the entire spectra deviates from a blackbody. The near-blackbody spectra at $t = 1$ d have temperatures $T \sim (2.8 - 3.6) \times 10^3$ K and surficial radii $R \sim (1.8 - 2.4) \times 10^{15}$ cm.

The Doppler modification shifts the spectra to higher frequencies in the direction $(\mu,\phi) = (\mu_P,\phi_P)$ (i.e. when the bulk of the ejecta approaches the observer) and to lower frequencies in the direction $(\mu,\phi) = (-\mu_P,\phi_P-\pi)$ (i.e. when the bulk of the ejecta recedes from the observer). The spectral peaks lie near the boundary between the optical and IR regions ($\nu \sim 4 \times 10^{14}$ Hz), so even small shifts can have a large impact on the optical and IR light curves.

\begin{figure*}
\gridline{
\fig{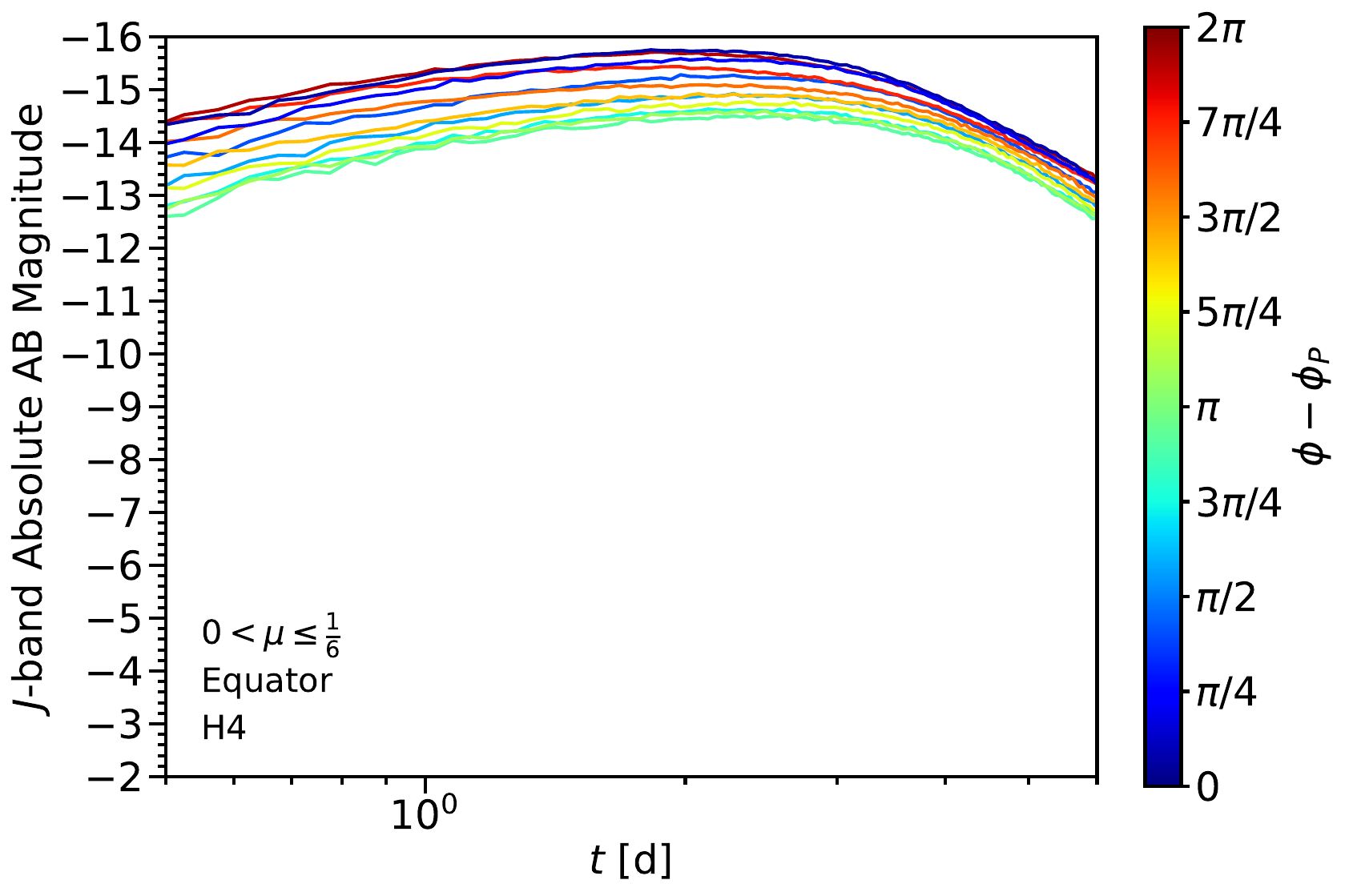}{0.49\textwidth}{(a) $J$-band}
\fig{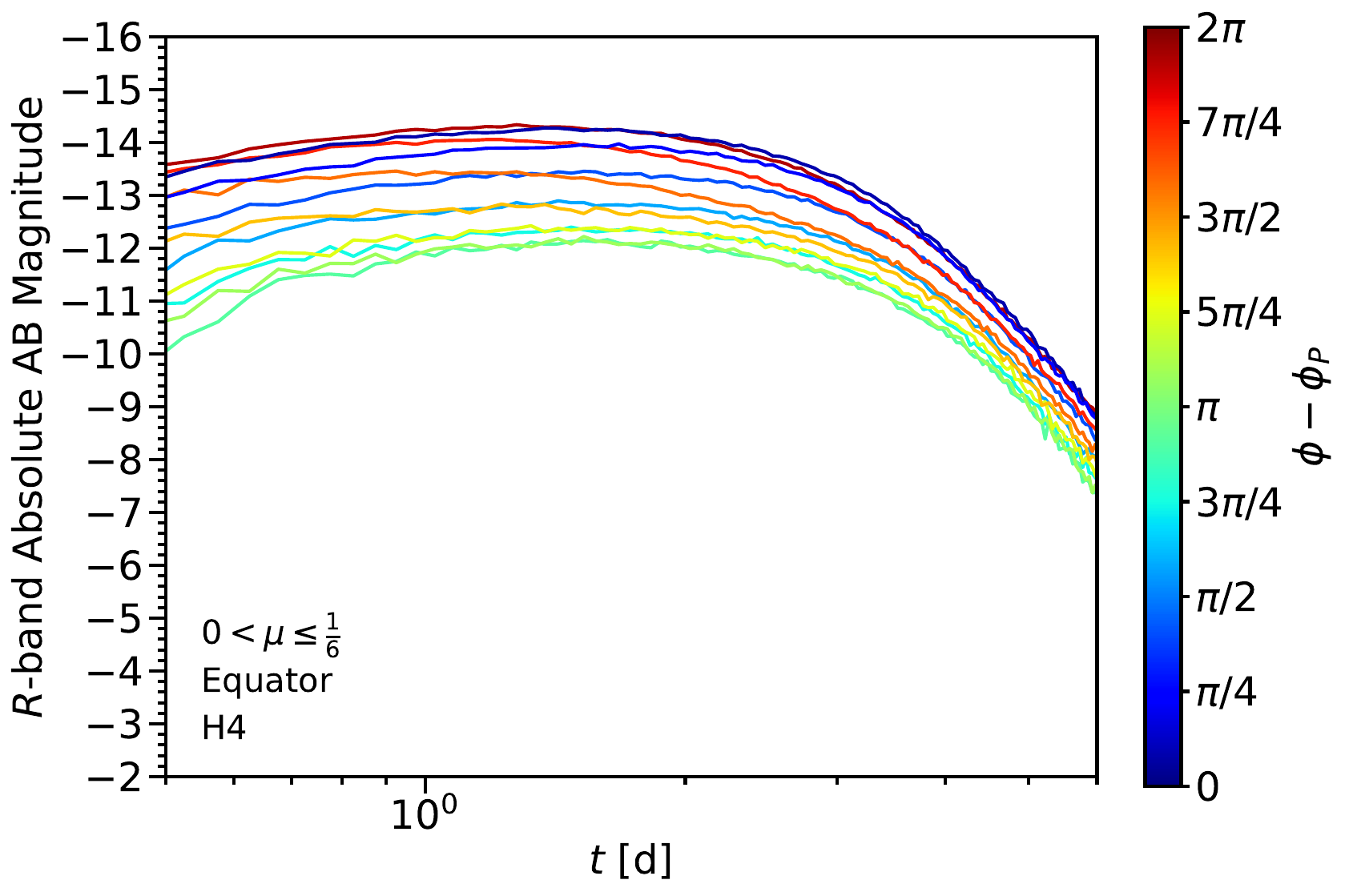}{0.49\textwidth}{(b) $R$-band}
}
\gridline{
\fig{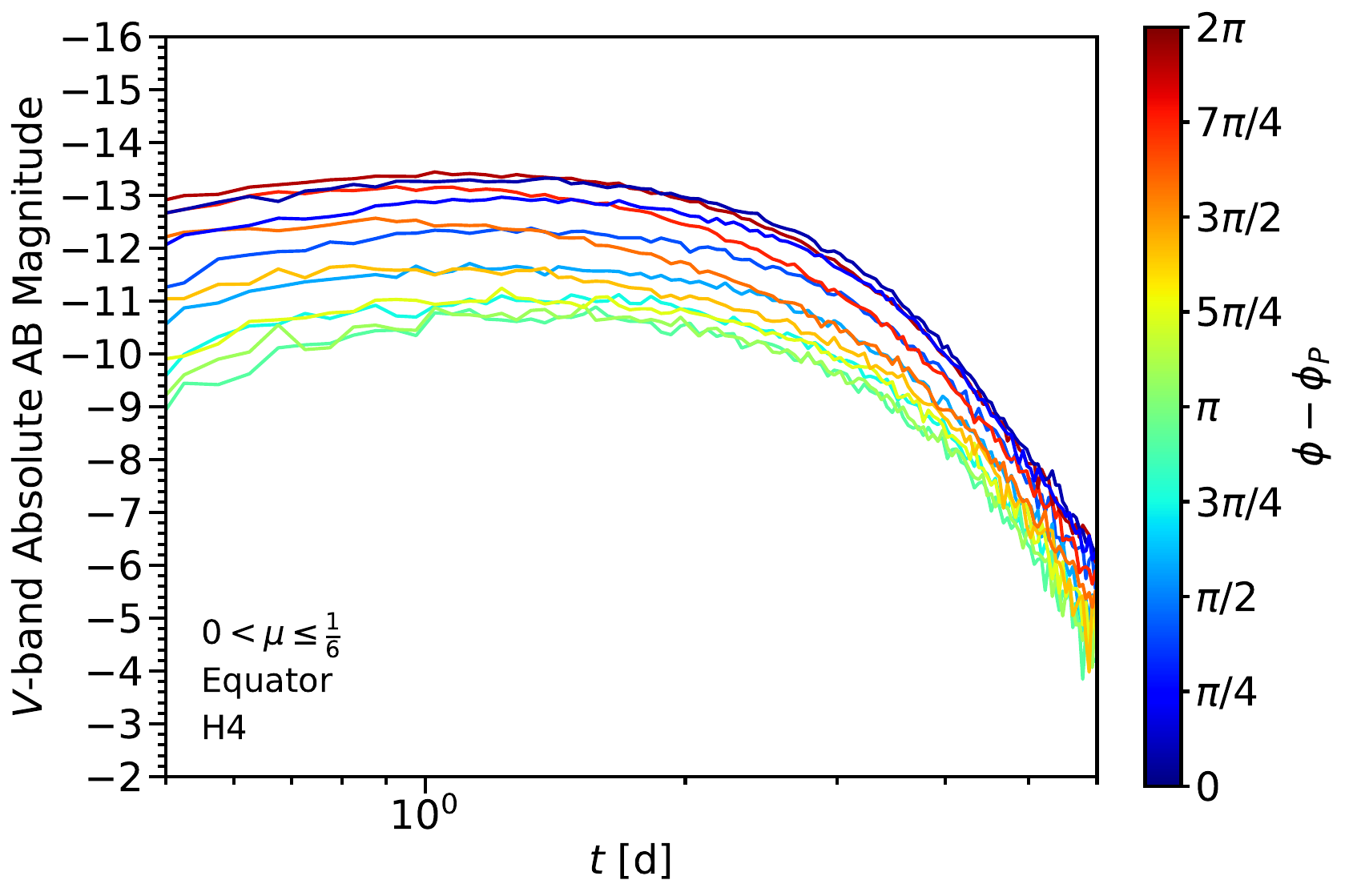}{0.49\textwidth}{(c) $V$-band}
\fig{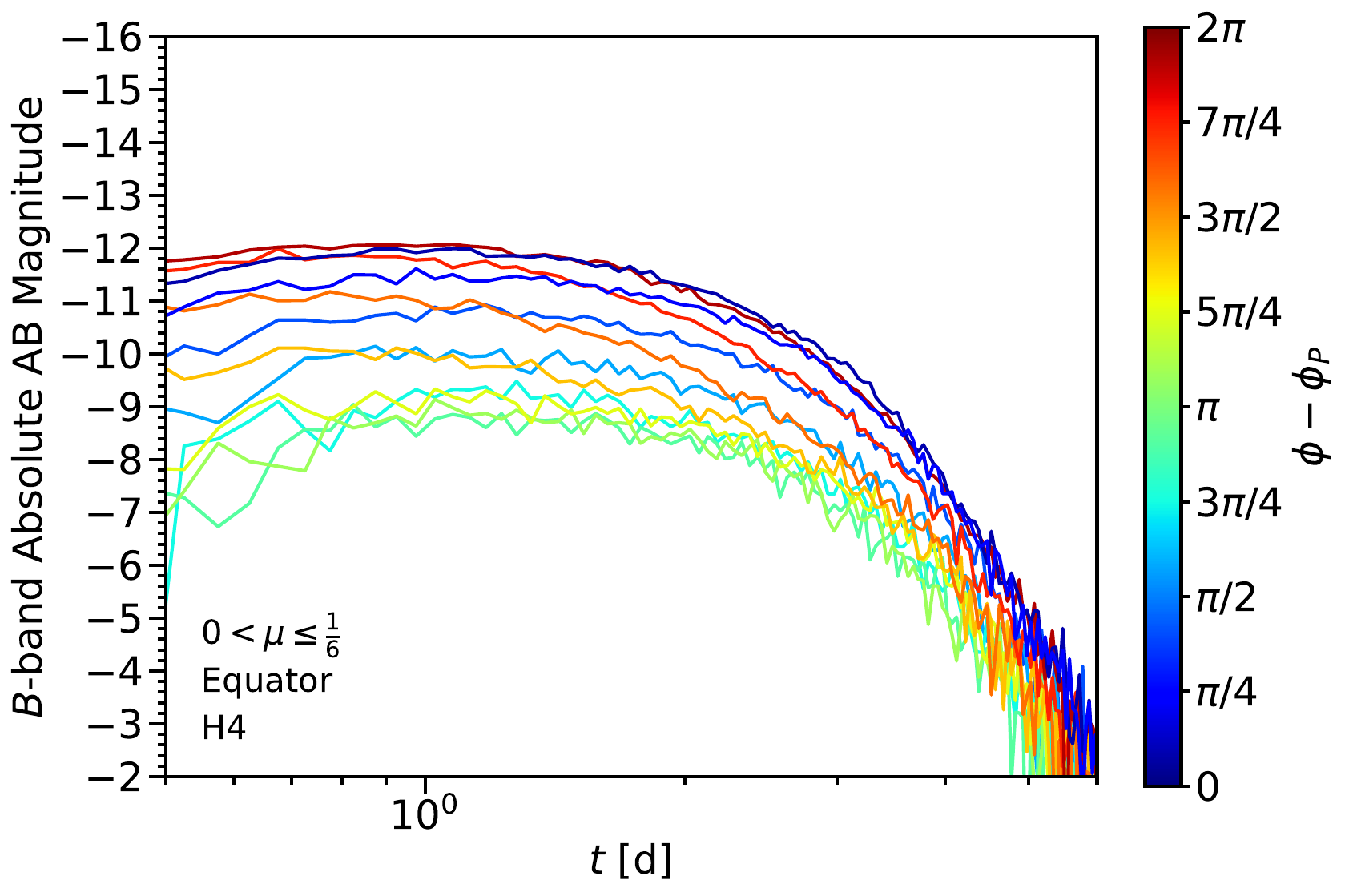}{0.49\textwidth}{(d) $B$-band}
}
\caption{Isotropic-equivalent broadband light curves for the heating model H4. The magnitudes are Absolute AB magnitudes. The polar viewing angle is near the equator, $0 < \mu \leq \frac{1}{6}$. The panels correspond to the (a) $J$-band, (b) $R$-band, (c) $V$-band, and (d) $B$-band, representative of the four observable frequency intervals. The color bar shows the centers of the 
$\phi - \phi_P$ bins. In each panel, the brightest curve corresponds to the azimuthal direction $\phi_P \simeq 5.2$ of the total momentum and the dimmest curve corresponds to the opposite direction $\phi_P - \pi$. The light curves are roughly the same for $\mu \rightarrow -\mu$ because the ejecta are roughly symmetric about $z = 0$ after the active rotation in stage (1) (Section \ref{subsec:numerical_relativity}).}
\label{fig:broadband_light_curves}
\end{figure*}

Figure \ref{fig:broadband_light_curves} shows broadband light curves for the heating model H4. The figure presents four filter bands representative of the four main observable frequency intervals: the $J$-band (IR), the $R$-band (optical/IR), the $V$-band (optical), and the $B$-band (optical/UV). The IR band is the brightest, has the smallest variation with viewing angle, and exhibits the largest and slowest rise to peak. The optical/UV band has the opposite characteristics. For instance, the $J$-band peaks in the range $\sim -14.4$ to $-15.8$ in $\sim 2 - 3$ days, and the $V$-band peaks in the range $\sim -10.6$ to $-13.4$ in $\sim 1 - 2$ days. In Appendix \ref{sec:broadband_light_curves}, Figures \ref{fig:broadband_light_curves_IR} - \ref{fig:broadband_light_curves_optical_UV} show a larger set of filters and polar angles.

The broadband light curves exhibit the same viewing angle dependence as the bolometric light curves due to the same two effects. Importantly, the Doppler modification shifts part of the spectra from the IR into the optical in the direction $(\mu,\phi) = (\mu_P,\phi_P)$ and from the optical to the IR in the direction $(\mu,\phi) = (-\mu_P,\phi_P-\pi)$. The optical bands thus show a larger variation between these two directions. For instance, in the direction $(\mu,\phi) = (\mu_P,\phi_P)$, the $V$-band rises to a magnitude of $\sim -13.4$; this is currently too faint for a detection, but it raises the prospects of an optical observation.

\begin{figure}
\includegraphics[width=0.47\textwidth]{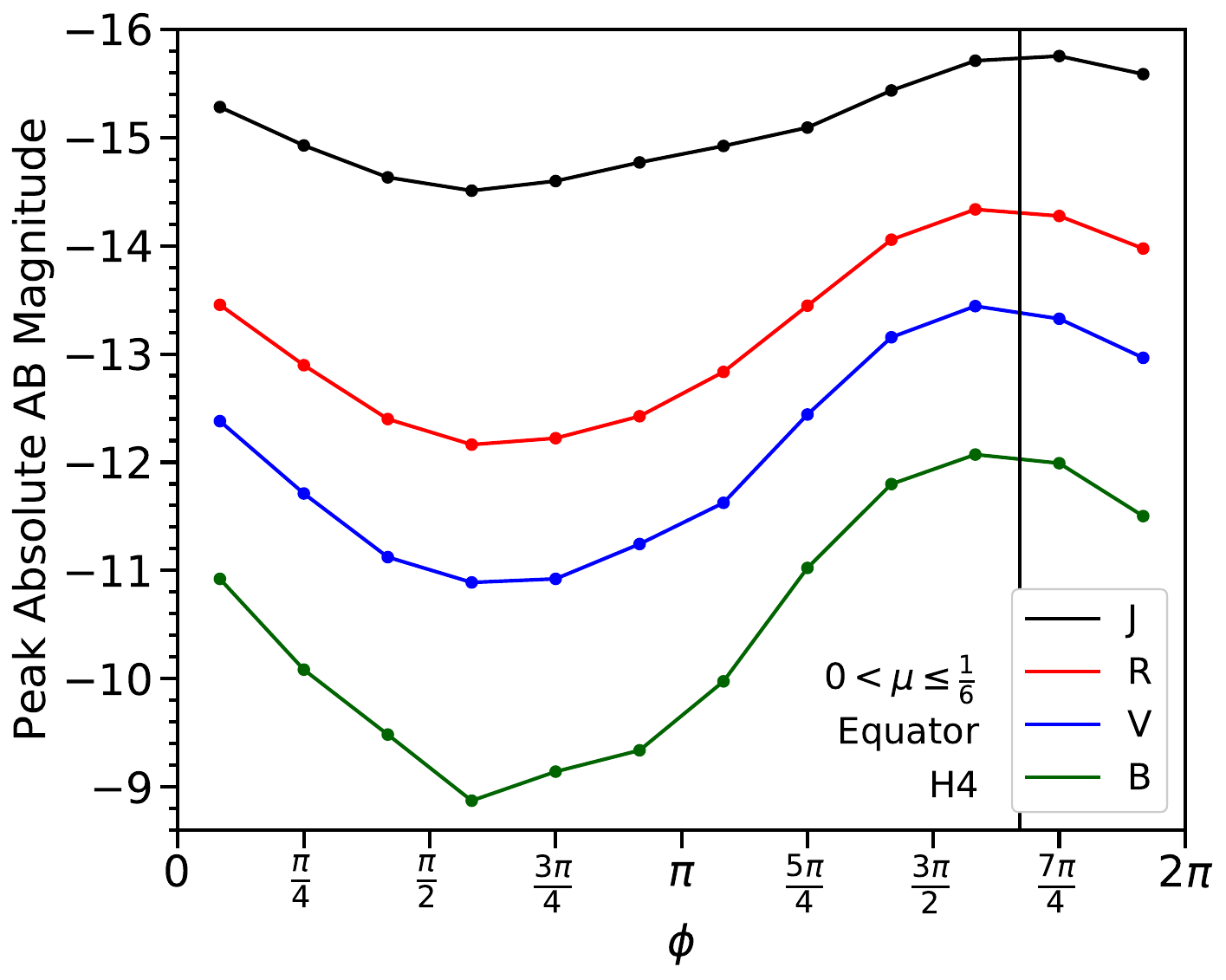}
\caption{
Peak magnitudes of the broadband light curves for the heating model H4. The magnitudes are Absolute AB magnitudes. The polar viewing angle is near the equator, $0 < \mu \leq 1/6$. The colors show the $J$-band (black), $R$-band (red), $V$-band (blue), and $B$-band (green). The black vertical line shows the azimuthal direction $\phi = \phi_P$ of the total momentum.
}
\label{fig:broadband_minima_vs_phi-equator}
\end{figure}

Figure \ref{fig:broadband_minima_vs_phi-equator} shows the peak magnitudes of the broadband light curves when viewed along the equator, which roughly isolates the Doppler modification effect, as discussed previously for Figure \ref{fig:Lpeak_plots}, Panels (b) and (d). In particular, the curves clearly distill the effect of Doppler shifting. In Appendix \ref{sec:light_curve_peaks}, Figures \ref{fig:Jmin_grids} and \ref{fig:Vmin_grids} show the overall trends in the minima of the $J$ and $V$ bands.

The light curves and spectra expand on the results of previous work. We can directly compare the results for H0 with previous studies that used the ballistic approximation \citep{roberts11,tanaka14,fernandez17}, since the ejecta evolves similarly in both cases (Section \ref{subsec:results_hydrodynamics}). \citet{tanaka14} found that the UVOIR luminosity is a factor of $\sim 2$ brighter in the polar direction than the direction of the total momentum. In Figure \ref{fig:bolometric_light_curves}, comparing the brightest curve in Panel (a) to the thick black curve in Panel (b), we similarly find a factor $\sim 2 - 3$ difference. However, with heating (H4), the trend is inverted (Figure \ref{fig:bolometric_light_curves}, Panels c and d): the luminosity is a factor of $\sim 2$ brighter in the direction of the total momentum than the polar direction. In the model H4, the mass and opening angle of the ejecta are comparable to those of the 2D geometric model BHNS\textunderscore DYN in \citet{kawaguchi20a}. They find that the $J$-band has a peak magnitude $\sim -16.5$ along the pole and $\sim -16$ along the equator, both at $\sim 2$ days. In Figure \ref{fig:broadband_light_curves}, we find slightly dimmer values of $\sim -15$ to $-15.6$ along the pole and $\sim -14.4$ to $-15.8$ along the equator, where the spread in each case quantifies the variation with $\phi$. They also find that the $R$-band is dimmer and declines at an earlier time; our results find similar trends.

\section{Discussion}
\label{sec:discussion}

In this paper, we calculated the kilonova signatures from the unbound dynamical ejecta of a BH-NS merger. 
The data for the isotropic-equivalent bolometric and broadband light curves for three of the stage (2) heatings models (H0, H3, and H4) is provided in a .tar.gz package with the online version of this publication. 
We examined a binary in which initially the BH spin is misaligned with the binary orbital plane. We performed GRSPH simulations with a parameterized r-process heating model to study the evolution of the ejecta morphology. We then performed 3D MCRT simulations with a parameterized analytic opacity model, designed to mimic the line opacities of nuclei past the second r-process peak, to study the dependence of the emission on viewing angle. 
We obtained several results:
\begin{enumerate}[1., labelindent=0pt, leftmargin=*, itemsep=0pt]

    \item The unbound dynamical ejecta is initially flattened, directed, and largely confined to a plane. This post-merger ejecta plane differs from the initial binary orbital plane of the NR simulation (Section \ref{subsec:numerical_relativity} and Appendix \ref{sec:ejecta_plane}), and is set by the orbital angular momentum at the instant of merger. We performed an active coordinate transformation to rotate the ejecta to lie in the $xy$-plane to optimize the radiative transfer simulation resolution (Appendix \ref{sec:active_rotation}). This plane serves as a convenient reference for the viewing angle, which we parameterize with the polar direction cosine $\mu = \cos\theta$ and the azimuthal angle $\phi$.

    \item The presence of r-process heating modifies the structure of the tidal ejecta considerably (Figures \ref{fig:hydrodynamic_evolution} and \ref{fig:column_density}). It smooths the small scale inhomogeneities, isotropizes the momentum in the rest frame, inflates the ejecta into a more spherical shape, and mildly accelerates the ejecta to higher velocities. The ejecta retains a bulk, directed motion with characteristic velocity $v \sim 0.2 c$. The direction $(\mu_P,\phi_P)$ of the total momentum is largely insensitive to heating.

    \item The light curves vary significantly with the viewing angle (Figure \ref{fig:Lpeak_plots} and Figure \ref{fig:Lpeak_grids}). This effect is less pronounced for ejecta with r-process heating since their morphology becomes more spherical. For all polar angles, $\mu$, the light curves are brightest for the azimuthal direction $\phi = \phi_P$ aligned with the ejecta total momentum and dimmest in the opposite direction. The light curves are generally brighter from the poles $\mu \sim \pm 1$ than the equator $\mu \sim 0$; however, for realistic levels of heating, the light curves in the direction $\phi = \phi_P$ become brighter from the equator than the poles. The variation with viewing angle can be roughly explained by two effects: projected area and Doppler enhancement/reduction.

    \item The observed spectral intensity lies primarily in the IR, peaking near 1~micron (Figure \ref{fig:spectra}). The spectra have a near-blackbody shape at early times and deviate from it near and after peak, first at longer wavelengths then at shorter ones. The optical flux is on the exponential Wien part of the blackbody, and so is significantly affected by Doppler shifting, being enhanced when the ejecta moves towards the observer. These results are based on transport calculations with an analytic opacity function that replicates lanthanide-like opacities.
    
    \item The broadband light curves differ considerably between the IR and optical bands (Figure \ref{fig:broadband_light_curves} and Figures \ref{fig:broadband_light_curves_IR} -- \ref{fig:broadband_light_curves_optical_UV}). The IR bands are brighter, have a smaller viewing angle variation, and peak at slightly later times compared to the optical/UV bands. The larger variation with orientation in the optical is due to the Doppler boosting, which enhances the optical emission in the direction $(\mu_P,\phi_P)$. From this angle, the peak R-band magnitude is $\gtrsim -14$, about a magnitude brighter than a comparable spherical model \citep{barnes13}, raising the prospects for the optical detection of lanthanide-rich dynamical ejecta in BH-NS mergers.
    
\end{enumerate}

Our study only examined the unbound component of the dynamical ejecta. A more comprehensive treatment would also consider the fallback material, disk winds, and a jet. The neutrino emission and winds from a disk likely have little impact on the structure or composition of the 
unbound dynamical component \citep{fernandez15,fernandez17, roberts17}. 
Since the dynamical ejecta in BH-NS mergers lies outside the polar regions, it is also likely unaffected by a jet, in contrast to the situation in NS-NS mergers \citep{klion20}. The presence of a lanthanide-poor disk wind can contribute bluer emission, although this optical light will be obscured for certain viewing angles by the overlying dynamical ejecta \citep{kasen15}. In our BH-NS merger model H4, the dynamical ejecta only subtends a solid angle of $\Omega/4\pi \approx 0.1$, suggesting that the disk wind will be visible for $\sim 90\%$ of orientations, significantly greater than that expected for NS-NS mergers. 

Our simulations used an approximate $\gamma$-law EOS and a parameterized radioactive heating rate that did not evolve the composition. To improve upon this, one should use an EOS that accounts for dense matter effects and heating rates derived from detailed nuclear reaction rate calculations, as differences in heating due to the ejecta composition or input nuclear physics can significantly affect the kilonova signatures \citep{barnes20}.

Our radiation transport calculation used an analytic opacity function. A more physical model with bound-bound line opacities derived from atomic structure calculations may modify our predicted light curves quantitatively. Synthetic spectra calculated using realistic opacities exhibit broad spectral features \citep{barnes13}, which could be potentially useful diagnostics of the orientation. For equatorial viewing angles, the spectral features will all be systematically blueshifted (redshifted) when the bulk motion of the kilonova is towards (away from) the observer.

The merger model studied here set the initial BH spin at an angle $\iota = 60^\circ$ with respect to the initial binary orbital angular momentum.  As a result, the plane of the dynamical ejecta was inclined relative to the BH spin plane. The ejecta from post-merger disk winds is thus likely to be misaligned with the dynamical ejecta, although this may have only subtle effects on the kilonova properties. For BH-NS mergers with aligned spin, the ejecta plane and BH spin plane will coincide, and the general structure of the dynamical ejecta should be similar to that studied here. In all cases the GW and kilonova signals vary with viewing angle in a correlated fashion, which should be taken into account in joint analysis and detectibility estimates.

\acknowledgments
{
We thank the Reviewer for helpful comments. 
This research used resources of the National Energy Research Scientific Computing Center, a Department of Energy Office of Science User Facility supported by the Office of Science of the U.S. Department of Energy under Contract No. DE-AC02-05CH11231. 
This research was supported in part by the U.S. Department of Energy, Office of Science, Office of Nuclear Physics, under contract number DE-AC02-05CH11231 and DE-SC0017616, by a SciDAC award DE-SC0018297, and by the Gordon and Betty Moore Foundation through Grant GBMF5076. 
DK acknowledges support from the Simons Foundation Investigator program under award number 622817. 
FF gratefully acknowledges support from the NSF through grant PHY-1806278, from the DOE through grant DE-SC0020435, and from NASA through grant 80NSSC18K0565. 
This collaborative work was supported in part by the NSF Physics Frontier Center N3AS under cooperative agreement \#2020275. 
DP was supported by Australian Research Council grant FT130100034. 
SD thanks the Yukawa Institute for Theoretical Physics at Kyoto University. Discussions during the YITP-T-19-07 International Molecule-type Workshop ``Tidal Disruption Events: General Relativistic Transients" were useful to complete this work.
}

%



\software{
\textsc{sedona} \citep{kasen06},
\textsc{phantom} \citep{price18,liptai19},
\textsc{spec} \citep{spec},
numpy \citep{harris20},
matplotlib \citep{hunter07}
}



\appendix

\section{Schwarzschild Metric in 3+1 Form}
\label{sec:3_plus_1}

We modeled the gravity of the post-merger BH using the Schwarzschild metric. In this appendix, we write the Schwarzschild metric in 3+1 form and express some relevant quantities in that framework.

The Schwarzschild spacetime is globally hyperbolic and the Schwarzschild coordinates $x^\mu = (t,x^i)$ are adapted to a 3+1 foliation \citep{gourgoulhon07,baumgarte10}, i.e. the time coordinate $t$ is global and foliates the spacetime into spacelike hypersurfaces $\Sigma_t$ on which we define the spatial coordinates $x^i$. The metric in these coordinates thus has a 3+1 form with lapse function $\alpha = \left( 1 - \frac{2M}{r} \right)^{1/2}$, shift vector $\beta^i = 0$, and spatial metric $\gamma_{ij} = \operatorname{diag} \left[ \left( 1 - \frac{2M}{r} \right)^{-1}, r^2, r^2 \sin^2\theta \right]$ induced on $\Sigma_t$. The unit one-form that gives the direction of the hypersurfaces is $n_a = -\left( 1 - \frac{2M}{r} \right)^{1/2} (dt)_a$, where $(dt)_a$ is the exterior derivative of $t$. The unit timelike vector field $n^a = g^{ab} n_a = \left( 1 - \frac{2M}{r} \right)^{-1/2} (\partial_t)^a$ is normal to the hypersurfaces, $g_{ab} n^a (\partial_i)^b = 0$. Eulerian observers are defined as the observers with four-velocity $n^a$, who thus perform measurements in the adapted basis $(e_\mu)^a = \lbrace n^a, (\partial_i)^a \rbrace$.

The four-velocity $u^a$ of a timelike curve $\eta$ can be written in a coordinate basis as $u^a = u^\mu (\partial_\mu)^a$, with components $u^\mu = \frac{dx^\mu}{d\tau} \equiv \frac{d(x^\mu \circ \eta)}{d\tau}$ where $\tau$ is the proper time. In the basis of an Eulerian observer, the four-velocity is
\begin{equation}
    u^a = \Gamma \left[ n^a + \bar{v}^i (\partial_i)^a \right] ,
\end{equation}
where 
$\Gamma \equiv -g_{ab} u^a n^b = \left( 1 - \frac{2M}{r} \right)^{1/2} u^t$ is the local Lorentz factor measured by the Eulerian observer, which can also be written as $\Gamma = (1 - \gamma_{ij} \bar{v}^i \bar{v}^j)^{-1/2}$ due to the normalization $u^a u_a = -1$, and $\bar{v}^i$ is the 3-velocity measured by the Eulerian observer, which can be obtained from the definition $\Gamma \bar{v}_i \equiv g_{ab} u^a (\partial_i)^b$. The four-velocity can also be written in terms of the coordinate 3-velocity $v^i \equiv \frac{dx^i}{dt} = \frac{u^i}{u^t}$ as
\begin{equation}
    u^a = \Gamma \left( 1 - \frac{2M}{r} \right)^{-1/2} \left[ (\partial_t)^a + v^i (\partial_i)^a \right] .
\end{equation}
The 3-velocities $\bar{v}^i$ and $v^i$ are thus related by
\begin{equation}
    v^i = \left( 1 - \frac{2M}{r} \right)^{1/2} \bar{v}^i .
\end{equation}

\section{Ejecta Plane}
\label{sec:ejecta_plane}

In Section \ref{subsec:numerical_relativity}, we noted that the post-merger ejecta is roughly concentrated in a plane that differs from the orbital plane at the beginning of the NR simulation. In this appendix, we present a method to calculate the optimal orientation of this plane using the particles extracted from the NR snapshot.

Let $M$ be the Schwarzschild spacetime of the post-merger BH. Let $\Sigma_t$ be the spatial hypersurface defined at the time $t = t_\mathrm{nr,f}$. Let $(r,\theta,\phi)$ be the coordinates of the pre-merger orbital plane. Let $(\beta,\alpha)$ be the polar and azimuthal angles that define a direction. We perform a passive rotation to a new coordinate system $x'^{\mu'}(x^\nu)$ with $z'$-axis in the direction $(\beta,\alpha)$. The new coordinate $z'$ is related to the old coordinates $(r,\theta,\phi)$ by
\begin{equation}
    z' = r \left[ \sin\beta \sin\theta \cos(\phi - \alpha) + \cos\beta \cos\theta \right] .
\end{equation}
The unit one-form $N_a = \mathcal{N}^{-1/2} (dz')_a$ describes a family of constant $z'$ planes, where $(dz')_b$ is the exterior derivative of $z'$ and the normalization $\mathcal{N} = g^{ab} (dz')_a (dz')_b$ is given by
\begin{equation}
    \mathcal{N} = 1 - \frac{2M}{r} \left[ \frac{z'}{r} \right]^2 .
\end{equation}
The vector field normal to these planes is $N^a = g^{ab} N_b$. The outward radial direction is given by the unit vector field 
$(e_r)^a = \left( 1 - \frac{2M}{r} \right)^{1/2} (\partial_r)^a$. At the location of a particle $i$, let $\Omega_{(i)}$ be the inner product of these two vector fields, i.e. $\Omega_{(i)} \equiv \left. N_a (e_r)^a \right|_{(i)}$, which in coordinates becomes
\begin{equation}
    \Omega_{(i)} = \left( 1 - \frac{2M}{r_{(i)}} \right)^{1/2} \mathcal{N}_{(i)}^{-1/2} \left[ \frac{z'_{(i)}}{r_{(i)}} \right] .
\end{equation}
This inner product encapsulates the degree to which the outward radial vector of particle $i$ lies in the plane of constant $z'$ that intersects it; if it lies in the plane then $\Omega_{(i)} = 0$, and as it becomes more orthogonal $\Omega_{(i)}$ takes a larger value. The plane that optimally accommodates all the particles has its $z'$-axis in the direction $(\beta^\mathrm{opt}, \alpha^\mathrm{opt}) = \argmin_{(\beta,\alpha)} \norm{\Omega}^2$, where $\norm{\cdot}$ is the $L^2$-norm over the particles $i = 1, \hdots, N$. For the model M14M5S9I60, we find $(\beta^\mathrm{opt}, \alpha^\mathrm{opt}) = (0.861, 1.86)$.

\section{Active Rotation}
\label{sec:active_rotation}

In Section \ref{subsec:numerical_relativity}, we performed an active rotation on the particles to align the post-merger plane of the ejecta with the $xy$-plane. In this appendix, we outline the details of that transformation.

Let $M$ be the Schwarzschild spacetime of the post-merger BH. Let $(r,\theta,\phi)$ be the coordinates of the pre-merger orbital plane. Let $(\beta,\alpha)$ be the polar and azimuthal angles that define a direction. Let $\psi$ be an active rotation (i.e. a diffeomorphism) that rotates the point $(\beta,\alpha)$ to the point $\theta = 0$ (i.e., the $z$-axis). The source point $p \in M$ has coordinates $x^\mu = x^\mu(p)$ and the target point $\psi(p)$ has coordinates $\tilde{x}^{\tilde{\mu}} = (x^\mu \circ \psi)(p)$. In rectangular coordinates, the transformation is
\begin{align}
    \tilde{t} &= t , \\
    \begin{pmatrix}
        \tilde{x} \\
        \tilde{y} \\
        \tilde{z}
    \end{pmatrix}
    & =
    \begin{pmatrix}
        \cos(-\beta) & 0 & \sin(-\beta) \\
        0 & 1 & 0 \\
        -\sin(-\beta) & 0 & \cos(-\beta)
    \end{pmatrix}
    \begin{pmatrix}
        \cos(-\alpha) & -\sin(\alpha) & 0 \\
        \sin(-\alpha) & \cos(-\alpha) & 0 \\
        0 & 0 & 0
    \end{pmatrix}
    \begin{pmatrix}
        x \\
        y \\
        z
    \end{pmatrix} .
\end{align}
In spherical coordinates, the transformation is thus
\begin{align}
    \tilde{t} &= t , \\
    \tilde{r} &= r , \\
    \cos\tilde{\theta} &= \sin\beta \sin\theta \cos{(\phi - \alpha)} + \cos\beta \cos\theta , \\
    \tan\tilde{\phi} &= \frac{\sin{(\phi - \alpha)}}{\cos\beta \cos{(\phi - \alpha)} - \sin\beta \cot\theta} .
\end{align}
We note that the transformation has the same expression as a passive rotation to a new coordinate system $x'^{\mu'}(x^\nu)$ with $z'$-axis in the direction $(\beta,\alpha)$, as in Appendix \ref{sec:ejecta_plane}.

The 4-velocity at the source point $p$ is $u^a_p = u^\mu \left. (\partial_\mu)^a \right|_p$ and the 4-velocity at the target point $\psi(p)$ is $\tilde{u}^a_{\psi(p)} = \psi_*(u^a_p) = \tilde{u}^{\tilde{\mu}} \left. (\partial_{\tilde{\mu}})^a \right|_{\psi(p)}$, where $\psi_*$ is the pushforward of $\psi$. 
The 4-velocity transforms as $\tilde{u}^{\tilde{\mu}}_{\psi(p)} = \left(\frac{\partial(x^\nu \circ \psi)}{\partial x^\nu}\right)_p u^\nu_p$, or explicitly
\begin{align}
    \tilde{u}^{\tilde{t}} &= u^t , \\
    \tilde{u}^{\tilde{r}} &= u^r , \\
    \tilde{u}^{\tilde{\phi}} &= \left(\frac{\partial\tilde{\theta}}{\partial\theta}\right)_p u^\theta + \left(\frac{\partial\tilde{\theta}}{\partial\phi}\right)_p u^\phi , \\
    \tilde{u}^{\tilde{\theta}} &= \left(\frac{\partial\tilde{\phi}}{\partial\theta}\right)_p u^\theta + \left(\frac{\partial\tilde{\phi}}{\partial\phi}\right)_p u^\phi ,
\end{align}
where
\begin{align}
    \left(\frac{\partial\tilde{\theta}}{\partial\theta}\right)_p &= - \frac{\sin\beta \cos\theta \cos(\phi-\alpha) - \cos\beta \sin\theta}{\sin\tilde{\theta}} , \\
    \left(\frac{\partial\tilde{\theta}}{\partial\phi}\right)_p &= \frac{\sin\beta \sin\theta \sin(\phi-\alpha)}{\sin\tilde{\theta}} , \\
    \left(\frac{\partial\tilde{\phi}}{\partial\theta}\right)_p &= - \cos^2\tilde{\phi} \frac{\sin\beta \csc^2\theta \sin(\phi-\alpha)}{[\cos\beta \cos(\phi-\alpha) - \sin\beta \cot\theta]^2} , \\
    \left(\frac{\partial\tilde{\phi}}{\partial\phi}\right)_p &= \cos^2\tilde{\phi} \frac{\cos\beta - \sin\beta \cot\theta \cos(\phi-\alpha)}{[\cos\beta \cos(\phi-\alpha) - \sin\beta \cot\theta]^2} .
\end{align}

\section{Broadband Light Curves}
\label{sec:broadband_light_curves}

\begin{figure*}
\gridline{
\fig{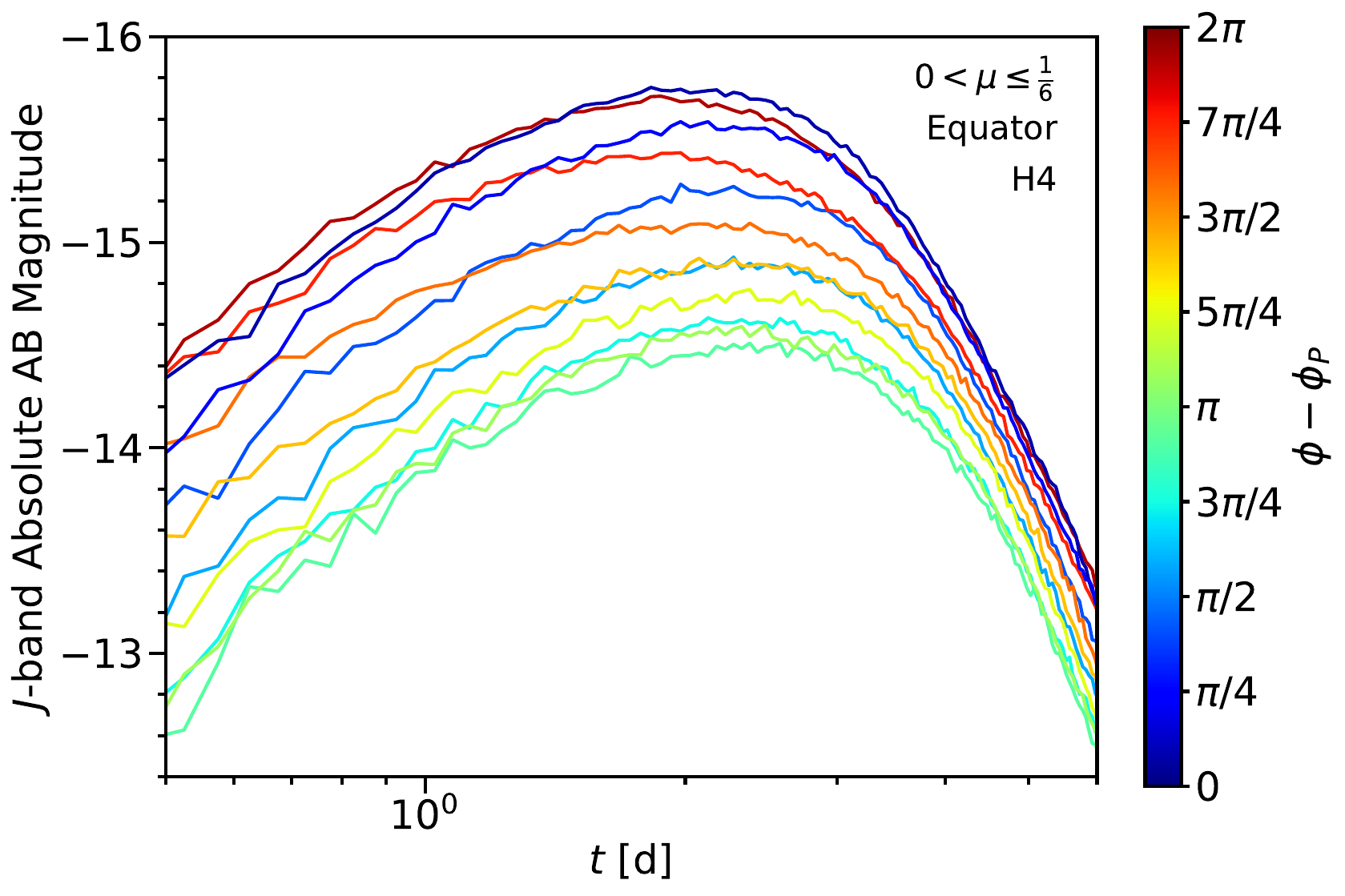}{0.49\textwidth}{(a) $J$-band, $0 < \mu \leq \frac{1}{6}$ (Equator)}
\fig{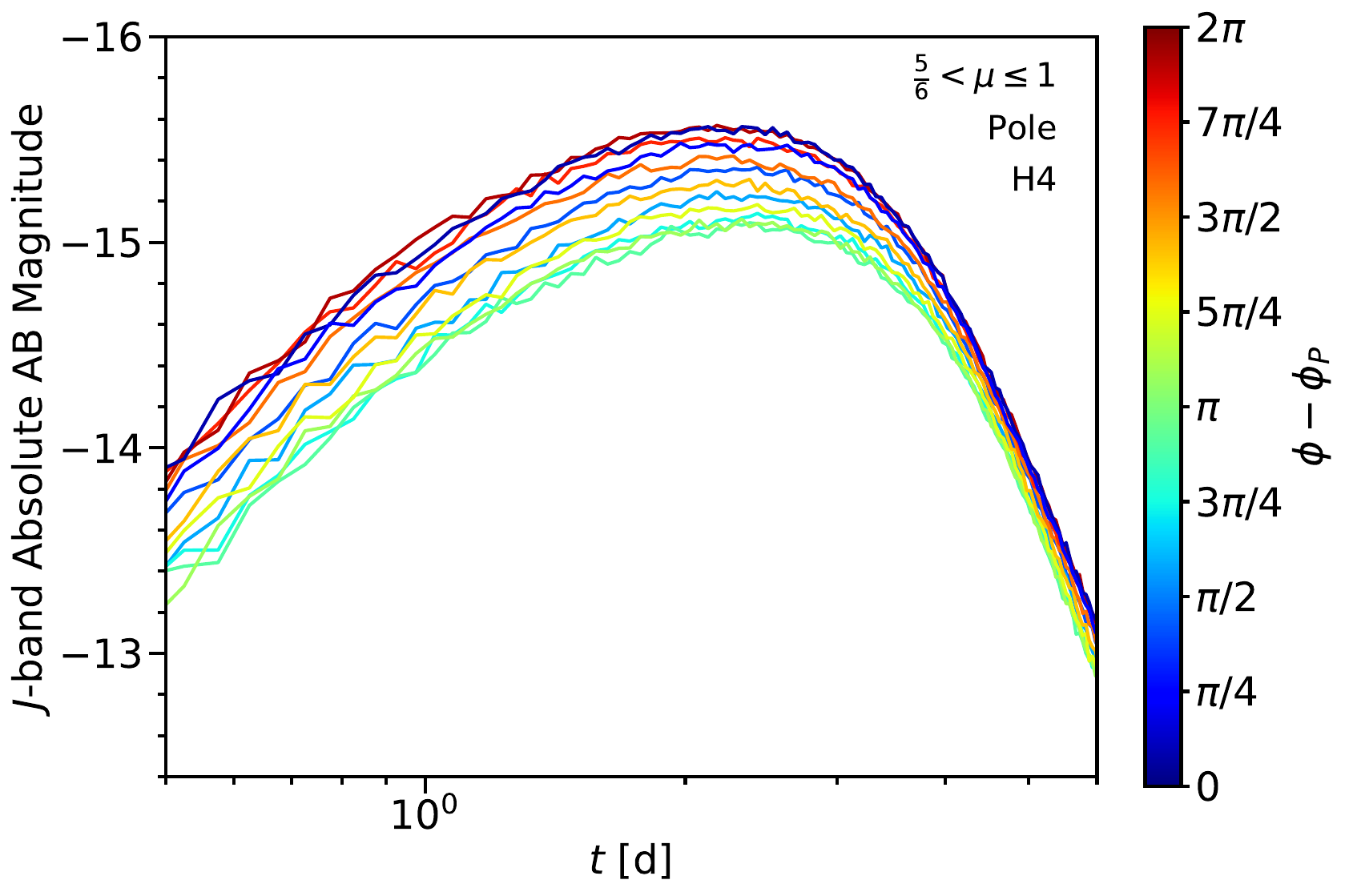}{0.49\textwidth}{(b) $J$-band, $\frac{5}{6} < \mu \leq 1$ (Pole)}
}
\gridline{
\fig{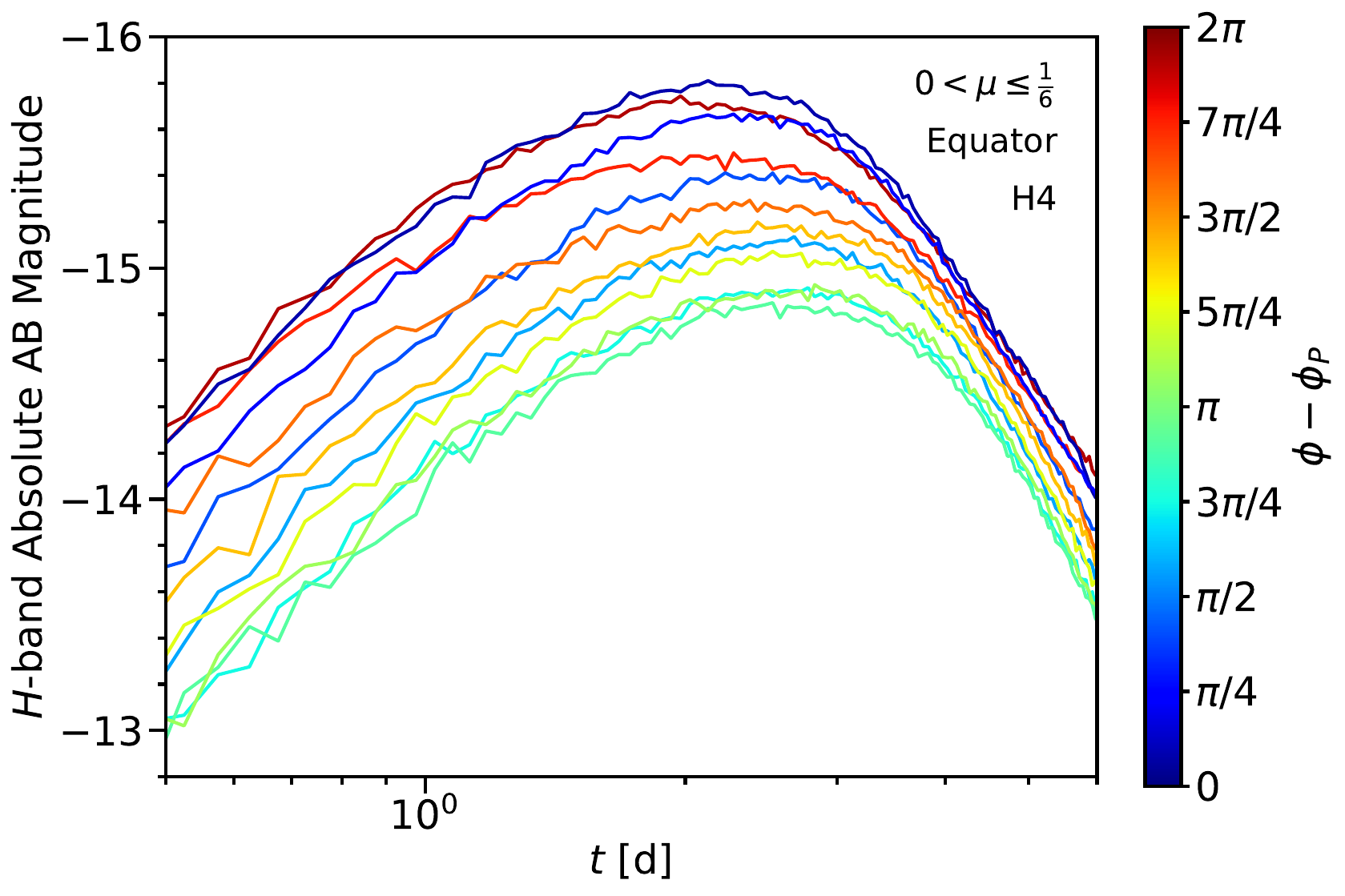}{0.49\textwidth}{(c) $H$-band, $0 < \mu \leq \frac{1}{6}$ (Equator)}
\fig{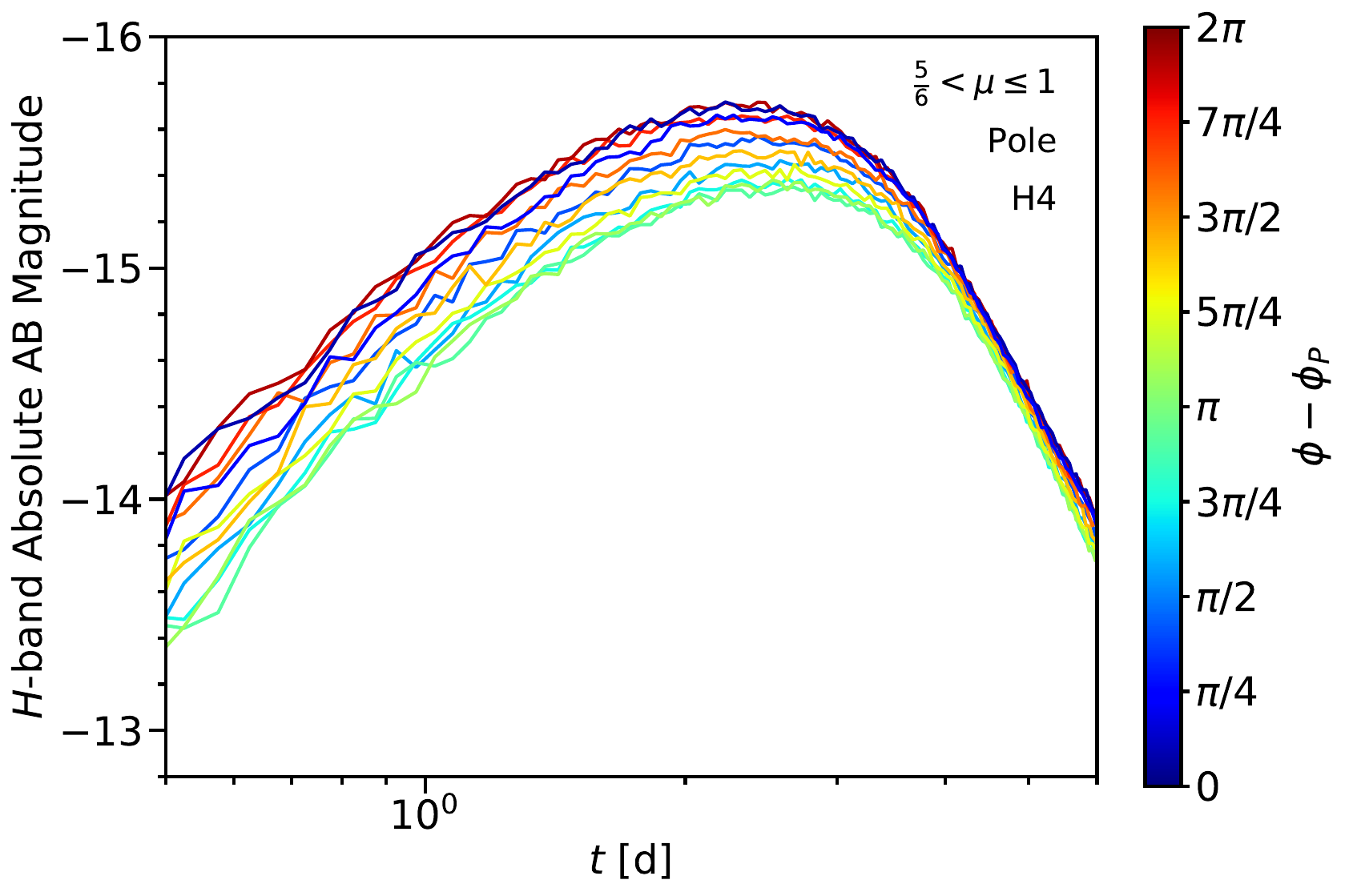}{0.49\textwidth}{(d) $H$-band, $\frac{5}{6} < \mu \leq 1$ (Pole)}
}
\gridline{
\fig{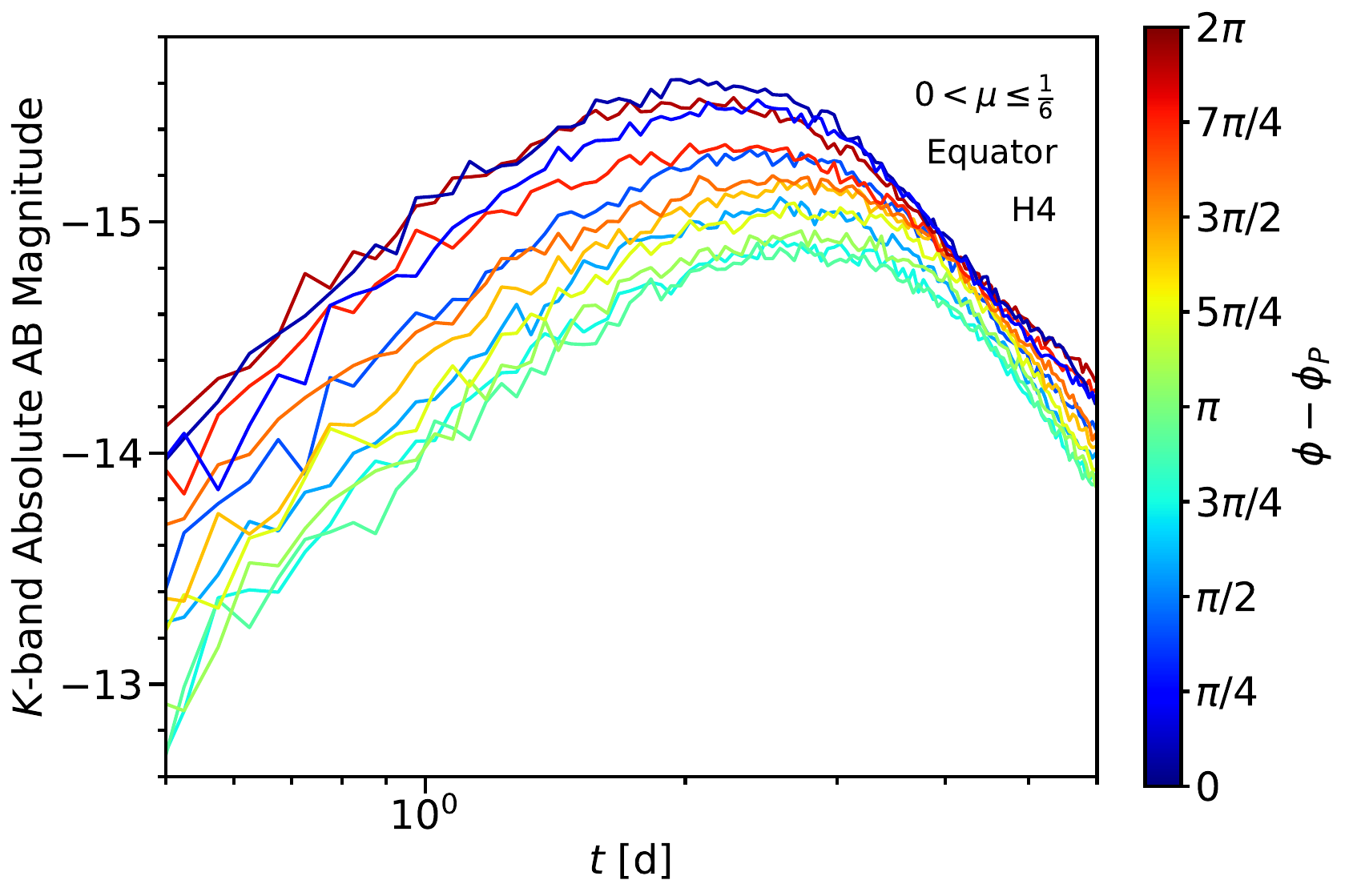}{0.49\textwidth}{(c) $K$-band, $0 < \mu \leq \frac{1}{6}$ (Equator)}
\fig{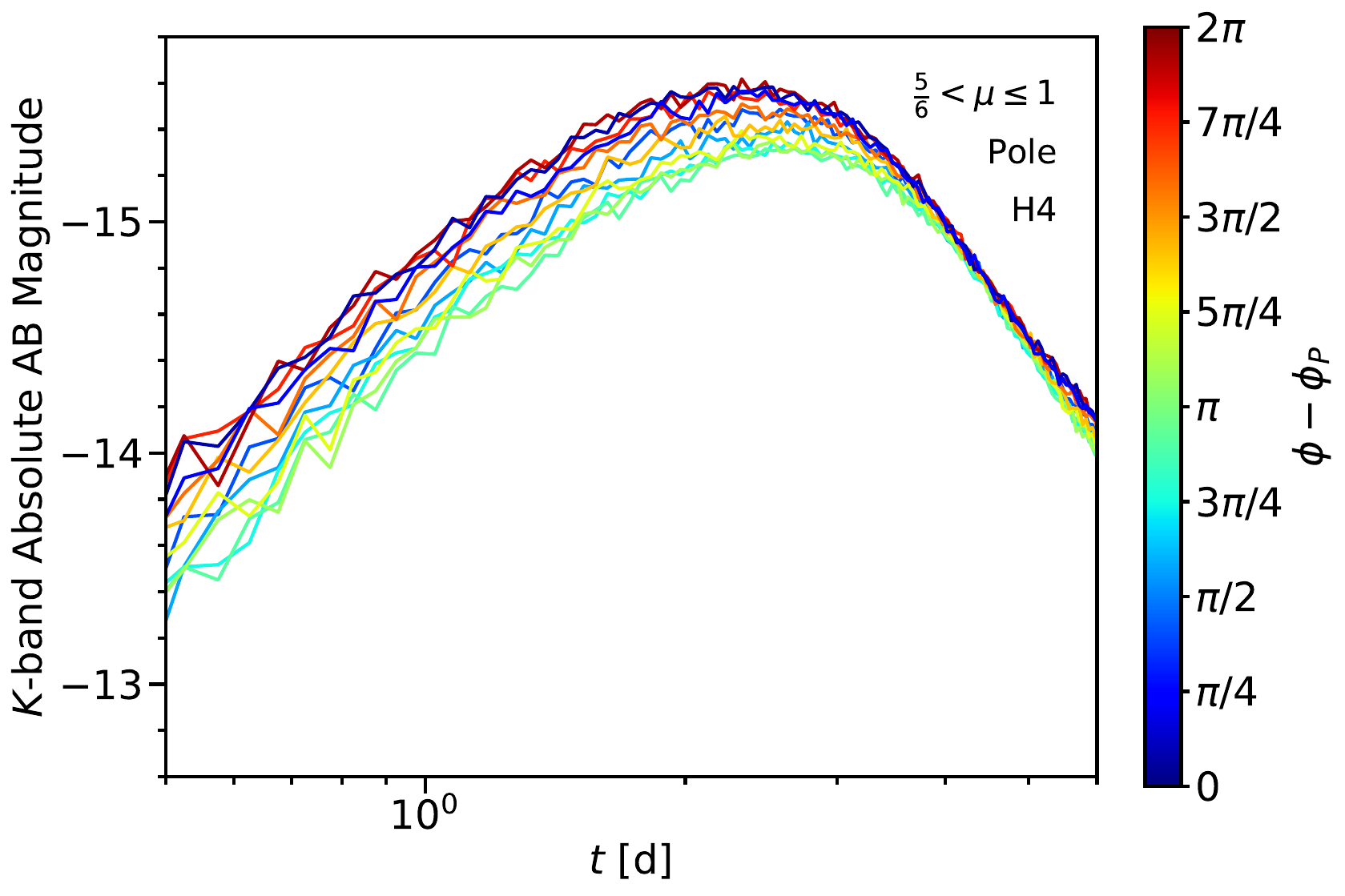}{0.49\textwidth}{(d) $K$-band, $\frac{5}{6} < \mu \leq 1$ (Pole)}
}
\caption{Isotropic-equivalent broadband light curves in the IR bands for the heating model H4. The magnitudes are Absolute AB magnitudes. The rows correspond to the $J$ (top), $H$ (middle), and $K$ (bottom) bands. The columns correspond to different bins of $\mu = \cos\theta$, where $\theta$ is the polar angle. The color bar shows the centers of the 
$\phi - \phi_P$ bins. In each panel, the brightest curve corresponds to the azimuthal direction $\phi_P \simeq 5.2$ of the total momentum and the dimmest curve corresponds to the opposite direction $\phi_P - \pi$. The light curves are roughly the same for $\mu \rightarrow -\mu$ because the ejecta are roughly symmetric about $z = 0$ after the active rotation in stage (1) (Section \ref{subsec:numerical_relativity}).}
\label{fig:broadband_light_curves_IR}
\end{figure*}

\begin{figure*}
\gridline{
\fig{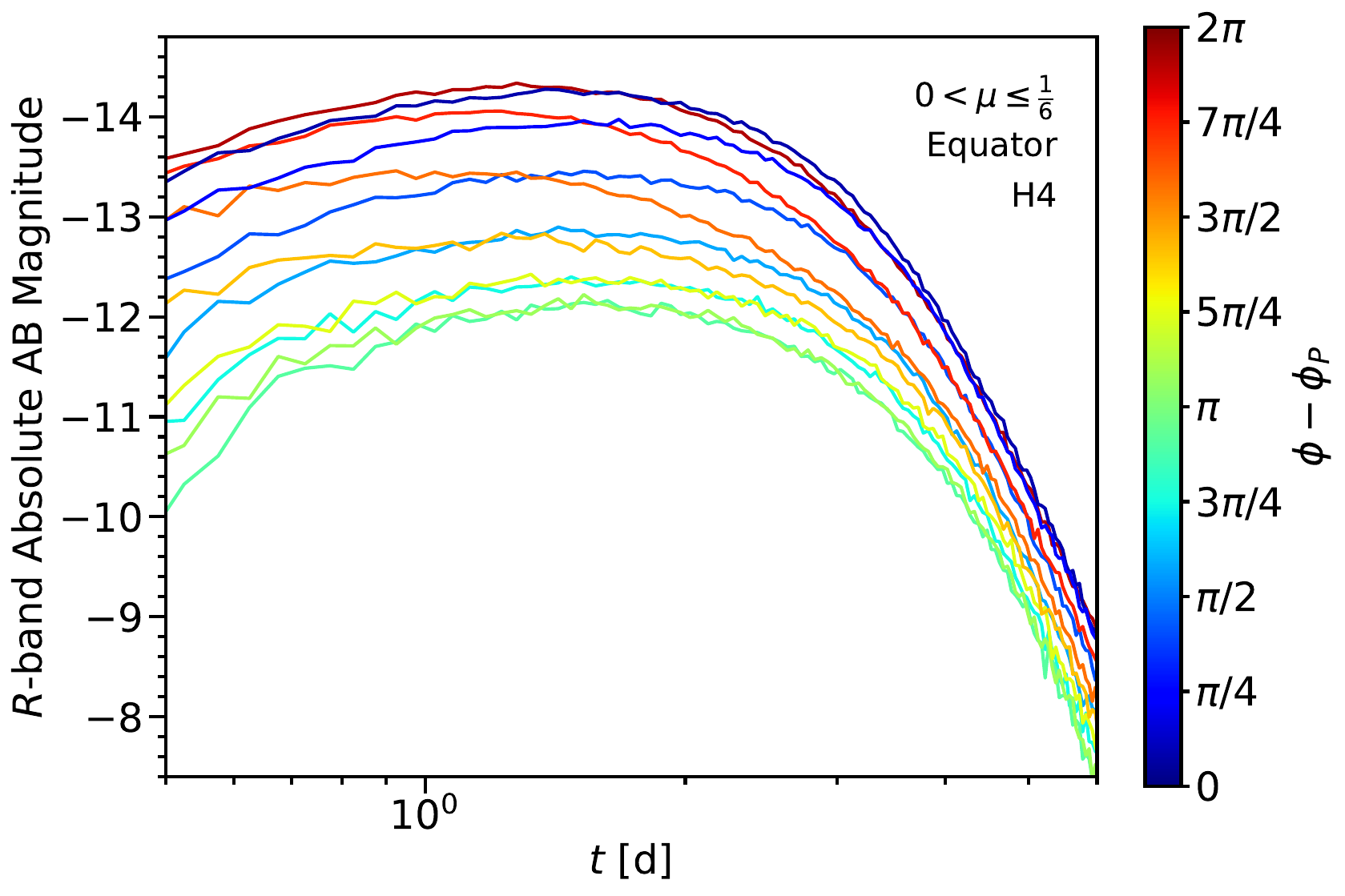}{0.49\textwidth}{(c) $R$-band, $0 < \mu \leq \frac{1}{6}$ (Equator)}
\fig{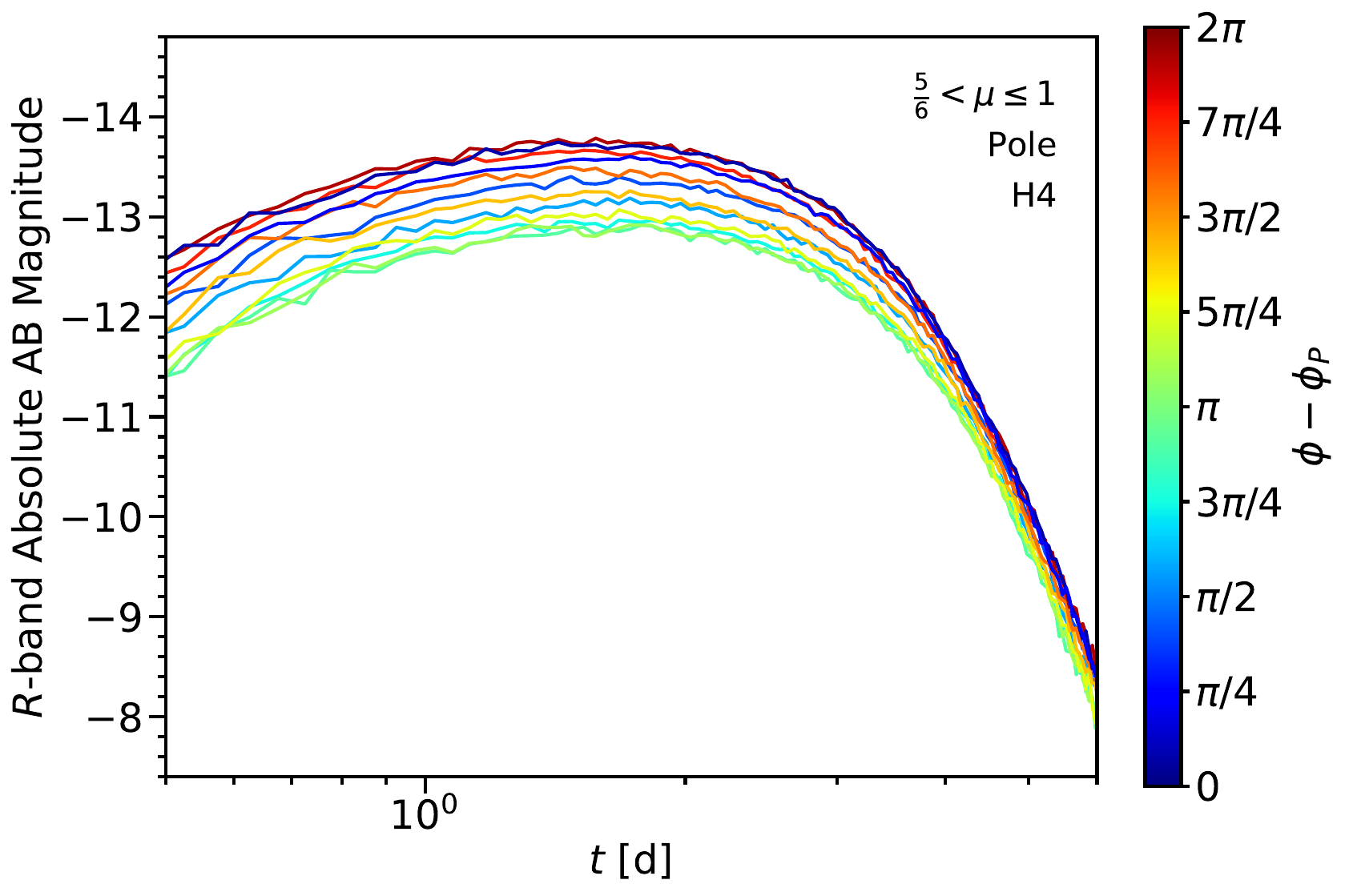}{0.49\textwidth}{(d) $R$-band, $\frac{5}{6} < \mu \leq 1$ (Pole)}
}
\gridline{
\fig{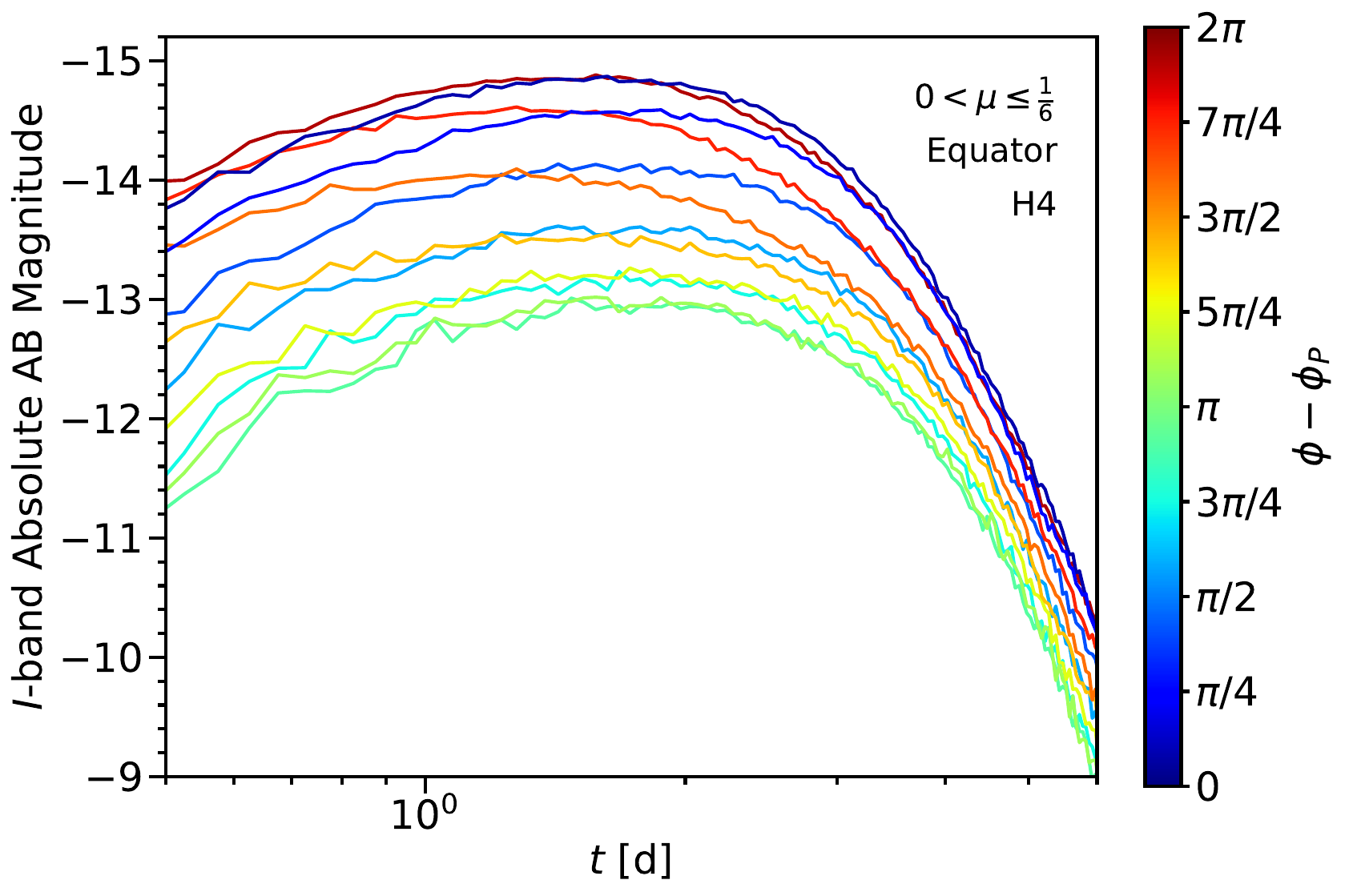}{0.49\textwidth}{(c) $I$-band, $0 < \mu \leq \frac{1}{6}$ (Equator)}
\fig{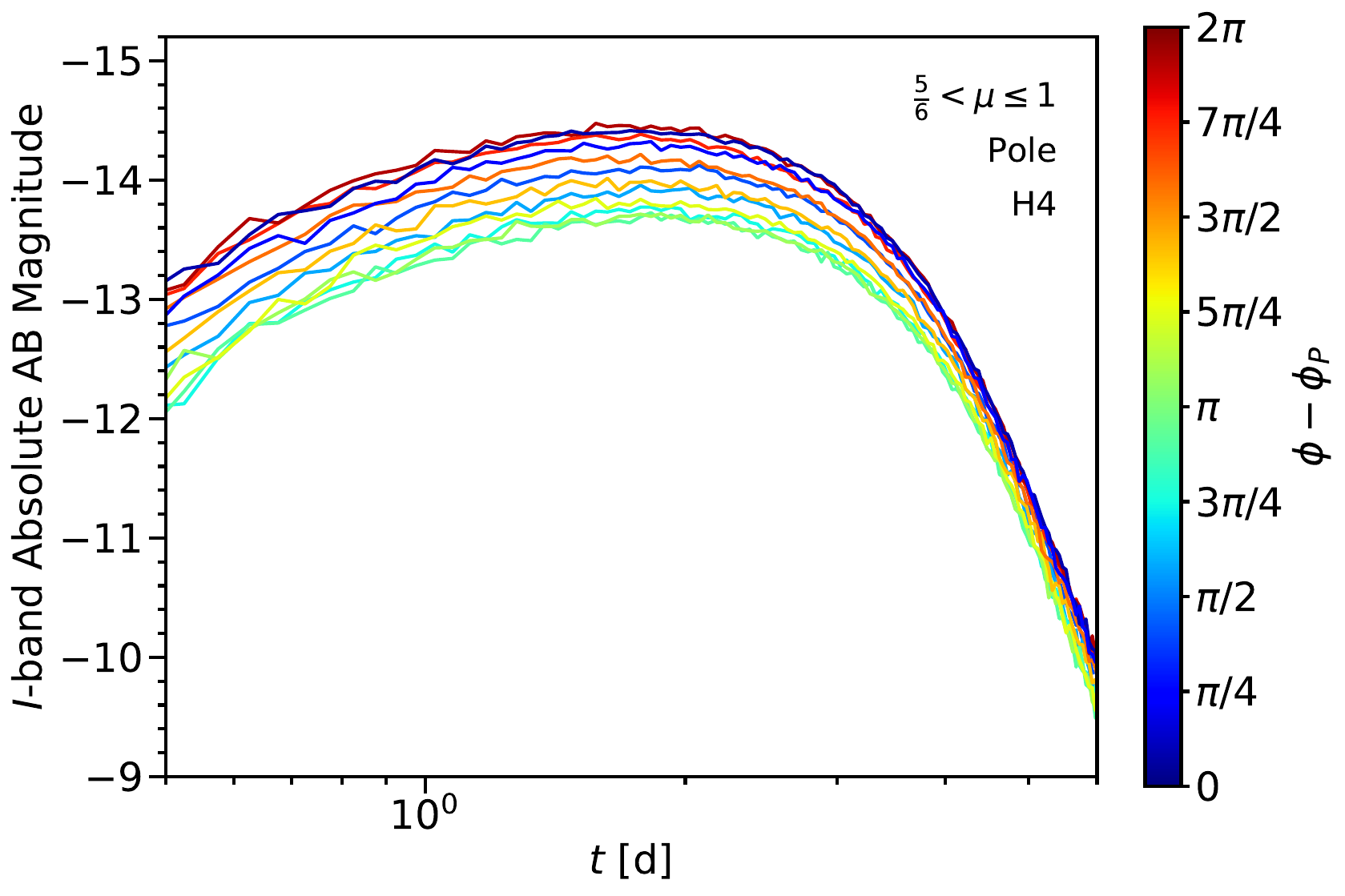}{0.49\textwidth}{(d) $I$-band, $\frac{5}{6} < \mu \leq 1$ (Pole)}
}
\caption{Isotropic-equivalent broadband light curves in the optical/IR bands for the heating model H4. The magnitudes are Absolute AB magnitudes. The rows correspond to the $R$ (top) and $I$ (bottom) bands. The columns correspond to different bins of $\mu = \cos\theta$, where $\theta$ is the polar angle. The color bar shows the centers of the 
$\phi - \phi_P$ bins. In each panel, the brightest curve corresponds to the azimuthal direction $\phi_P \simeq 5.2$ of the total momentum and the dimmest curve corresponds to the opposite direction $\phi_P - \pi$. The light curves are roughly the same for $\mu \rightarrow -\mu$ because the ejecta are roughly symmetric about $z = 0$ after the active rotation in stage (1) (Section \ref{subsec:numerical_relativity}).}
\label{fig:broadband_light_curves_optical_IR}
\end{figure*}

\begin{figure*}
\gridline{
\fig{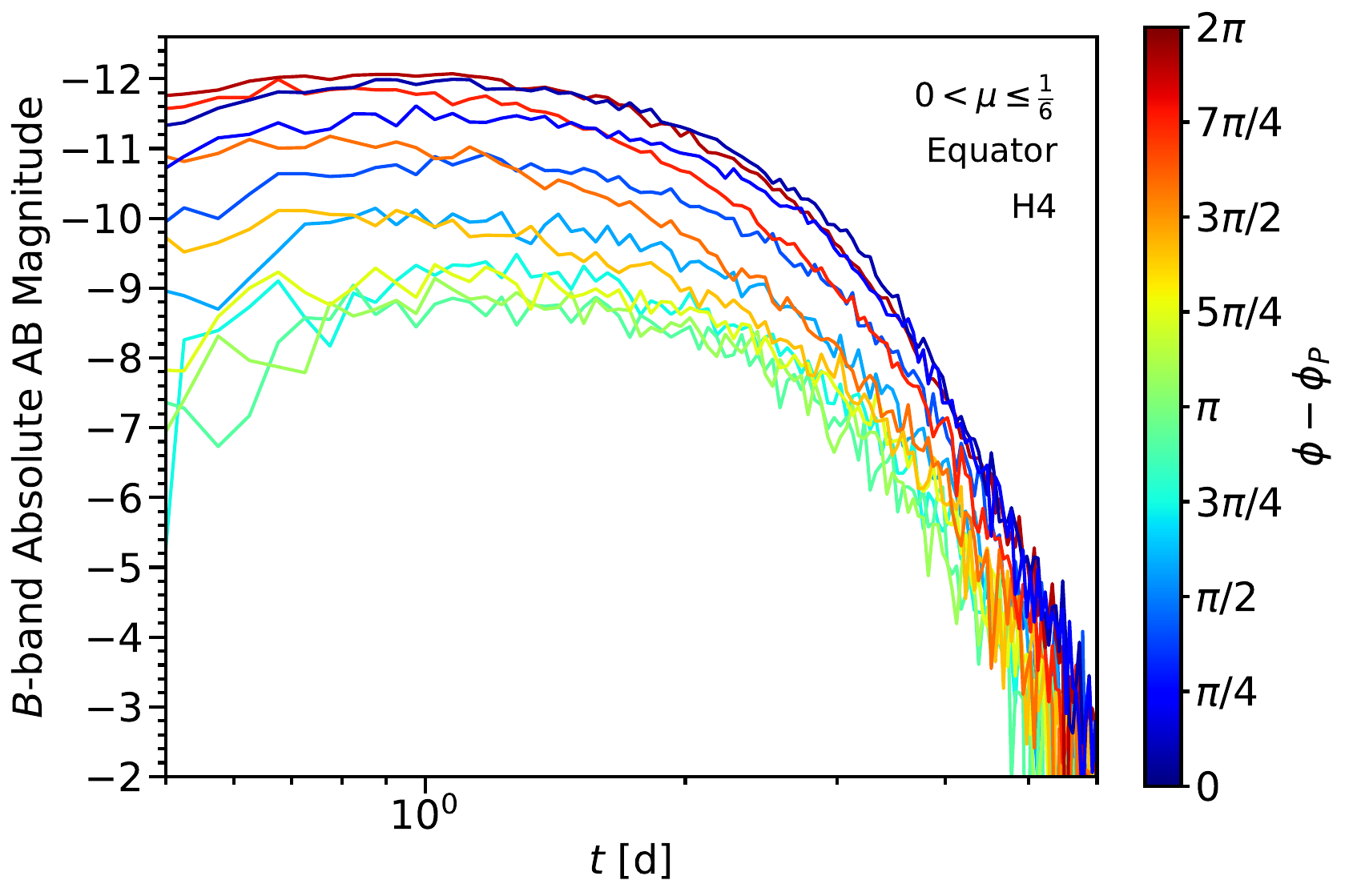}{0.49\textwidth}{(a) $B$-band, $0 < \mu \leq \frac{1}{6}$ (Equator)}
\fig{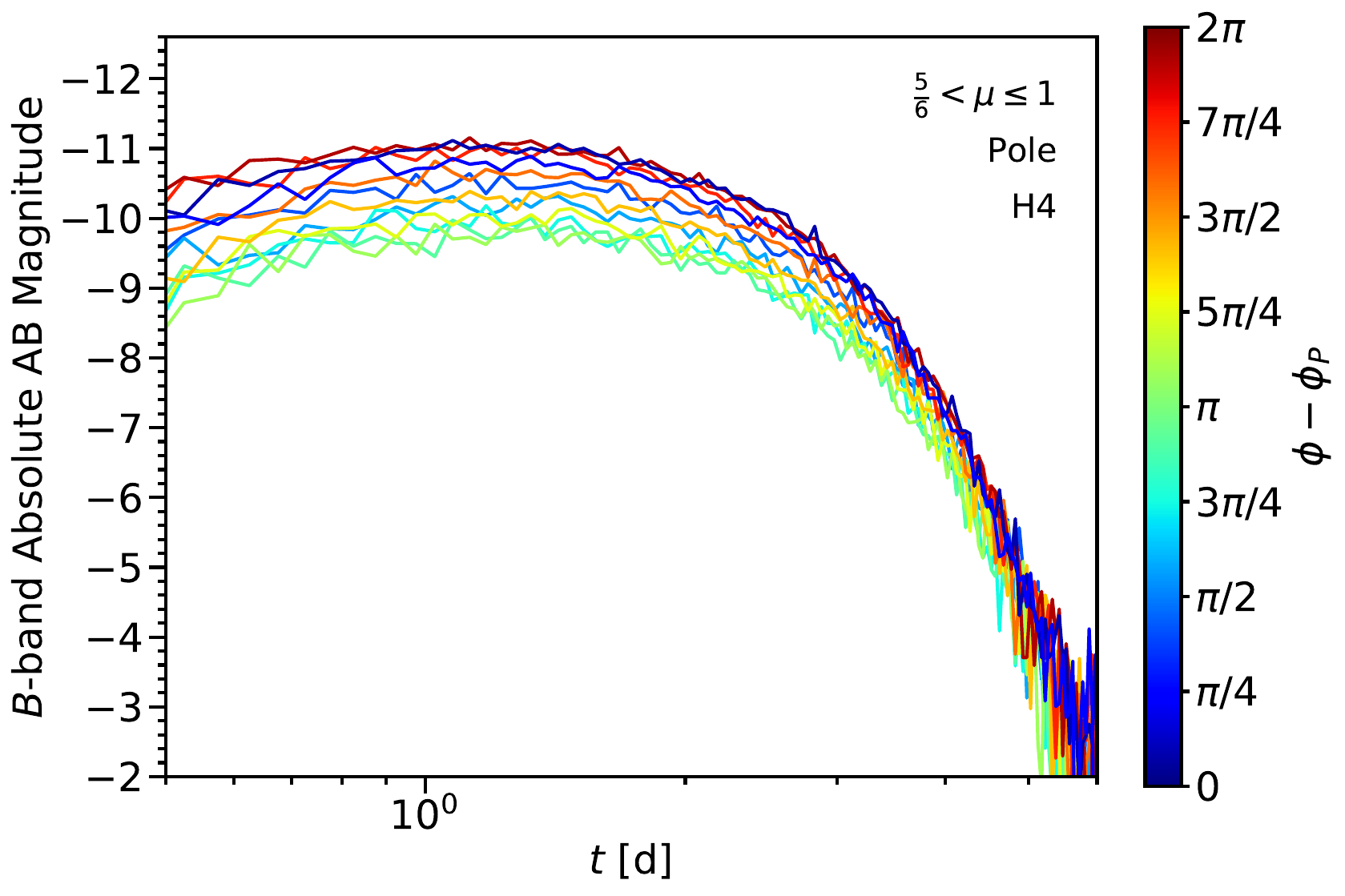}{0.49\textwidth}{(b) $B$-band, $\frac{5}{6} < \mu \leq 1$ (Pole)}
}
\gridline{
\fig{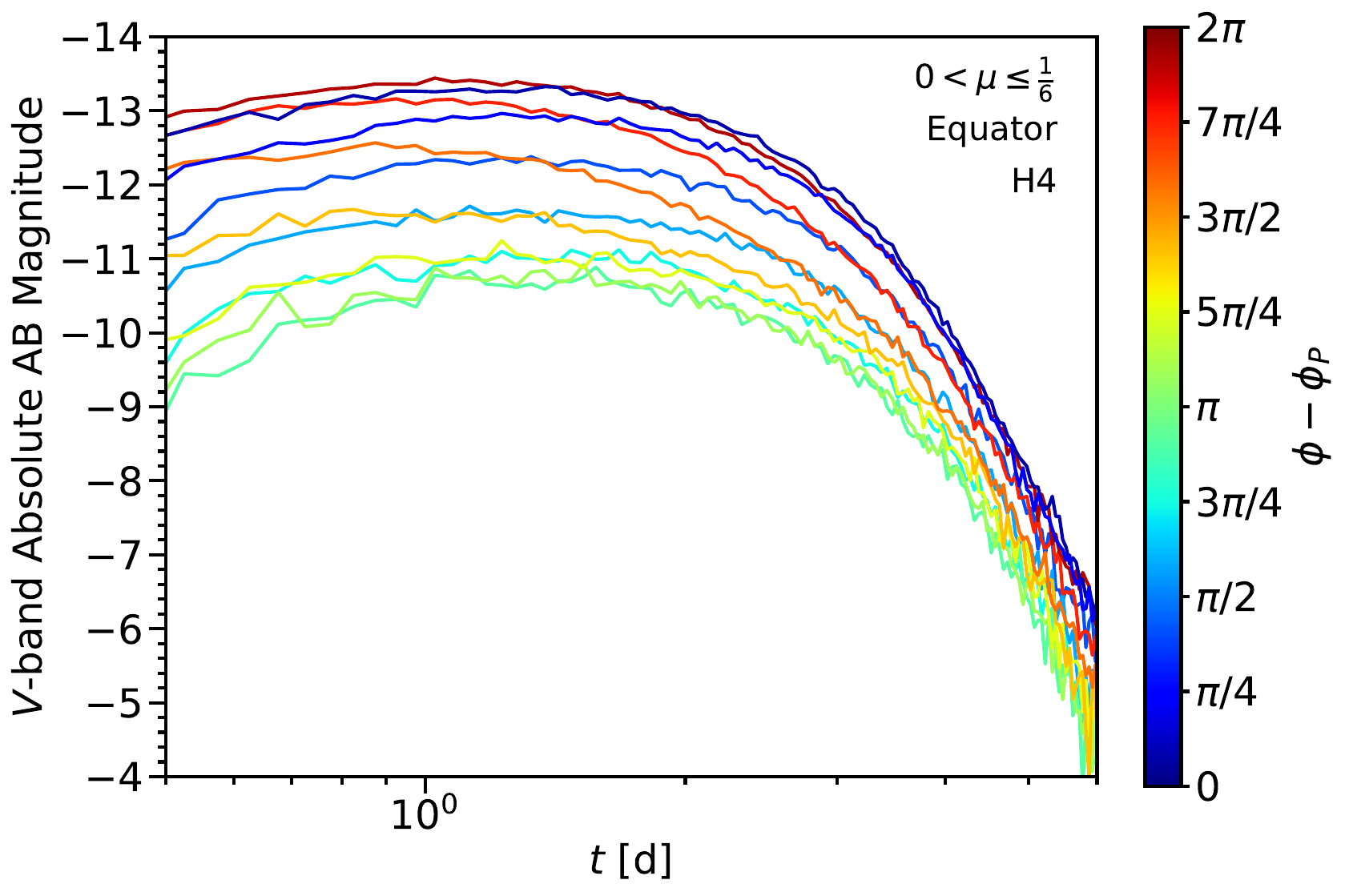}{0.49\textwidth}{(c) $V$-band, $0 < \mu \leq \frac{1}{6}$ (Equator)}
\fig{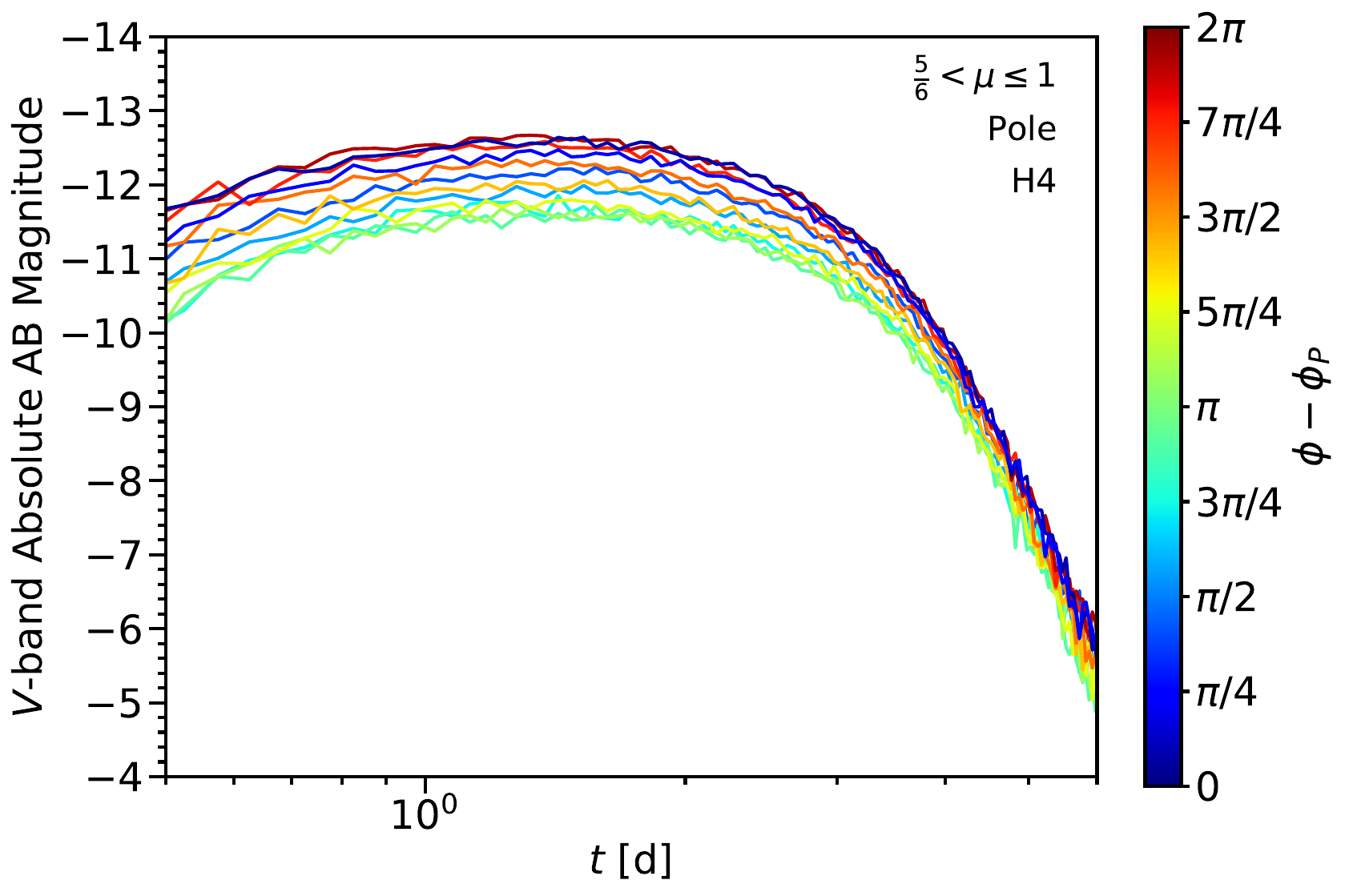}{0.49\textwidth}{(d) $V$-band, $\frac{5}{6} < \mu \leq 1$ (Pole)}
}
\caption{Isotropic-equivalent broadband light curves in the optical/UV bands for the heating model H4. The magnitudes are Absolute AB magnitudes. The rows correspond to the $B$ (top) and $V$ (bottom) bands. The columns correspond to different bins of $\mu = \cos\theta$, where $\theta$ is the polar angle. The color bar shows the centers of the 
$\phi - \phi_P$ bins. In each panel, the brightest curve corresponds to the azimuthal direction $\phi_P \simeq 5.2$ of the total momentum and the dimmest curve corresponds to the opposite direction $\phi_P - \pi$. The light curves are roughly the same for $\mu \rightarrow -\mu$ because the ejecta are roughly symmetric about $z = 0$ after the active rotation in stage (1) (Section \ref{subsec:numerical_relativity}).}
\label{fig:broadband_light_curves_optical_UV}
\end{figure*}

In this appendix, we present the broadband light curves for a larger set of filters and polar angles. The figures show broadband light curves in IR bands (Figure \ref{fig:broadband_light_curves_IR}), optical/IR bands (Figure \ref{fig:broadband_light_curves_optical_IR}), and optical/UV bands (Figure \ref{fig:broadband_light_curves_optical_UV}).

\section{Light Curve Peaks}
\label{sec:light_curve_peaks}

\begin{figure*}
\gridline{
\fig{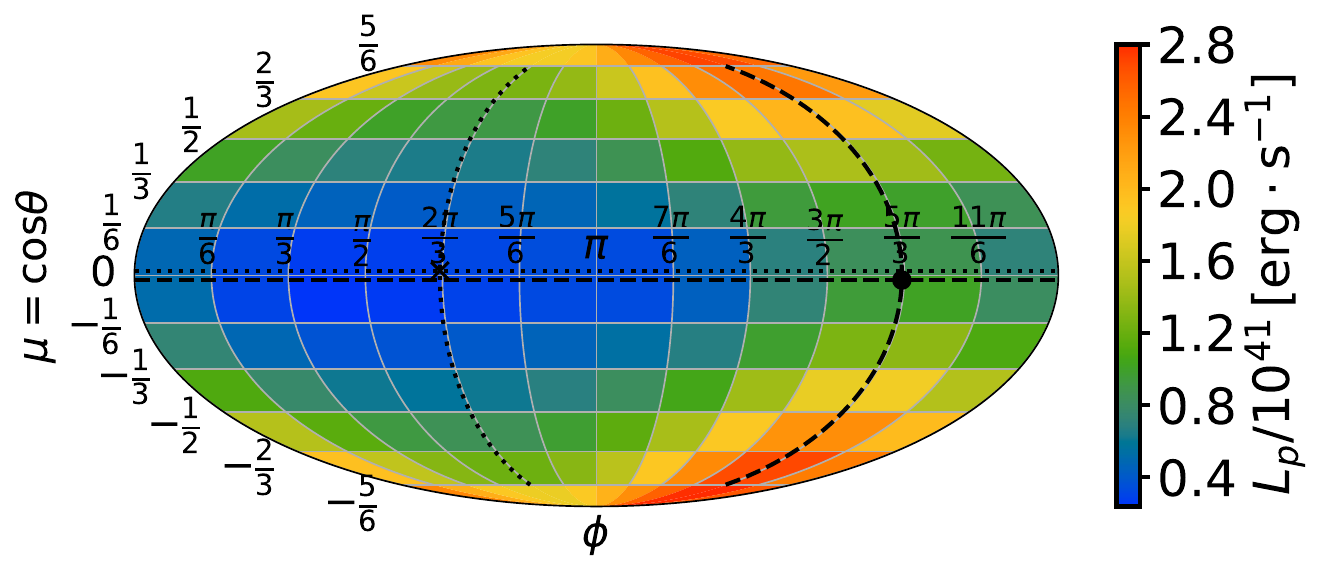}{0.32\textwidth}{(a) $L_p$, H0}
\fig{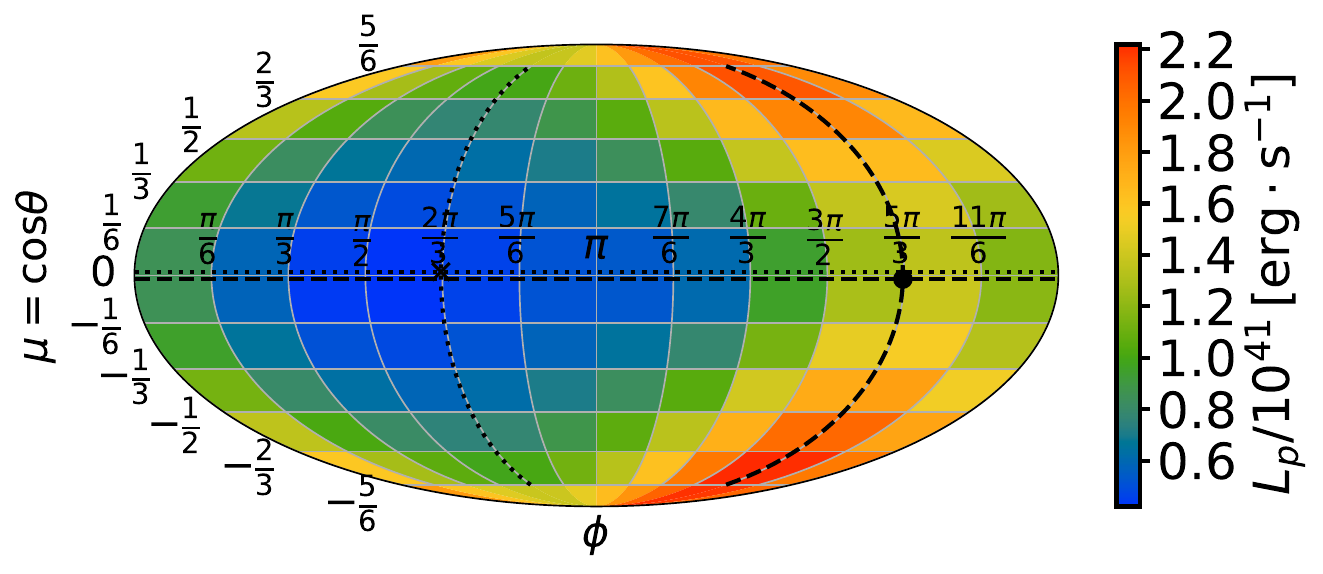}{0.32\textwidth}{(b) $L_p$, H3}
\fig{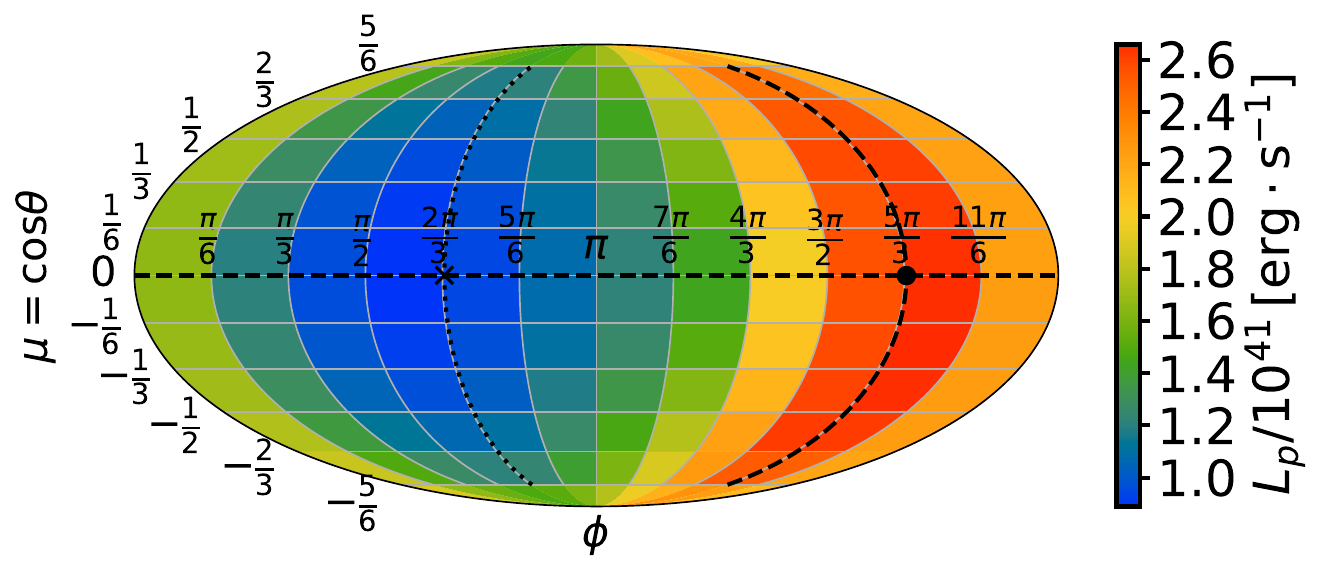}{0.32\textwidth}{(c) $L_p$, H4}
}
\gridline{
\fig{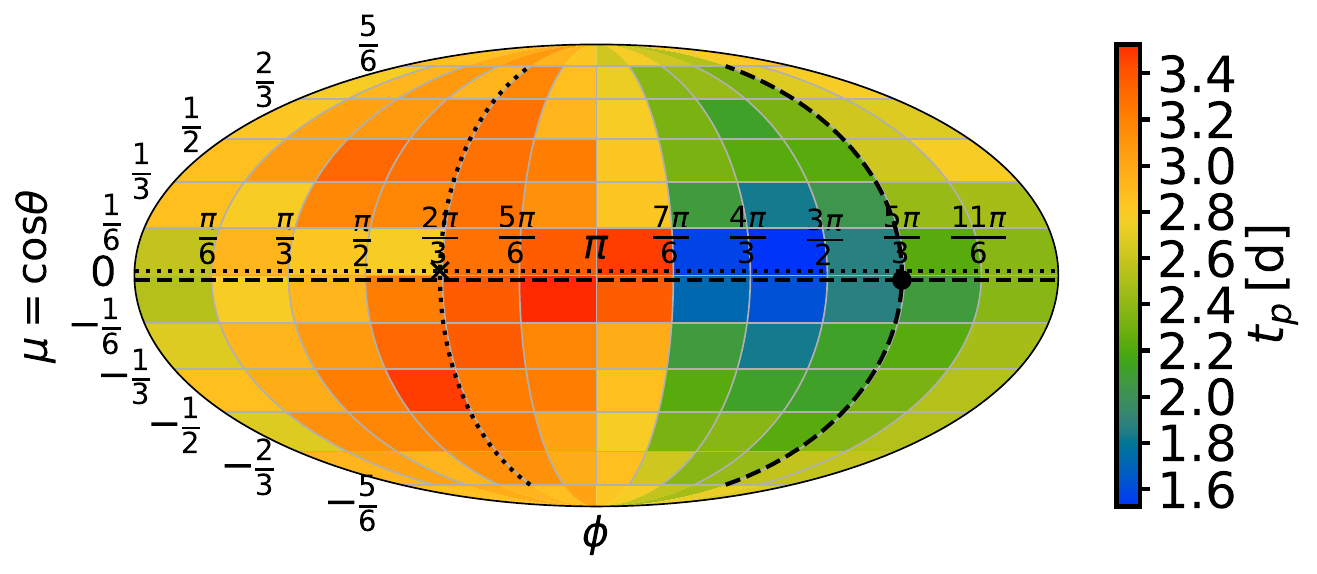}{0.32\textwidth}{(d) $t_p$, H0}
\fig{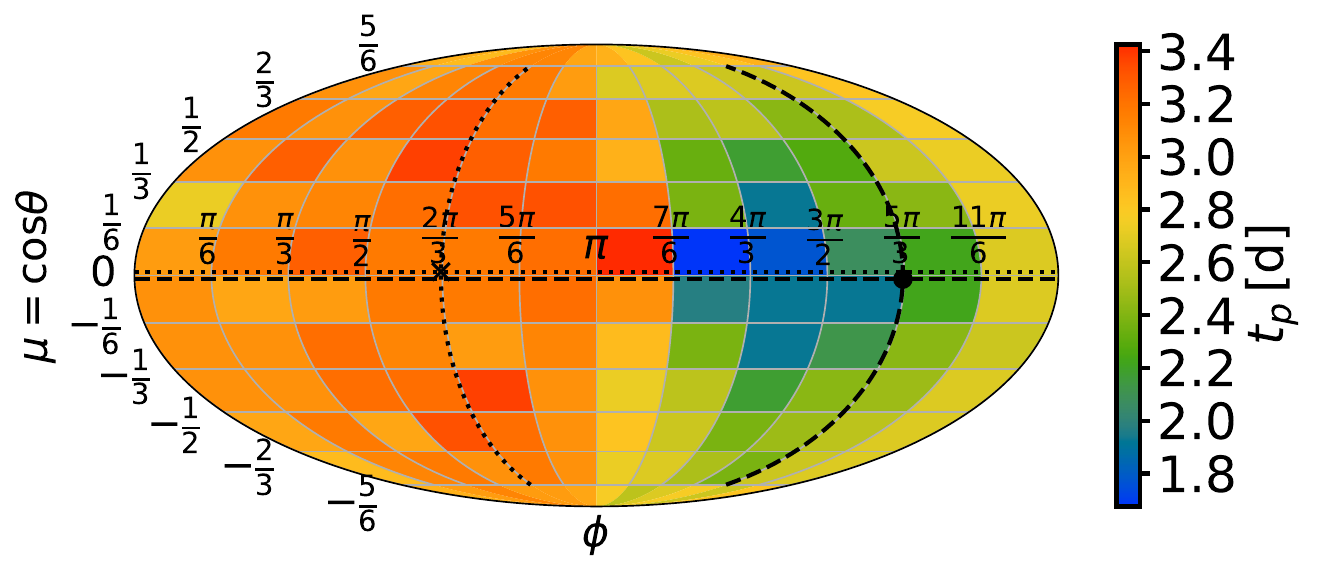}{0.32\textwidth}{(e) $t_p$, H3}
\fig{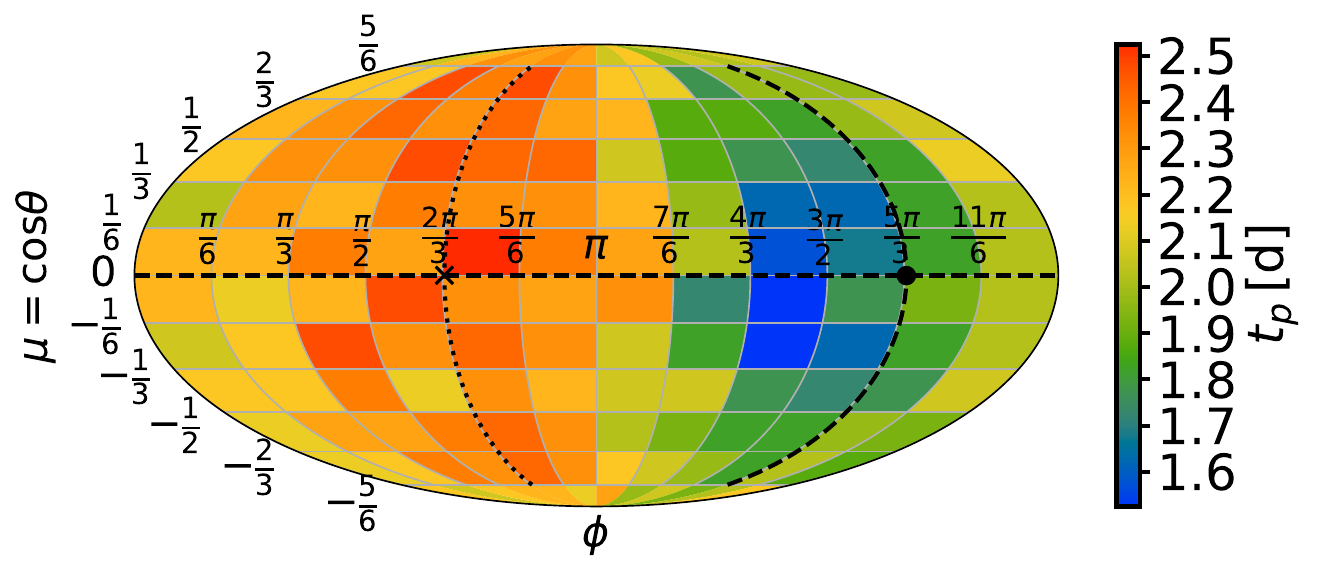}{0.32\textwidth}{(f) $t_p$, H4}
}
\caption{
The peak luminosities and times of the isotropic-equivalent bolometric light curves at each viewing angle. 
The rows show the peak luminosity $L_p$ (top) and time-to-peak $t_p$ (bottom). The columns show the stage (2) heating models H0 (left), H3 (middle), and H4 (right) (Table \ref{tab:heating_models}). The black dot shows the direction $(\mu_P, \phi_P) \simeq (-0.030, 5.2)$ of the total momentum $P_i = \sum_{j=1}^N m S_{(j)i}$, and the black cross shows the opposite direction.}
\label{fig:Lpeak_grids}
\end{figure*}

\begin{figure*}
\gridline{
\fig{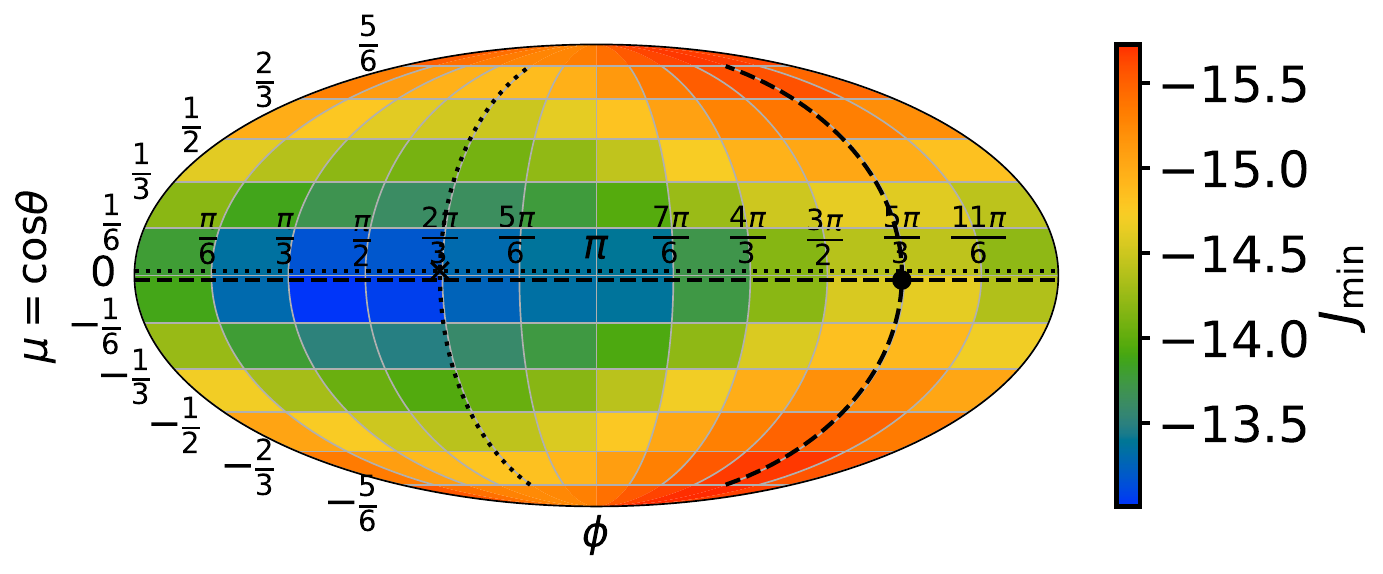}{0.32\textwidth}{(a) $J_\mathrm{min}$, H0}
\fig{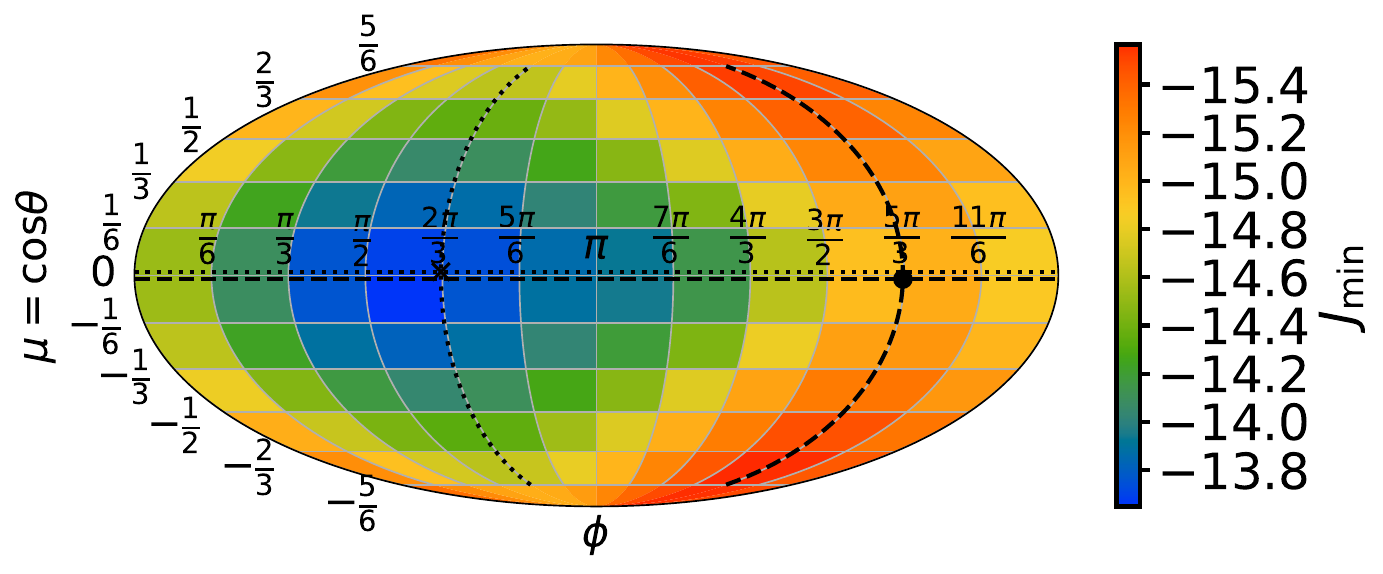}{0.32\textwidth}{(b) $J_\mathrm{min}$, H3}
\fig{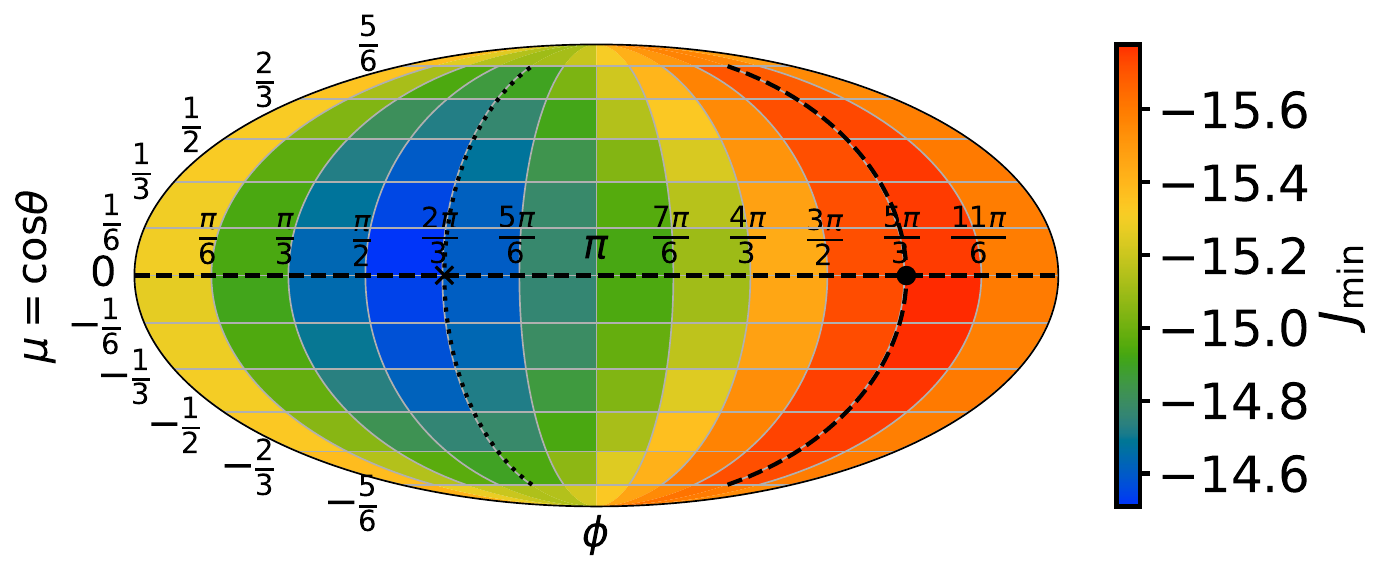}{0.32\textwidth}{(c) $J_\mathrm{min}$, H4}
}
\gridline{
\fig{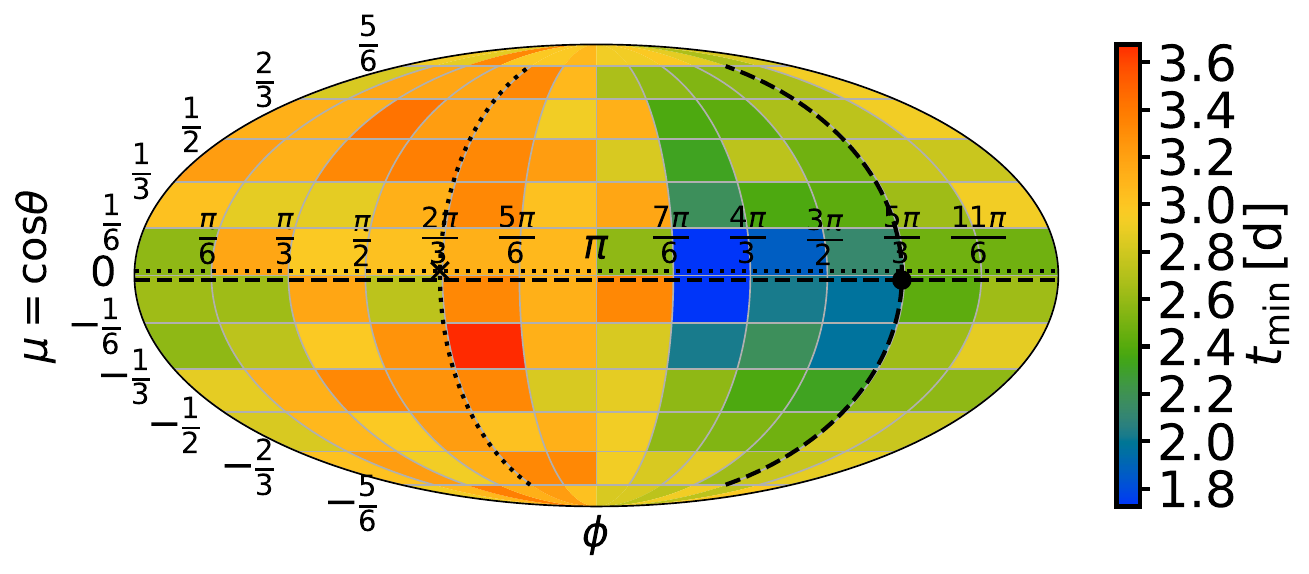}{0.32\textwidth}{(d) $t_\mathrm{min}$, H0}
\fig{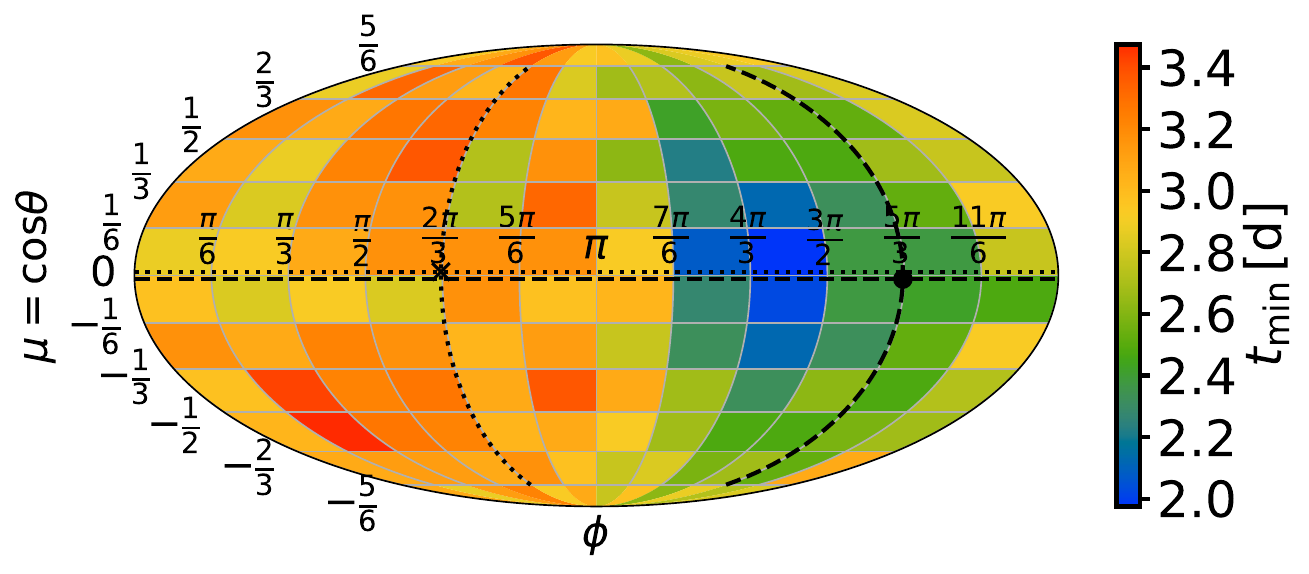}{0.32\textwidth}{(e) $t_\mathrm{min}$, H3}
\fig{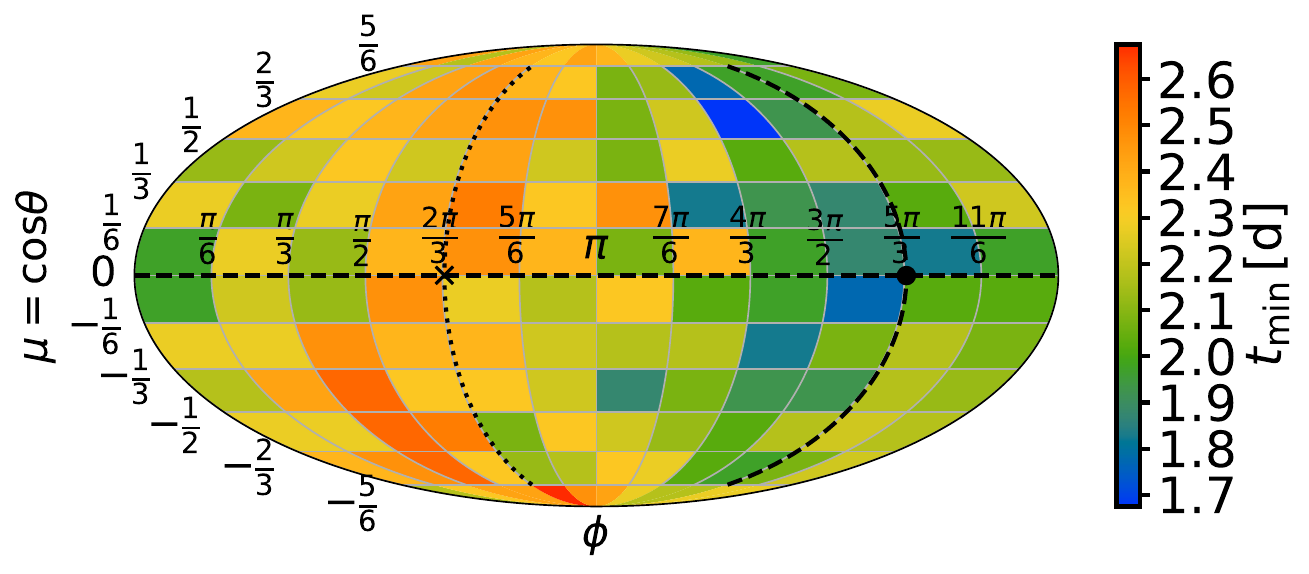}{0.32\textwidth}{(f) $t_\mathrm{min}$, H4}
}
\caption{
The peak magnitudes and times of the $J$-band light curves at each viewing angle. 
The magnitudes are Absolute AB magnitudes. The rows show the peak $J$-band magnitude $J_\mathrm{min}$ (top) and time-to-peak $t_\mathrm{min}$ (bottom). The columns show the stage (2) heating models H0 (left), H3 (middle), and H4 (right) (Table \ref{tab:heating_models}). The black dot shows the direction $(\mu_P, \phi_P) \simeq (-0.030, 5.2)$ of the total momentum $P_i = \sum_{j=1}^N m S_{(j)i}$, and the black cross shows the opposite direction.}
\label{fig:Jmin_grids}
\end{figure*}

\begin{figure*}
\gridline{
\fig{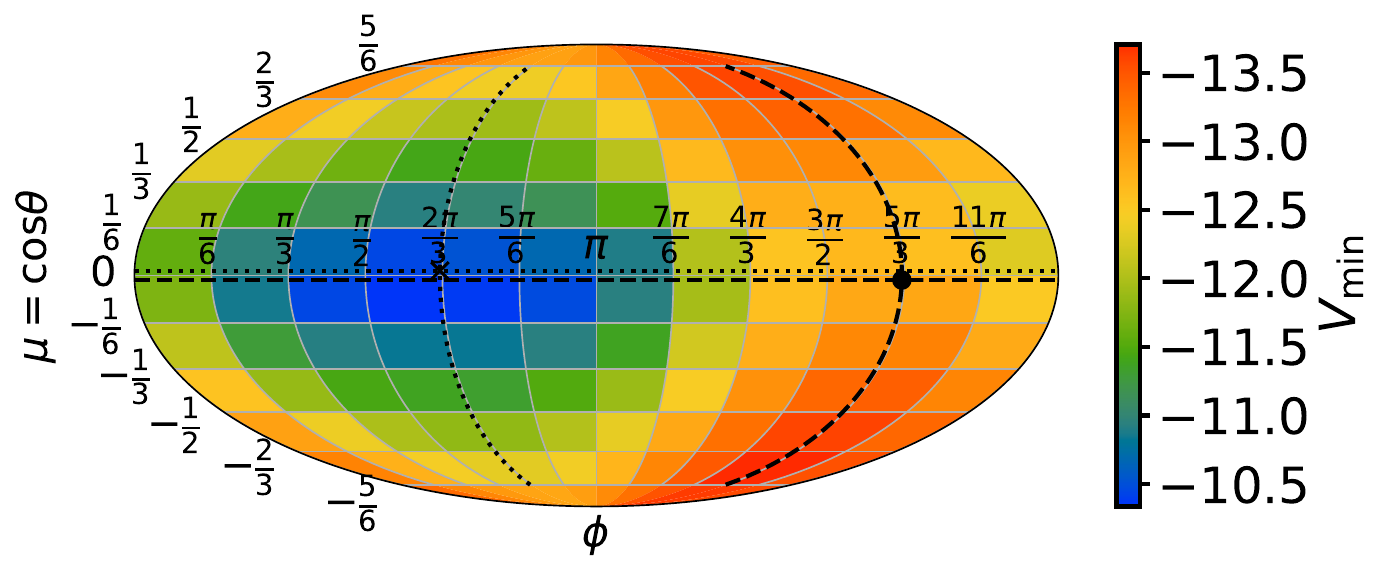}{0.32\textwidth}{(a) $V_\mathrm{min}$, H0}
\fig{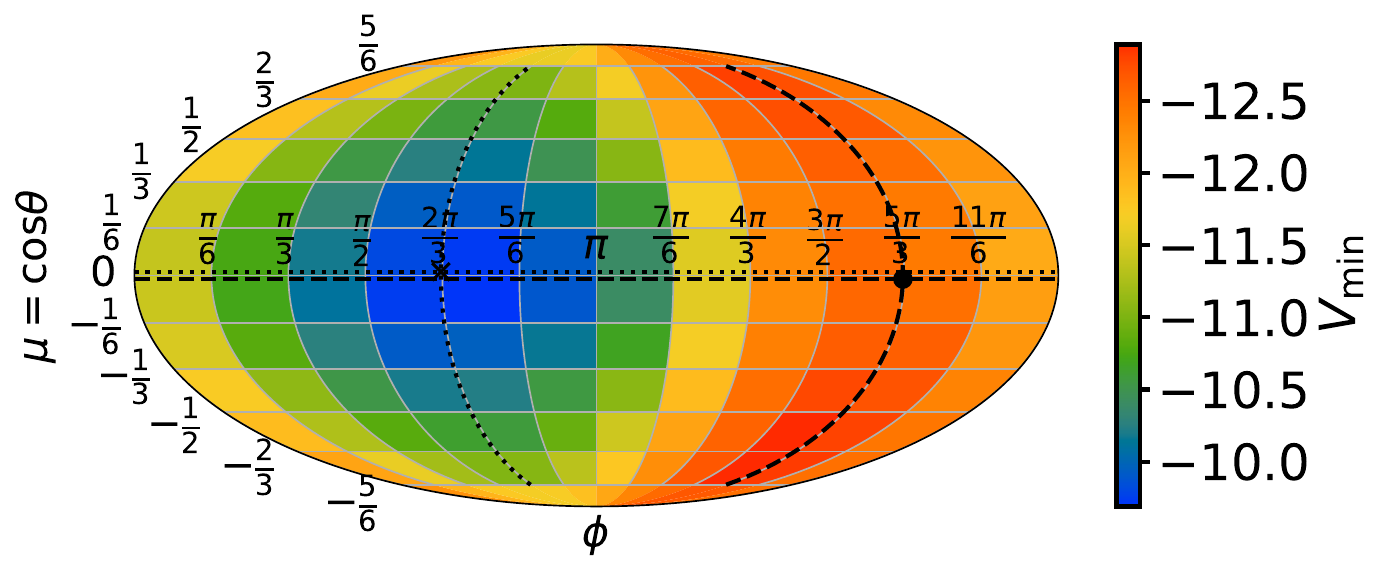}{0.32\textwidth}{(b) $V_\mathrm{min}$, H3}
\fig{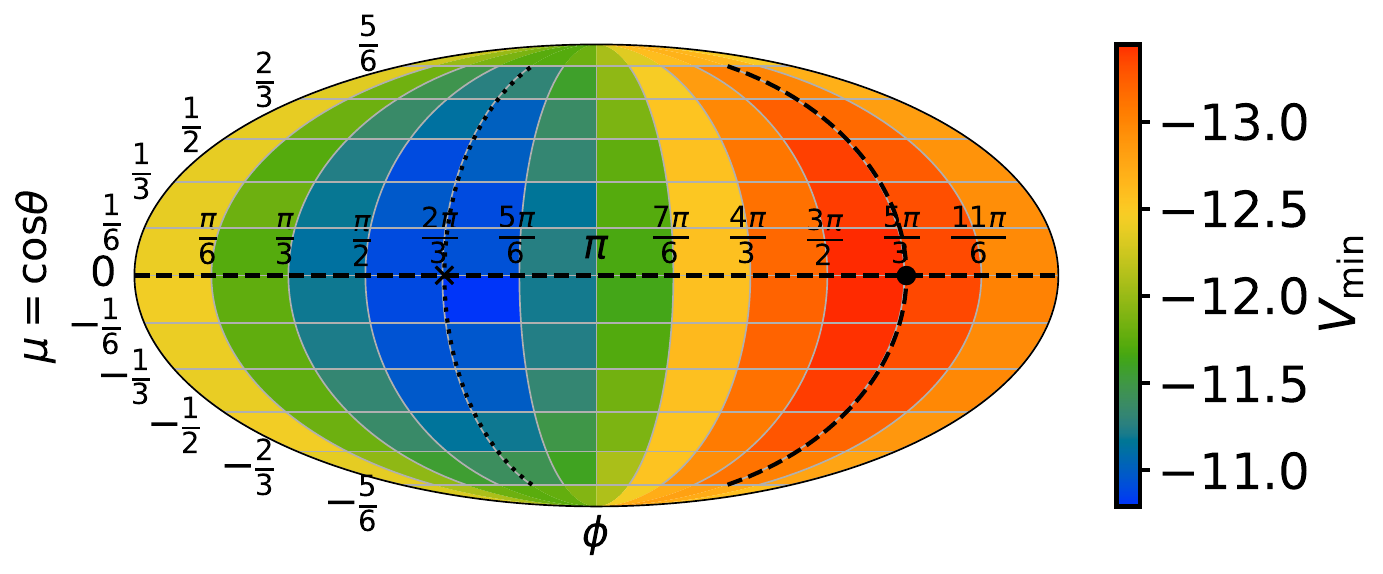}{0.32\textwidth}{(c) $V_\mathrm{min}$, H4}
}
\gridline{
\fig{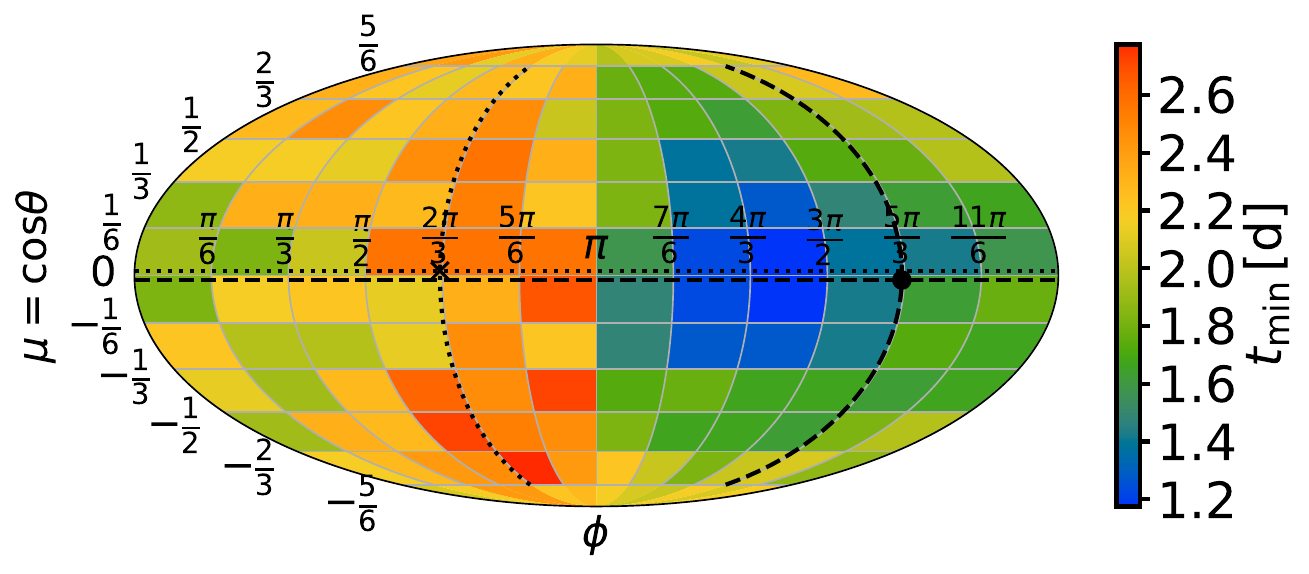}{0.32\textwidth}{(d) $t_\mathrm{min}$, H0}
\fig{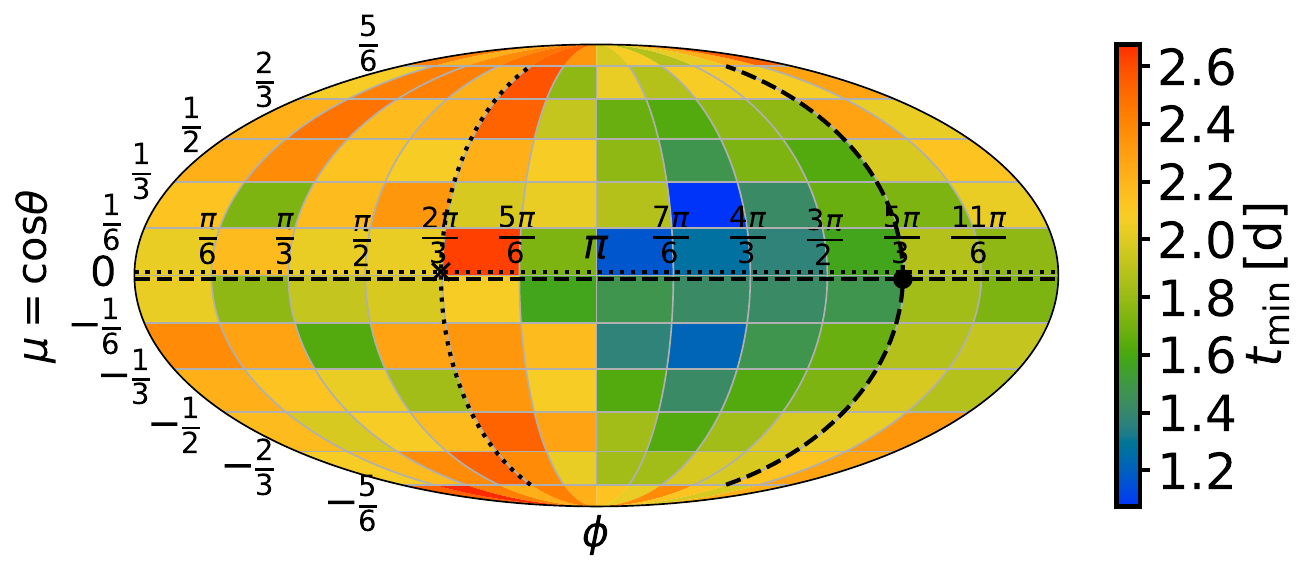}{0.32\textwidth}{(e) $t_\mathrm{min}$, H3}
\fig{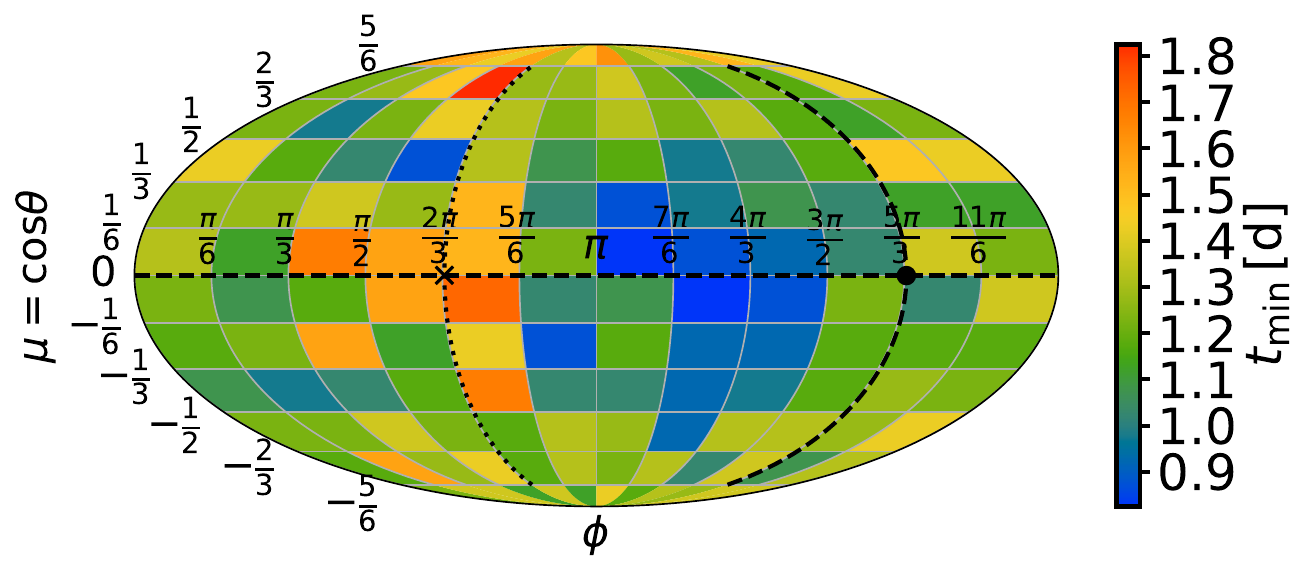}{0.32\textwidth}{(f) $t_\mathrm{min}$, H4}
}
\caption{
The peak magnitudes and times of the $V$-band light curves at each viewing angle. 
The magnitudes are Absolute AB magnitudes. The rows show the peak $V$-band magnitude $V_\mathrm{min}$ (top) and time-to-peak $t_\mathrm{min}$ (bottom). The columns show the stage (2) heating models H0 (left), H3 (middle), and H4 (right) (Table \ref{tab:heating_models}). The black dot shows the direction $(\mu_P, \phi_P) \simeq (-0.030, 5.2)$ of the total momentum $P_i = \sum_{j=1}^N m S_{(j)i}$, and the black cross shows the opposite direction.}
\label{fig:Vmin_grids}
\end{figure*}

In this appendix, we present the peaks of the bolometric and broadband light curves. The figures show the peaks of the bolometric light curves (Figure \ref{fig:Lpeak_grids}), the peak magnitudes of the $J$-band light curves (Figure \ref{fig:Jmin_grids}), and the peak magnitudes of the $V$-band light curves (Figure \ref{fig:Vmin_grids}).

\bibliographystyle{aasjournal}
\bibliography{references} 



\end{document}